\documentclass[11pt]{article}
\usepackage[final]{acl}
\usepackage{times}

\usepackage[T1]{fontenc}
\usepackage[utf8]{inputenc} 
\usepackage{lipsum} 

\usepackage{etoolbox}
\newbool{includeappendix}
\setbool{includeappendix}{true}

\ifdefined\isoverfull
\else
\fi

\usepackage{xcolor} 

\definecolor{my-full-blue}{HTML}{1F77B4}

\definecolor{my-full-orange}{HTML}{FF7F0E}

\definecolor{my-full-green}{HTML}{2CA02C}

\definecolor{my-full-red}{HTML}{d62728}

\definecolor{my-full-purple}{HTML}{9467bd}

\colorlet{my-blue}{my-full-blue!30}
\colorlet{my-orange}{my-full-orange!30}
\colorlet{my-green}{my-full-green!30}
\colorlet{my-red}{my-full-red!30}
\colorlet{my-purple}{my-full-purple!30}

\usepackage{listings}

\usepackage{textcomp}

\usepackage{xcolor}

\usepackage[scaled=0.8]{beramono}

\definecolor{ckeyword}{HTML}{7F0055}
\definecolor{ccomment}{HTML}{3F7F5F}
\definecolor{cstring}{HTML}{2A0099}

\lstdefinestyle{numbers}{
	numbers=left,
	framexleftmargin=20pt,
	numberstyle=\tiny,
	firstnumber=auto,
	numbersep=1em,
	xleftmargin=2em
}

\lstdefinestyle{layout}{
	frame=none,
	captionpos=b,
}

\lstdefinestyle{comment-style}{
	morecomment=[l]//,
	morecomment=[s]{/*}{*/},
	commentstyle={\color{ccomment}\itshape},
}

\lstdefinestyle{string-style}{
	morestring=[b]",%
	morestring=[b]',%
	stringstyle={\color{cstring}},
	showstringspaces=false,%
}

\lstdefinestyle{keyword-style}{
	keywordstyle={\ttfamily\bfseries},
	morekeywords={
		function,
		constructor,
		int,
		bool,
		return,
		returns,
		uint
	},
	morekeywords = [2]{},
	keywordstyle = [2]{\text},
	sensitive=true,
}

\lstdefinestyle{input-encoding}{
	inputencoding=utf8,
	extendedchars=true,
	literate=
	{ℝ}{$\reals$}1%
	{→}{$\rightarrow$}1%
	{α}{$\alpha$}1%
	{β}{$\beta$}1%
	{λ}{$\lambda$}1%
	{θ}{$\theta$}1%
	{ϕ}{$\phi$}1%
}

\lstdefinestyle{escaping}{
	moredelim={**[is][\color{blue}]{\%}{\%}},
	escapechar=|,
	mathescape=true
}

\lstdefinestyle{default-style}{
	basicstyle=\fontencoding{T1}\ttfamily\footnotesize,
	style=numbers,
	style=layout,
	style=comment-style,
	style=string-style,
	style=keyword-style,
	style=input-encoding,
	style=escaping,
	tabsize=2,
	upquote=true
}

\lstdefinelanguage{BASIC}{
	language=C++,
	style=default-style
}[keywords,comments,strings]%

\usepackage{microtype}
\usepackage{inconsolata}
\usepackage{graphicx}
\usepackage{subcaption}
\usepackage{booktabs} 

\usepackage{hyperref}

\usepackage[T1]{fontenc}
\usepackage{amsfonts}       
\usepackage{nicefrac}       
\usepackage{url}
\usepackage{xurl}
\usepackage{xcolor}
\usepackage{xspace}
\usepackage{etoolbox}
\usepackage{fontawesome}
\usepackage{enumitem}
\usepackage{amsmath,amssymb}
\usepackage{array}
\usepackage{multirow}
\usepackage{wrapfig}
\usepackage{adjustbox}
\usepackage{siunitx}
\usepackage{caption}
\usepackage{fontawesome}
\usepackage{stackengine}
\usepackage{svg}
\usepackage{mathtools}
\usepackage{xspace}
\usepackage{wrapfig}
\usepackage{makecell}
\usepackage{longtable}
\usepackage[most]{tcolorbox}
\usepackage{diagbox}
\usepackage{ulem}
\usepackage[capitalize,noabbrev]{cleveref}

\usepackage{colortbl}

\usepackage{tikz}

\usetikzlibrary{calc,decorations,decorations.pathmorphing}
\usetikzlibrary{positioning,fit,arrows}
\usetikzlibrary{decorations.markings}
\usetikzlibrary{shapes,shapes.geometric}
\usetikzlibrary{shadows,patterns,snakes}
\usetikzlibrary{backgrounds,decorations.pathreplacing,calligraphy,automata}
\usetikzlibrary{intersections}
\usetikzlibrary{angles,quotes}
\usetikzlibrary{plotmarks}
\usetikzlibrary{patterns}
\usetikzlibrary{arrows.meta}
\usetikzlibrary{shapes.misc}
\usetikzlibrary{chains}
\usetikzlibrary{shapes.callouts}

\usepackage{amsthm}
\usepackage{stmaryrd}
\usepackage{algorithm}
\usepackage{algpseudocode}

\usepackage{tikz}

\theoremstyle{plain}

\theoremstyle{definition}

\theoremstyle{remark}

\newcolumntype{x}[2]{S[table-format=#1.#2,table-auto-round]}
\newcolumntype{M}{>{$}l<{$}}

\renewcommand{\emph}[1]{\textit{#1}}

\newcommand{\bench}{\textsc{SABER-Math}\xspace}

\newcommand{\numina}{\textsc{NuminaMath-1.5}\xspace}

\DeclareRobustCommand*\bluecircled[1]{\tikz[baseline=(char.base)]{
		\node[shape=circle, fill=cyan!70!blue, text=white, inner sep=0.3pt, font=\sffamily\bfseries] (char) {#1};}}

\definecolor{acceptblue}{HTML}{6494EA}
\hypersetup{citecolor=acceptblue}

\definecolor{lightred}{HTML}{ffcbc7}
\definecolor{gemini}{HTML}{4285F4}
\definecolor{claude}{HTML}{f3e9d7}
\definecolor{deepseek}{HTML}{FADA4B}
\definecolor{qwen}{HTML}{FA574B}
\definecolor{oai}{HTML}{10a37f}
\definecolor{LightGreen}{HTML}{CCFFCC}

\lstdefinestyle{promptstyle}{
	breaklines=true,
	basicstyle=\scriptsize\ttfamily,
	numbers=none,
	language={},
	framextopmargin=0pt,
	framexbottommargin=0pt,
	breakindent=0pt,
	showspaces=false,
	keywordstyle=\bfseries,
	showstringspaces=false,
	columns=fullflexible,
	keepspaces=true,
	mathescape=false,
	texcl=false,
	escapechar=,
	escapeinside={(*@}{@*)},
	morekeywords={Answer},
	moredelim=**[is][\bfseries]{!!}{!!},
	literate={|}{{\char`\|}}1
}

\newtcblisting{prompt}[2][]{
	arc=3pt,
	outer arc=3pt,
	width=\linewidth,
	left=1mm,
	top=0mm,
	bottom=0mm,
	title=#2,
	colback=lightred!5!white,
	colframe=black,
	fonttitle=\bfseries,
	listing only,
	listing options={style=promptstyle},
	breakable,
	#1
}

\newtcblisting{smallprompt}[2][]{
    arc=3pt, outer arc=3pt,
    width=0.9\linewidth,
    left=1mm,
    top=0mm,
    bottom=0mm,
    title=#2,
    colback=lightred!5!white,
    colframe=black,
    fonttitle=\bfseries,
    listing only,
    listing options={style=promptstyle},
    breakable,
    #1
}

\newtcblisting{response}[2][]{
    arc=3pt, outer arc=3pt,
    width=0.9\linewidth,
    left=1mm,
    top=0mm,
    bottom=0mm,
    title=#2,
    colback=lightred!5!white,
    colframe=oai,
    fonttitle=\bfseries,
    listing only,
    listing options={style=promptstyle},
    breakable,
    #1
}

\newcommand{\bifront}[1]{\underline{#1}}
\newcommand{\crossfront}[1]{\textit{#1}}

\usepackage{amsmath,amsfonts,bm}

\def\eqref#1{equation~\ref{#1}}

\def\1{\bm{1}}

\DeclareMathAlphabet{\mathsfit}{\encodingdefault}{\sfdefault}{m}{sl}
\SetMathAlphabet{\mathsfit}{bold}{\encodingdefault}{\sfdefault}{bx}{n}

\usepackage[capitalize]{cleveref}

\crefformat{section}{\S#2#1#3}

\crefrangeformat{section}{\S#3#1#4\crefrangeconjunction\S#5#2#6}

\crefmultiformat{section}{\S#2#1#3}{\crefpairconjunction\S#2#1#3}{\crefmiddleconjunction\S#2#1#3}{\creflastconjunction\S#2#1#3}

\newcommand{\crefrangeconjunction}{--}

\crefname{listing}{Lst.}{listings}
\crefname{line}{Lin.}{Lin.}
\crefname{appendix}{App.}{App.}

\newcommand{\app}[1]{%
	\ifbool{includeappendix}{\cref{#1}}{the appendix}%
}
\newcommand{\App}[1]{%
	\ifbool{includeappendix}{\cref{#1}}{The appendix}%
}

\title{\bench: An Automated Reranking Benchmark for Mathematical Information Retrieval}

\author{
	\textbf{Nikolay Georgiev\textsuperscript{1,*}},
	\textbf{Maria Drencheva\textsuperscript{1,*}},
	\textbf{Kseniia Ibragimova\textsuperscript{1,\dag}},
	\\
	\textbf{Ivo Petrov\textsuperscript{1}},
	\textbf{Dimitar I. Dimitrov\textsuperscript{1}},
	\textbf{Martin Vechev\textsuperscript{1,2}}
	\\
	\\
	\textsuperscript{1}INSAIT, Sofia University ``St. Kliment Ohridski'',
	\textsuperscript{2}ETH Zurich
	\\
	\small{
		\textsuperscript{*}Equal contribution. \quad
		\textsuperscript{\dag}Work done during an internship at INSAIT, Sofia University ``St. Kliment Ohridski''.
	}
	\\
	\small{
		\textbf{Correspondence:} \href{mailto:ivo.petrov@insait.ai}{ivo.petrov@insait.ai}
	}
	\\\\
	{\normalfont\small
		\raisebox{-0.16em}{\includegraphics[height=1em]{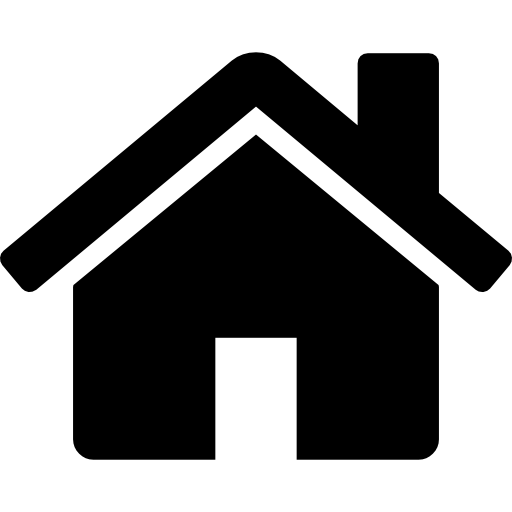}}~
		\href{https://sabermath.insait.ai}{Website}
		\quad
		\raisebox{-0.16em}{\includegraphics[height=1em]{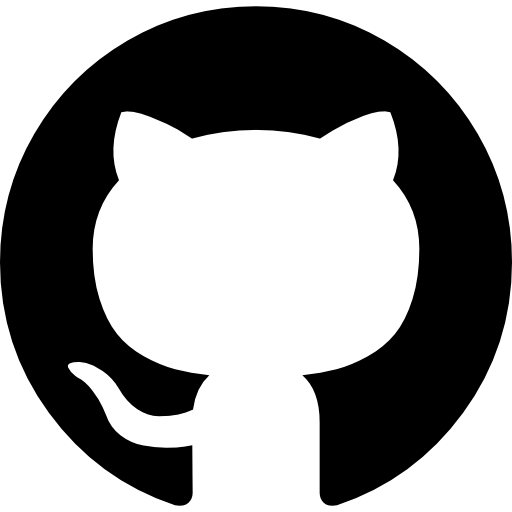}}~
		\href{https://github.com/insait-institute/sabermath/}{GitHub}
		\quad
		\raisebox{-0.16em}{\includegraphics[height=1em]{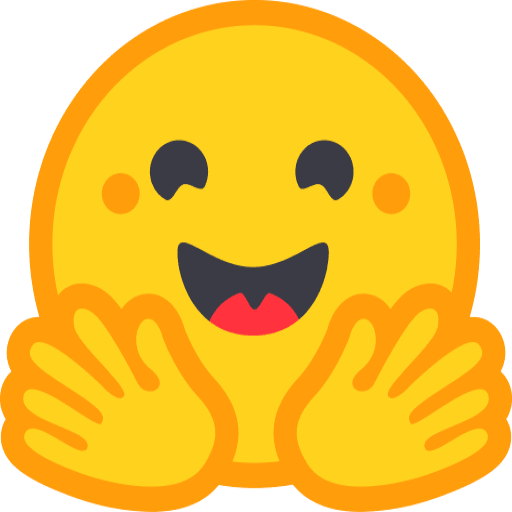}}~
		\href{https://huggingface.co/collections/INSAIT-Institute/saber-math}{Hugging Face}
	}
}

\begin{document}
	\maketitle

\begin{abstract}
	
	As agentic AI systems tackle more complex mathematical tasks, they increasingly rely on information retrieval (IR) to search problem databases, theorem libraries, and educational resources. However, choosing the right retriever remains difficult, as it is infeasible to isolate its effect on downstream performance. On the other hand, existing retrieval-specific benchmarks often fail to capture fine-grained mathematical relevance, penalizing systems that retrieve useful documents. We address this gap with \bench, the first fully automated reranking benchmark for mathematical IR, built without expert annotation. Starting from 283K high-school-level problems with solutions, it builds challenging tasks in three steps: (i) LLMs extract concise solution summaries and topics for each problem; (ii) candidates are discovered per query using ontology-based topic similarity and lexical overlap between solution summaries; and (iii) a Swiss-style LLM preference tournament assigns every candidate a graded relevance rating. Over these judged candidate pools, we evaluate retrievers across all major paradigms, from lexical and math-specific systems to modern embedding, late-interaction, and reranking models. Reasoning embedders lead, followed by general-purpose models, both far outperforming classical and math-specific baselines. Yet, all still struggle in symbol-heavy domains such as Algebra and Calculus. Importantly, we show that generic IR benchmarks, such as MTEB, track mathematical performance broadly but are less discriminative among the strongest models, highlighting the need for domain-specific evaluation.
	
\end{abstract}
\begin{figure*}[t]
	\centering
	\hspace*{-0.22cm}%
	\includegraphics[width=1.025\linewidth]{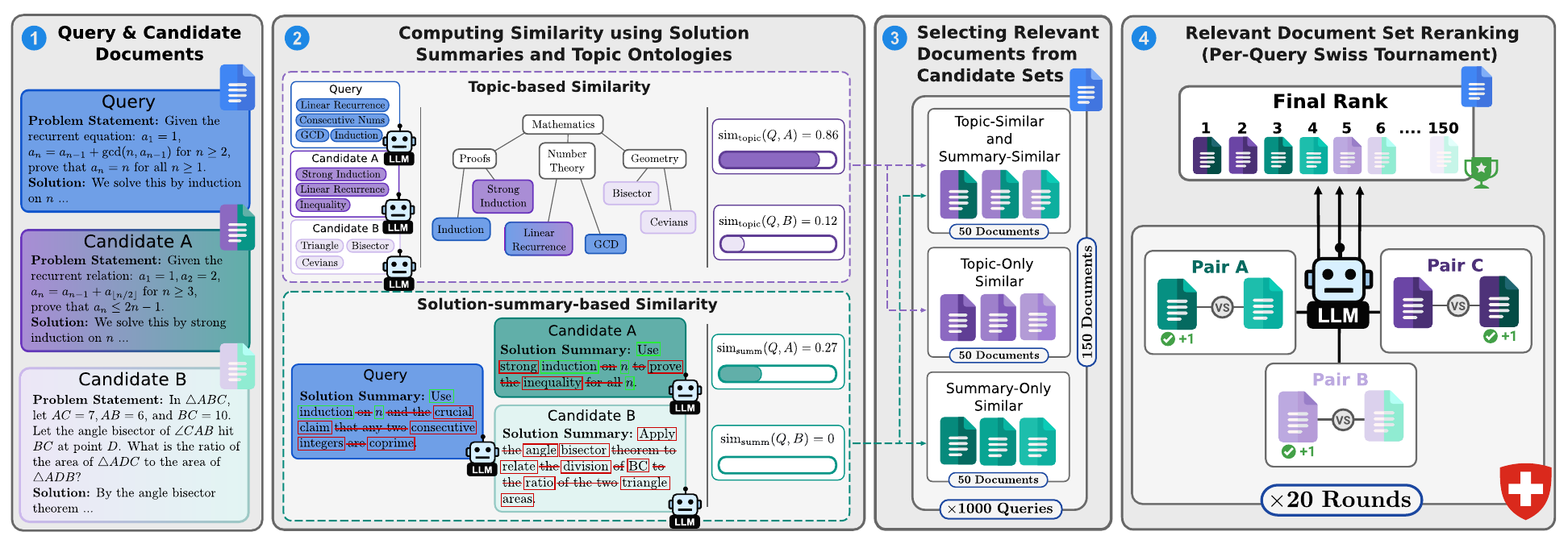}
	\vspace{-2em}
	\caption{Overview of \bench construction. First, we source a large mathematical corpus of problems and their solutions \bluecircled{1}. Using LLM annotation, we compute two separate relevance signals for every document pair based on topics and solution summaries \bluecircled{2}. We then select 1000 target query documents, each with 150 relevant documents split evenly across relevance signals \bluecircled{3}. Finally, we rank the candidates per query using an LLM-as-a-judge in a 20-round Swiss tournament based on a	 Bradley-Terry model \bluecircled{4}. Throughout the figure our query documents are in blue, saturated purple denotes topic relevance, saturated teal denotes solution-summary relevance, and muted colors denote irrelevant documents.  }
	\label{fig:overview}
	\vspace{-1em}
\end{figure*}

\vspace{-3mm}
\section{Introduction}
\vspace{-1mm}

Large language models are increasingly used in mathematical reasoning systems, serving both as natural-language assistants and components of formal reasoning pipelines~\citep{reasoningsurvey,reasoningsurvey2}. To improve their effectiveness and reliability, modern agentic systems often supplement proof generation with retrieval of relevant mathematical knowledge. Information retrieval (IR)~\citep{irsurvey} helps these systems find analogous examples, relevant theorems, solution patterns, formal premises, and definitions that can guide multi-step reasoning, making retrieval a core component of mathematical AI systems.

\paragraph{Information retrieval for mathematics}

Retrieval has become important in both formal theorem proving and in informal mathematical problem solving. In formal settings, IR is used to expose relevant definitions, lemmas, declarations~\citep{yang2023leandojo,shen2025real,kozyrev2025rocqstar}, library code~\citep{gao2024semantic,shen2025real}, and previously written proofs to assist a prover or autoformalizer~\citep{liu2025rethinking}. In informal settings, it is used to ground answers in trusted mathematical material and to provide analogous examples, solution schemas, or theorem-level context~\citep{levonian2023retrieval, yang2024analogy}. Although these systems differ in what they retrieve and how they retrieve it, they share a common requirement: retrieved documents should be useful for mathematical reasoning, rather than simply textually similar to the query. Yet, such systems rarely justify their choice of retrieval framework, making rigorous evaluation of mathematical retrieval necessary for building more effective mathematical agents.

\paragraph{Challenges of evaluating mathematical retrieval}
Evaluating mathematical retrieval is challenging because related mathematical problems are relatively infrequent, and their relevance is often not captured by surface similarity. Two problems may use different notation, wording, or terminology while relying on the same method, e.g., an extremal argument, invariant, or auxiliary construction. Conversely, problems with similar objects or formulas may require very different reasoning. Prior work~\citep{bright,xiao2024rar,mathnet} has therefore relied on expert annotation of semantic relevance, but such annotation is costly, hard to scale, and prone to annotator disagreement. Moreover, to keep relevance assessment tractable, many benchmarks use narrow relevance definitions and low-granularity labels~\citep{bright,mirb,mathnet}. This penalizes genuinely useful retrievals that do not match the benchmark's specific notion of relevance. All in all, existing methods thus remain hard to scale to larger corpora and diverse mathematical domains, while preserving a fine-grained notion of relevance.

\paragraph{SABER-Math}
To address this gap, we introduce \textbf{\bench}, the \textbf{S}calable \textbf{A}utomated \textbf{BE}nchmark for \textbf{R}eranking in \textbf{Math}. To our knowledge, it is the first fully-automated benchmark for fine-grained evaluation of mathematical information retrieval without expert annotation.

\paragraph{Scalable relevant problem pair discovery}
To address the issues of rare relevant problem pairs, \bench starts from a large corpus of $283$K high-school and olympiad-style problem-solution pairs and uses two scalable complementary relevance signals to find relevant problem pairs. First, we assign each problem topics in a mathematical ontology and use topic overlap to find problems involving related concepts, theorems, or domains. Second, we extract short LLM-generated summaries of each solution's core idea and use lexical overlap between summaries to find problems with similar solution methods. These signals are not used as final relevance labels. Instead, we use them only for candidate selection, identifying $1000$ query problems across Algebra, Geometry, Number Theory, Combinatorics, and Calculus, which we pair each with $150$ candidate documents spanning different forms of mathematical relatedness.

\paragraph{Automated relevance estimation}
After constructing each query's candidate set, \bench assigns fine-grained relevance labels for each candidate using pairwise LLM judgments. To make this scalable, we avoid exhaustive all-pairs comparison and instead use a Swiss-style tournament, which prioritizes comparisons between candidates of similar estimated relevance. We aggregate the resulting preferences with a Bradley--Terry model to obtain continuous relevance scores, which serve as ground-truth ratings for a reranking task that serves as our benchmark.

\paragraph{Experimental results}
We evaluate a range of IR systems, spanning lexical and neural bi-encoders, as well as modern late-interaction and cross-encoder systems, ranging from early encoders to recent state-of-the-art systems. Recent reasoning-trained embedding and reranker models substantially outperform everything else, with the ReasonEmbed/Reranker open model family~\citep{reasonembed} ranking highest across all domains. Among newer systems, we find that performance on general-purpose IR benchmarks such as MTEB is not a reliable predictor for mathematical performance, demonstrating the need for our work. We show results reliably transfer to a downstream deduplication task, and provide a counterfactual analysis of models' reliance on mathematical notation, showing that weaker retrievers over-rely on formulas without capturing their meaning, which hurts their performance.

\paragraph{Key contributions} Our contributions include:
\vspace{-1mm}
\begin{itemize}[noitemsep]
	\item \bench, a scalable automated benchmark for mathematical information retrieval.
	\item An extensive evaluation of classical, specialized, bi- and cross-encoder models, showing that unlike ours general IR benchmarks do not fully capture mathematical IR performance.
	\item Domain-level and counterfactual analyses, identifying weaknesses in symbolic domains.
	\item Publicly-available code and data to support reproducibility and further research.
\end{itemize}
\vspace{-1mm}

\vspace{-4mm}
\section{Related Work} \label{sec:related}
\vspace{-1mm}
In this section, we discuss related work on evaluating information retrieval, its applications in mathematical reasoning, and the use of LLMs as data processors and judges.

\vspace{-2mm}
\paragraph{General-purpose retrieval benchmarks}
General-purpose retrieval benchmarks such as BEIR~\citep{beir} and MTEB~\citep{mteb} have become standard for comparing lexical, dense, and reranking models across domains and task types. These resources are useful for testing generalization and overall embedding quality, but they mainly measure general-purpose passage relevance: topical overlap, answer containment, or semantic rephrasing. When retrieval necessitates in-depth reasoning, it remains unclear how well general retrieval capabilities transfer to mathematical domains.

\vspace{-1mm}
\paragraph{Agentic workflows with retrieval}
Agentic LLM systems interleave planning, tool use~\citep{archon}, memory~\citep{park2023generative,memgpt}, and self-reflection~\citep{selfrefine}, 
with retrieval as a common tool for grounding reasoning in external evidence or domain context~\citep{irsurvey,reasoningsurvey,rag,webgpt}. In such workflows, retrieval quality matters, as irrelevant or incomplete context can derail subsequent reasoning, while well-targeted retrieval can support completion of harder tasks and reduce hallucinations~\citep{clasheval,hong2024gullible,yoran2024making}. \bench abstracts away the agentic scaffolding and directly evaluates retriever performance.

\paragraph{Retrieval in mathematical reasoning and theorem proving}
Mathematical retrieval spans a continuum from informal problem similarity to formal premise selection. In proof assistants, accurate retrieval has long been studied as premise selection for advancing the formal proof~\citep{irving2016deepmath,holist}. LLM-based provers continue this line: ReProver~\citep{yang2023leandojo} and REAL-Prover~\citep{shen2025real} retrieve a set of premises conditioned on the formal Lean proof state code. At the same time, newer systems increasingly retrieve over natural-language mathematical statements, even when the final goal is formal reasoning.
\citet{gao2024semantic} and \citet{liu2025rethinking} do so by informalizing mathlib4 statements for proving and autoformalization respectively, while Archon~\citep{archon} combines informal retrieval with formal search to translate informal arguments into machine-checkable developments. Beyond proof assistants, informal retrieval is also used to ground mathematical question answering, retrieve solution schemas or analogous problems, and construct mathematical knowledge graphs~\citep{levonian2023retrieval,dixit2024sbi,yang2024analogy,bian2025automathkg}. However, these works evaluate retrieval only on downstream performance, which conflates retriever quality with the generator's reasoning ability, prior knowledge, prompt sensitivity, and robustness to irrelevant context --- a model may succeed despite poor retrieval or fail despite useful retrieval. Our work therefore focuses on informal mathematical IR as a standalone problem.

\paragraph{Mathematical retrieval benchmarks and relevance coverage}

Several recent benchmarks evaluate mathematical or reasoning-intensive retrieval more directly. BRIGHT~\citep{bright} evaluates mathematical relevance through a shared theorem or solving method, and MathNet~\citep{mathnet} studies retrieval of structurally similar or equivalent Olympiad problems.
However, these works rely on human expert annotation, which is costly and difficult to scale to larger corpora or diverse mathematical domains.
Other works provide automatic pipelines, with MIRB~\citep{mirb} unifying semantic statement, and formula retrieval, while RAR-b~\citep{xiao2024rar} converts reasoning datasets, including MATH~\citep{hendrycksmath} and GSM8K~\citep{gsm8k}, into retrieval over pooled answer candidates.
While these benchmarks improve scalability, their construction can still lead to partial relevance coverage. Pooling-based collections leave many query-document pairs unvalidated, so relevant items surfaced by new systems may be treated as irrelevant, potentially introducing false negatives and affecting model comparisons~\citep{buckley2004retrieval,sakai2008incomplete}. Recent work shows that missing judgments matter, but also that repairing them reliably is difficult~\citep{upadhyay2024patch,abbasiantaeb2024holes}.
This is especially acute in mathematics, where useful relationships are sparse and fine-grained, as two problems may share a common method without being equivalent.
\bench is designed with this problem in mind, as relevance is estimated on a continuous scale, and evaluates retrievers on a task where each document has been sufficiently validated.
\vspace{-1mm}
\paragraph{LLMs as data processors and judges}

LLMs are increasingly used to process data at scale, including annotation~\citep{bansal2023annotation,gilardi2023chatgpt}, preference estimation~\citep{thomas2024preferences}, and evaluation~\citep{zheng2023judging,upadhyay2024patch,wildbench}. In particular, strong LLM judges can align well with human judgments in open-ended tasks, but also exhibit known biases and therefore require task-specific validation~\citep{justiceorprejudice,selfpreferencebias,jasper,judgingthejudges,verbositybias}. Notably, individual pointwise scores can be noisy or non-transitive~\citep{liu2024aligning,arenahard}, making it difficult to reliably measure subjective criteria such as relevance~\citep{proofrank}. To address these issues, prior work has used pairwise comparisons, converting many local preferences into a global ranking that can reduce noise and mitigate bias~\citep{zheng2023judging,liu2024aligning,jasper}. In \bench, we adopt this method to generate fine-grained relevance scores, validated against human annotations, and show it is preferable to standard ordinal ranking methods. We also use LLMs for data processing.
\vspace{-1mm}

\section{Methodology}\label{sec:tech}
\vspace{-1mm}
In this section, we describe the construction of \bench, our reranking benchmark for evaluating IR in mathematics. \bench is designed to test whether retrieval systems can identify mathematically meaningful relevance between problems, such as shared concepts, reasoning patterns, or solution structure, rather than relying on surface-level textual similarity.

We begin by building \bench from a large corpus of high-school-level problems and solutions (\cref{sec:tech:data}). To identify related problem pairs at scale, we use two complementary automatically extracted relevance signals: topic overlap (\cref{sec:tech:topics}) and lexical similarity between short solution summaries (\cref{sec:tech:solsummaries}). We then construct reranking tasks by selecting a set of query and candidate problems of varying types and strengths of relatedness~(\cref{sec:tech:seclectbench}). Finally, we assign fine-grained relevance ratings to the candidates using an LLM-as-a-judge pairwise preference framework (\cref{sec:tech:rating}), which serve as reference labels for evaluating performance in \bench. All prompts are included in \cref{sec:app:prompts} and we provide information about the computational resources used to construct \bench in \cref{sec:app:compute}.
\vspace{-3mm}
\vspace{-1mm}
\subsection{Sourcing a Large Mathematical Corpus}\label{sec:tech:data}
Mathematical relevance is sparse, as few problems share non-trivial concepts, reasoning patterns, or solution structure. To identify them at scale, we build a 283K-sample corpus of mostly high-school problems and solutions across Algebra, Geometry, Number Theory, Combinatorics, and Calculus.

The corpus contains $\sim150,000$ AoPS forum problems with community solutions, providing a broad but potentially lower-quality data source. Thus, we augment it with further $\sim35,000$ problems from national and international competitions (IMO, RMM, USAMO), including higher-quality solutions with deeper structure. We further add $\sim 115,000$ non-synthetic problems from the olympiad split of \numina.
We perform deduplication across the full dataset using DataTrove's MinHash pipeline~\citep{datatrove}, retaining duplicated samples from the official sources, Numina, and AoPS in that order. This removed $~17,000$ examples, leaving us with a total of $283,000$ total problem-solution pairs. \cref{sec:app:data} gives further data collection details, along with a discussion of quality control.



\subsection{Scoring Relevance using Extracted Topics}\label{sec:tech:topics}
Overlapping objects, theorems, subfields and topics provide a strong signal of mathematical relevance beyond textual similarity, despite potentially different phrasing across documents. In this section, we explain how we capture this notion by constructing a MathWorld-derived ontology~\citep{MathWorld} (\cref{sec:tech:ontology}), assigning topics to problems (\cref{sec:tech:ontologymatching}), and computing pairwise relevance from topic sets using the hierarchy.


\subsubsection{Mathematical Ontology}\label{sec:tech:ontology}

We represent each problem by a set of mathematical topics and theorems, which provide good signals of mathematical relatedness. Because related topics may differ lexically while remaining close in the subject hierarchy, we use a fixed ontology tree rather than treating topics as a flat set.


\paragraph{Ontology tree construction}

%
%

We build our ontology from Wolfram's MathWorld~\citep{MathWorld}, whose entries provide broad coverage of high-school, olympiad, and university mathematics and are linked hierarchically across mathematical areas, concepts, and theorems. We represent each MathWorld topic as a node and each hierarchical link as a directed edge from a more general topic to a more specific one, yielding a directed acyclic graph (DAG). This graph is not generally a tree: a topic can appear along multiple root-to-topic paths because mathematical concepts often belong to several areas. For example, \textit{Pascal's Triangle} appears under both a combinatorial branch, $\text{Discrete Mathematics} \rightarrow \text{Combinatorics}$, and a number-theoretic branch, $\text{Number Theory} \rightarrow \text{Special Numbers}$, reflecting different uses and contexts. To preserve this information while obtaining a tree structure, we duplicate nodes with multiple root-to-topic paths. This lets us disambiguate topic assignments by context when annotating problems from our corpus. The conversion increases the ontology size from $\sim 10.5K$ to $\sim 17.1K$ nodes, while the maximum root-to-node depth remains 5.


\subsubsection{Scoring Relevance based on Ontologies}\label{sec:tech:ontologymatching}

\paragraph{Automatic topic assignment} To calculate problem pairs relevance signals, we need to first extract relevant ontology topics for each problem in our mathematical corpus. Given our ontology's size, presenting it to an LLM in full is infeasible. Instead, we do a top-down traversal of the ontology per problem. At each visited ontology node we rely on \textsc{GPT-OSS-120B}~\citep{oss} to do local relevance scoring. The model is called with the problem statement, solution, and the ontology labels of the node and its children, and is asked to assign each candidate a relevance score, which we then normalize across nodes. We retain candidates whose normalized scores exceed a threshold $\tau = 0.85$ and recursively apply the same procedure to their descendants. By having the parent node as a reference, we allow the algorithm to terminate early when the children are substantially less relevant than their parent, preventing redundant computation. Finally, we keep only the most specific assigned topics by removing topic ancestors.




\paragraph{Topic similarity over ontologies}
To calculate the relevance signal for a pair of problems with annotated topics, we optimally match the two problems topic sets based on topic similarity. Our final relevance signal is given by the average similarity score within this optimal match. We measure topic similarity using Lin's information-content similarity \citep{linsim}, where two ontology nodes are more similar if their lowest common ancestor is more informative. We estimate information content from topic frequencies over our problem corpus. Concretely, let $c(t)$ be the number of annotations for node $t$ or any of its descendants, and $N$ be the total number of topic annotations. We define
\begin{equation}
	p(t)=\frac{c(t)}{N},
	\qquad
	IC(t)=-\log p(t).
\end{equation}
For two nodes $t_1$ and $t_2$, we compute the similarity:
\begin{equation}
	\operatorname{sim}_{\mathrm{Lin}}(t_1,t_2)
	=
	\frac{
		2IC(\operatorname{LCA}(t_1,t_2))
	}{
		IC(t_1)+IC(t_2)+\varepsilon
	},
\end{equation}
where $\operatorname{LCA}(t_1,t_2)$ is the nodes' lowest common ancestor in the ontology tree and $\varepsilon=10^{-12}$ is used for numerical stability. This assigns high similarity to topics sharing a rare, specific ancestor, and low similarity to topics sharing only broad ancestors, capturing both exact and nearby ontology overlap.

\paragraph{Topic set matching}
Given the node-level similarities above, we lift them to problem-level using Best-Match Average (BMA)~\citep{bma}. For problems $x$ and $y$ with topic sets $T_x=\{t_1,\ldots,t_m\}$ and $T_y=\{u_1,\ldots,u_n\}$, we define:
\begin{equation}
	\resizebox{0.91\columnwidth}{!}{$\displaystyle
		\begin{aligned}
			\operatorname{sim}_{\mathrm{topic}}(x,y)
			&=
			\frac{1}{|T_x|+|T_y|}
			\Bigg(
			\sum_{t\in T_x}\max_{u\in T_y}
			\operatorname{sim}_{\mathrm{Lin}}(t,u)
			\\
			&\qquad\qquad
			+
			\sum_{u\in T_y}\max_{t\in T_x}
			\operatorname{sim}_{\mathrm{Lin}}(t,u)
			\Bigg).
		\end{aligned}$}
\end{equation}
BMA matches each topic to its closest counterpart in the other problem and averages these matches in both directions. This rewards broad conceptual alignment while penalizing unmatched topics, which is useful when problems combine several mathematical concepts of unequal importance. We use $\operatorname{sim}_{\mathrm{topic}}$ as the topic-based relevance signal for selecting candidate pairs in \cref{sec:tech:seclectbench}.

\subsection{Scoring Relevance using Short Solution Summaries}\label{sec:tech:solsummaries}

While topic relevance captures an important aspect of overall mathematical relevance, problems may share underlying solution strategies despite belonging to completely different topics, or conversely, require substantially different methods even when they share topics. To capture this complementary notion of relevance, we use LLMs to extract concise natural-language summaries from problem solutions, retaining the core methodology while abstracting away formatting and presentation details. We then estimate the methodological relevance using these extracted summaries at scale via the Jaccard similarity~\citep{jaccard}.

\paragraph{Automatic summary extraction}
For each problem-solution pair in our corpus, we use \textsc{GPT-OSS-120B} to extract a summary of the solution's core idea. We prompt the model to produce summaries of at most 30 words in an imperative style, resulting in hint-like descriptions emphasizing the main method without reproducing the full solution, or including computational and algebraic details.

\paragraph{Summary-based relevance estimation}
For this signal, we measure the lexical overlap between the summaries of two problems using Jaccard similarity. For a problem $p$, let $S_p$ denote its extracted short solution summary. Let $T(s)$ be the set of unique words appearing in a sentence $s$. For a pair of problems $x$ and $y$, we define their summary-based relevance score as
\begin{equation}
    \operatorname{sim}_{\mathrm{summ}}(x,y)
    :=
    \frac{
    |T(S_x) \cap T(S_y)|
    }{
    |T(S_x) \cup T(S_y)|
    }.
\end{equation}
We use lexical rather than embedding-based similarity to avoid biasing \bench toward models used in the benchmark construction. Further, the choice of lightweight metrics, such as BMA and Jaccard also allows us to scale computation efficiently, as we require deriving the pairwise relevance signals across the entire corpus, the number of which grows quadratically ($\sim7.8\times10^{10}$ total pairs). Implementation details are found in \cref{sec:app:jaccard_details}.
\subsection{Selecting Queries and Document Sets}\label{sec:tech:seclectbench}
We construct \bench's query and document sets from our problem corpus using our topic- and solution-summary signals. A pair is topic-relevant if $\operatorname{sim}_{\mathrm{topic}}>\tau_{topic}$ and summary-relevant if $\operatorname{sim}_{\mathrm{summ}}>\tau_{summ}$, where we choose $\tau_{topic}\approx 0.858$ and $\tau_{summ}\approx 0.211$ to obtain a representative sample spanning multiple domains. More details on how we determined these values can be found in \cref{sec:app:thresholds}. To ensure a balanced and challenging benchmark, we require each query to have at least $K=50$ relevant documents in each of three categories: topic-only, summary-only, and both topic- and summary-relevant. We then sample $K$ documents uniformly from each category, giving $150$ documents per query. The final benchmark contains $1000$ queries, selected to uniformly span the five top-level ontology topics: Algebra, Geometry, Number Theory, Combinatorics, and Calculus and Analysis (see \cref{fig:piechart-benchmark}). In \cref{sec:app:signal_ablation}, we show that both signals contribute to finding relevant problems, with the candidates selected by both ranking highest under our final relevance ratings.
\subsection{Tournament-based Relevance Ratings}
\label{sec:tech:rating}
While our topic- and summary-based signals capture specific forms of mathematical relatedness, using them as final labels would over-penalize systems that surface valid connections outside these criteria. This is a well-documented problem in IR~\citep{buckley2004retrieval,sakai2008incomplete}, that we find especially acute in mathematics (See \cref{sec:related}). To obtain fine-grained ratings, we instead rate every candidate document with a pairwise LLM-as-a-judge framework. We aggregate these judgments with a Bradley--Terry model to produce continuous relevance scores, and use a Swiss-style tournament to reduce the computation.

\paragraph{Pairwise preference judging with an LLM-as-a-Judge}
We assign each candidate its final relevance rating using an LLM-as-a-judge framework~\citep{chan2023chateval,zheng2023judging,liu2024aligning,dubois2024length}. Since judging mathematical relevance requires comparing solution ideas, proof structure, and techniques, we use a strong and efficient reasoning model, \textsc{GPT-OSS-120B}, as our judge. Rather than eliciting absolute relevance scores, we ask the judge for pairwise preferences, following prior work showing that comparative judgments are more reliable than direct scoring~\citep{arenahard}. This finding is confirmed by our human evaluation in \cref{sec:app:human_evaluation}. For each query problem and its solution, we ask the judge to compare two candidate problems and their solutions and pick the one that is mathematically relevant to the query.

\paragraph{Bradley--Terry estimation of latent relevance}

To convert judge preferences into continuous relevance scores, similarly to \citet{jasper}, we fit a Bradley-Terry (BT) model separately for each query. Under this model, the probability that candidate $i$ is judged to be more relevant than candidate $j$ by the LLM is assumed to be:
$$P(i > j) = \frac{\pi_i}{\pi_i + \pi_j},$$
where $\pi_i$ is candidate $i$'s latent relevance score.
BT gives a consistent ordering \emph{within} each query. However, a candidate at a given rank may be highly informative for one query but merely the best of weak alternatives for another. We alleviate this by measuring nDCG per query on normalized scores, so that each query contributes a bounded amount to the final benchmark result.

\paragraph{Scaling pairwise comparisons with a Swiss-style tournament}
Exhaustive pairwise judging requires $\frac{n(n-1)}{2}$ LLM calls, which is prohibitively expensive for large reranking sets, and wastes calls on clearly unequal candidates without gaining much information. We therefore use a Swiss-style tournament to select an informative subset of comparisons: candidates are paired uniformly at random in the first round, and in each subsequent round every candidate is paired with an unjudged opponent whose cumulative win count is as close as possible. This concentrates judge calls on close comparisons, which are the most informative. We use $R=20$ tournament rounds, as we find this to be sufficient in \cref{sec:app:num_rounds}. Further details on how we run the tournament are provided in \cref{sec:app:swiss_details}. Finally, we fit the BT model by maximizing the likelihood of the obtained pairwise judgments. These scores are linearly rescaled to the range $[0,5]$ and used as the final relevance ratings for each query.

\section{Results}\label{sec:experimental}

In this section, we present our experimental findings. In \cref{sec:results:main}, we describe our main results, identifying the most effective retrievers for mathematics, and analyzing different task settings across mathematical domains. Then, in \cref{sec:results:math_vs_words}, we analyze the behaviour of retrievers with respect to mathematical and textual content. Finally, in \cref{sec:results:dedup}, we show that the model ordering induced by \bench{} transfers to a downstream deduplication task.
\subsection{Main Results and Key Takeaways} \label{sec:results:main}

\cref{tab:statement-full} reports the main \bench{} results for 42 models, ranging from lexical retrievers to neural bi-encoders, cross-encoders, and late-interaction systems. We report performance both in aggregate and across the five mathematical domains, with the full results in \cref{tab:main-results-ci}. All scores are nDCG@10 with exponential gain against our relevance ratings, a choice our benchmark is robust to (\cref{app:ndgccomp}). Alongside each score we report a median per-query time, measured under the standardized timing protocol of \cref{sec:app:timing}. This allows us to mark the quality--latency Pareto frontiers of the two major model classes, which is further analyzed in \cref{sec:app:latency_frontier}. To keep comparisons across method families legible, we group models by architecture and training data, following the taxonomy of \cref{sec:app:taxonomy}. Each system is run with its default task instructions; we detail how IR systems are run in \cref{sec:app:embedding_details} and study the  effect of instructions in \cref{sec:app:instruction_effect}.

Unless otherwise stated, we present results in the \emph{statement--full} setting: the query contains only a problem statement, while each document contains both a problem statement and its solution. This captures a central use case in mathematical reasoning, where given a new problem, a system retrieves solved examples with relevant concepts, theorems, or solution techniques that can guide downstream reasoning~\citep{archon,kozyrev2025rocqstar,dixit2024sbi}.

In \cref{tab:statement-full}, we also evaluate two complementary settings. In \emph{full--full} retrieval, both queries and documents include statements and solutions. This models retrieval over solved mathematical material, where a downstream system may ground answers in trusted sources, retrieve analogous examples, extract solution schemas, or construct mathematical knowledge graphs from problem-solution corpora~\citep{levonian2023retrieval,yang2024analogy,bian2025automathkg}. In \emph{statement--statement} retrieval, both queries and documents contain only problem statements. This captures search over large mathematical corpora that do not include human-readable proofs or solutions, including theorem--statement retrieval, informalized formal library search, and retrieval over libraries of mathematical declarations~\citep{yang2023leandojo,shen2025real,gao2024semantic}. Comparing these settings alongside the main one allows us to isolate how available context affects mathematical IR.

\begin{table*}[!t]
	\centering
	\footnotesize
	\caption{Main results of \bench{}. All numbers are reported as nDCG@10, except the median time column, and the MTEB Retrieval column, which reports the model's Retrieval score from the MTEB Multilingual leaderboard (-- where the model has no reported score). The Type column follows the taxonomy described in \cref{sec:results:main}. Systems on the quality--latency frontier are \bifront{underlined} for bi-encoders and \crossfront{italicized} for cross-encoders (\cref{sec:app:latency_frontier}).}
	\vspace{-0.5em}
	\label{tab:statement-full}
	\renewcommand{\arraystretch}{1.05}
	\setlength{\tabcolsep}{4pt}
	\newcommand{\beforerulepad}{\\[4pt]}
	\newcommand{\afterrulepad}{\rule{0pt}{\dimexpr\ht\strutbox+4pt\relax}}
	\newcommand{\tightmidrule}{\specialrule{\lightrulewidth}{0pt}{0pt}}
	\newcommand{\vlinepadl}{6pt}
	\newcommand{\vlinepadr}{3pt}
	\newcolumntype{R}{>{$}r<{$}}
	\newcolumntype{C}{>{$}c<{$}}
	\resizebox{\linewidth}{!}{
		\begin{tabular}{@{}l c RRRRRR@{\hspace{\vlinepadl}}|@{\hspace{\vlinepadr}}R@{\hspace{\vlinepadl}}|@{\hspace{\vlinepadr}}R@{\hspace{\vlinepadl}}|@{\hspace{\vlinepadr}}R@{\hspace{\vlinepadl}}||@{\hspace{\vlinepadr}}C@{}}
			\toprule
			& \multicolumn{1}{c}{\makecell{Type}}
			& \multicolumn{1}{c}{\makecell{Overall}}
			& \multicolumn{1}{c}{\makecell{Algebra}}
			& \multicolumn{1}{c}{\makecell{Geometry}}
			& \multicolumn{1}{c}{\makecell{Number\\Theory}}
			& \multicolumn{1}{c}{\makecell{Comb.}}
			& \multicolumn{1}{c@{\hspace{\vlinepadl}}|@{\hspace{\vlinepadr}}}{\makecell[c]{Calc./\\Analysis}}
			& \multicolumn{1}{c@{\hspace{\vlinepadl}}|@{\hspace{\vlinepadr}}}{\makecell[c]{Statement-\\Statement}}
			& \multicolumn{1}{c@{\hspace{\vlinepadl}}|@{\hspace{\vlinepadr}}}{\makecell[c]{Full-Full}}
			& \multicolumn{1}{c@{\hspace{\vlinepadl}}||@{\hspace{\vlinepadr}}}{\makecell[c]{Med. s/query}}
			& \multicolumn{1}{c}{\makecell[c]{MTEB\\Retrieval}} \beforerulepad
			\tightmidrule
			\afterrulepad
			\crossfront{ReasonReranker-Qwen3-32B-Rewrite}~\citep{reasonembed} & RTWC & \mathbf{0.741} & \mathbf{0.710} & 0.749 & \mathbf{0.740} & \mathbf{0.783} & \mathbf{0.732} & 0.638 & 0.796 & 117.06 & \text{---} \\
			\bifront{ReasonEmbed-Qwen3-8B-Rewrite}~\citep{reasonembed} & RTWB & 0.738 & \mathbf{0.710} & \mathbf{0.767} & 0.734 & 0.781 & 0.708 & 0.659 & 0.769 & 29.93 & \text{---} \\ 
			\crossfront{ReasonReranker-Qwen3-32B}~\citep{reasonembed} & RTC & 0.733 & 0.699 & 0.751 & 0.726 & 0.777 & 0.712 & 0.647 & \mathbf{0.800} & 85.44 & \text{---} \\
			\bifront{ReasonEmbed-Qwen3-8B}~\citep{reasonembed} & RB & 0.710 & 0.689 & 0.744 & 0.697 & 0.738 & 0.682 & \mathbf{0.685} & 0.795 & 1.751 & \text{---} \\ 
			\crossfront{Diver-GroupRank-32B}~\citep{grouprank} & RTC & 0.693 & 0.645 & 0.716 & 0.674 & 0.729 & 0.696 & 0.651 & 0.729 & 64.012 & \text{---} \\
			RaDeR-14B~\citep{rader} & RB & 0.692 & 0.669 & 0.727 & 0.682 & 0.714 & 0.665 & 0.677 & 0.741 & 3.042 & \text{---} \\
			\bifront{RaDeR-7B}~\citep{rader} & RB & 0.690 & 0.666 & 0.725 & 0.675 & 0.715 & 0.666 & 0.671 & 0.738 & 1.668 & \text{---} \\
			\crossfront{Qwen3-Reranker-4B}~\citep{qwen3embedding} & C & 0.683 & 0.649 & 0.709 & 0.672 & 0.712 & 0.671 & 0.653 & 0.730 & 1.167 & \text{---} \\
			\bifront{Diver-4B}~\citep{diver} & RB & 0.681 & 0.665 & 0.711 & 0.672 & 0.708 & 0.645 & 0.659 & 0.737 & 1.456 & \text{---} \\
			ReasonEmbed-Llama-3.1-8B~\citep{reasonembed} & RB & 0.664 & 0.641 & 0.695 & 0.652 & 0.678 & 0.650 & 0.643 & 0.751 & 1.95 & \text{---} \\ 
			\bifront{RaDeR-3B}~\citep{rader} & RB & 0.654 & 0.629 & 0.694 & 0.648 & 0.680 & 0.624 & 0.644 & 0.705 & 0.951 & \text{---} \\
			\crossfront{Qwen3-Reranker-0.6B}~\citep{qwen3embedding} & C & 0.652 & 0.629 & 0.690 & 0.638 & 0.667 & 0.634 & 0.621 & 0.717 & 0.957 & \text{---} \\
			SPLADE-Code-8B~\citep{splade} & SB & 0.650 & 0.610 & 0.689 & 0.644 & 0.694 & 0.618 & 0.634 & 0.693 & 5.649 & \text{---} \\
			Octen-8B~\citep{octen2025rteb} & B & 0.636 & 0.594 & 0.664 & 0.629 & 0.665 & 0.630 & 0.623 & 0.672 & 1.656 & 71.61 \\
			Gemini-2~\citep{geminiembedding} & AB & 0.628 & 0.599 & 0.647 & 0.622 & 0.656 & 0.614 & 0.603 & 0.658 & 2.409 & \text{---} \\
			Gemini-001~\citep{geminiembedding} & AB & 0.616 & 0.585 & 0.651 & 0.617 & 0.635 & 0.596 & 0.599 & 0.667 & 6.062 & 67.71 \\
			Qwen3-Embedding-4B~\citep{qwen3embedding} & B & 0.615 & 0.576 & 0.652 & 0.610 & 0.642 & 0.597 & 0.597 & 0.667 & 1.081 & 69.60 \\
			Qwen3-Embedding-8B~\citep{qwen3embedding} & B & 0.611 & 0.569 & 0.647 & 0.606 & 0.633 & 0.598 & 0.600 & 0.662 & 1.673 & 70.88 \\
			Rank1-32B~\citep{rank1} & RTC & 0.610 & 0.583 & 0.617 & 0.613 & 0.673 & 0.574 & 0.556 & 0.629 & 114.603 & \text{---} \\
			Harrier-27b~\citep{harrier} & B & 0.608 & 0.569 & 0.649 & 0.601 & 0.620 & 0.596 & 0.585 & 0.659 & 6.741 & 78.27 \\
			INF-Retriever-v1-Pro~\citep{infretriever} & RB & 0.594 & 0.573 & 0.618 & 0.585 & 0.607 & 0.592 & 0.564 & 0.660 & 2.800 & \text{---} \\
			KaLM-12B-2511~\citep{kalm} & B & 0.584 & 0.548 & 0.617 & 0.583 & 0.606 & 0.569 & 0.579 & 0.599 & 3.283 & 75.66 \\
			LLaMa-Nemotron-8B~\citep{nemotron} & B & 0.579 & 0.542 & 0.610 & 0.580 & 0.600 & 0.562 & 0.565 & 0.616 & 1.624 & 68.69 \\
			\bifront{Qwen3-Embedding-0.6B}~\citep{qwen3embedding} & B & 0.575 & 0.545 & 0.629 & 0.564 & 0.589 & 0.546 & 0.566 & 0.625 & 0.472 & 64.65 \\
			Harrier-0.6b~\citep{harrier} & B & 0.572 & 0.538 & 0.613 & 0.566 & 0.581 & 0.557 & 0.549 & 0.632 & 0.491 & 70.75 \\
			Jina-v5-Small~\citep{jina} & B & 0.565 & 0.521 & 0.611 & 0.557 & 0.589 & 0.545 & 0.549 & 0.611 & 0.884 & 64.88 \\
			\crossfront{Reason-ModernColBERT}~\citep{reasonir} & RL & 0.558 & 0.533 & 0.594 & 0.555 & 0.567 & 0.537 & 0.539 & 0.611 & 0.607 & \text{---} \\
			Text-Embedding-3-Large~\citep{openaiembedding} & AB & 0.557 & 0.535 & 0.571 & 0.560 & 0.574 & 0.552 & 0.540 & 0.594 & 4.646 & 59.32 \\
			Jina-v5-Nano~\citep{jina} & B & 0.530 & 0.488 & 0.568 & 0.532 & 0.551 & 0.510 & 0.519 & 0.572 & 0.682 & 63.26 \\
			Gemma-300m~\citep{embeddinggemma} & B & 0.527 & 0.505 & 0.559 & 0.522 & 0.543 & 0.504 & 0.511 & 0.585 & 0.537 & 62.49 \\
			Text-Embedding-3-Small~\citep{openaiembedding} & AB & 0.512 & 0.487 & 0.557 & 0.504 & 0.526 & 0.491 & 0.497 & 0.554 & 6.489 & 50.69 \\
			\bifront{BGE-m3}~\citep{bge-m3} & B & 0.511 & 0.484 & 0.549 & 0.502 & 0.518 & 0.500 & 0.505 & 0.563 & 0.399 & 54.59 \\
			ReasonIR-8B~\citep{reasonir} & RB & 0.507 & 0.495 & 0.511 & 0.500 & 0.534 & 0.489 & 0.513 & 0.617 & 2.361 & \text{---} \\
			Harrier-270m~\citep{harrier} & B & 0.498 & 0.470 & 0.545 & 0.491 & 0.512 & 0.469 & 0.497 & 0.557 & 0.682 & 66.38 \\
			INF-X-Retriever~\citep{infretriever} & RWB & 0.494 & 0.493 & 0.468 & 0.450 & 0.535 & 0.522 & 0.437 & 0.508 & 3.326 & \text{---} \\
			Multilingual-E5-Large~\citep{e5} & B & 0.488 & 0.455 & 0.529 & 0.471 & 0.500 & 0.477 & 0.508 & 0.545 & 0.543 & 53.71 \\
			Approach Zero~\citep{approachzero} & SL & 0.468 & 0.485 & 0.443 & 0.454 & 0.480 & 0.490 & 0.469 & 0.481 & 1.275 & \text{---} \\
			\bifront{BM25}~\citep{bm25} & SB & 0.416 & 0.405 & 0.448 & 0.392 & 0.429 & 0.393 & 0.426 & 0.437 & 0.018 & \text{---} \\
			\bifront{Jaccard}~\citep{jaccard} & SB & 0.412 & 0.381 & 0.448 & 0.397 & 0.442 & 0.383 & 0.448 & 0.469 & 0.014 & \text{---} \\
			TF-IDF & SB & 0.412 & 0.414 & 0.402 & 0.383 & 0.424 & 0.414 & 0.434 & 0.447 & 0.022 & \text{---} \\
			BERT~\citep{bert} & B & 0.357 & 0.369 & 0.342 & 0.344 & 0.389 & 0.335 & 0.416 & 0.429 & 0.486 & \text{---} \\
			RoBERTa~\citep{roberta} & B & 0.311 & 0.306 & 0.293 & 0.314 & 0.342 & 0.287 & 0.406 & 0.397 & 0.933 & \text{---} \\
			\bottomrule
		\end{tabular}
	}
	\vspace{-1.5em}
\end{table*}

\paragraph{The ReasonEmbed/Reranker family scores highest}

The open \textsc{ReasonEmbed/Reranker} family achieves the best aggregate performance and holds the top score in every individual domain. The reranker models edge out their embedding counterparts, as one would expect from the reranking design of our task, a finding that carries over to all model families. Pairing the \textsc{ReasonEmbed}-family models with a query rewriter further improves performance, aligning the queries better with the model's expected format. However, this design does not transfer universally, as the rewriter-based \textsc{INF-X-Retriever} scores lower than \textsc{INF-Retriever-v1-Pro}. On the other hand, traditional lexical retrievers, older contextual encoders, and specialized mathematical baselines all perform poorly, showing that they are no longer competitive. We note that even the best score in the main setting remains far from perfect retrieval, leaving room for progress in mathematical IR.

\paragraph{Reasoning IR methods achieve the best results}

Beyond the \textsc{ReasonEmbed} family, training on reasoning data is a strong factor for higher performance on \bench{}. For instance, \textsc{ReasonEmbed-Qwen3-8B}, \textsc{RaDeR-7B}, and \textsc{Diver-4B} outperform \textsc{Octen-8B}, the strongest general-purpose bi-encoder we evaluate, while being no larger than it. Therefore, training on reasoning-heavy context appears to improve the representation of mathematical content, which is less prevalent in general-purpose training sets.

\paragraph{Strong LLM backbones matter too}

Where IR systems are fine-tuned from an underlying LLM, the quality of that backbone matters just as much. Within a single family, the \textsc{Qwen} and \textsc{LLaMa} versions of \textsc{ReasonEmbed} share a training recipe and a parameter count, yet differ by $0.046$ on the strength of their backbone alone. The same pattern holds across families, as the \textsc{LLaMa}-based \textsc{ReasonIR-8B} and the \textsc{ModernBERT}-based \textsc{Reason-ModernColBERT} both score lower than the generalist \textsc{Qwen3-Embedding-4B}, despite being trained specifically with reasoning data. Since the top-scoring systems are built on stronger LLMs, backbone progress also becomes key for retrieval development.

\paragraph{Generic retrieval benchmarks cannot solely predict performance on mathematical IR}

Using our main scores, we measure Pearson and Spearman rank correlations between \bench{} and MTEB multilingual IR~\citep{mteb} scores as $0.77$ and $0.82$, where scores are available. However, restricting the analysis to models released from 2025 onwards reduces correlations to $0.62$ and $0.73$, suggesting that general benchmarks are not reliable predictors of mathematical IR performance, especially as models become more capable. For example, \textsc{Harrier-27b} ranks 1st on MTEB, while being behind the Octen, Gemini, and Qwen models on \bench{}. This analysis goes beyond outliers, as removing the $n=1,2$ models with the largest rank gaps still yields Spearman correlations of $0.77$ and $0.79$.

\paragraph{IR systems struggle more on symbolic tasks}

For the strongest systems, Geometry and Combinatorics tend to be easier, while more symbolic-heavy domains like Algebra, Number Theory, and Calculus are more challenging. For example, \textsc{ReasonReranker-Qwen3-32B-Rewrite} achieves $0.749$ on Geometry and $0.783$ on Combinatorics, but only $0.710$ on Algebra. This pattern is consistent with the intuition that IR systems are worse at aligning and understanding mathematical formulas vs natural language text.

\paragraph{Stronger retrievers benefit more from additional information}

Additional context helps stronger systems, which can represent it, i.e. \textsc{ReasonReranker-Qwen3-32B-Rewrite} improves from $0.638$ to $0.741$ to $0.796$ across statement-statement, statement-full and full-full, whereas weaker ones like \textsc{BERT} fall from $0.416$ to $0.357$ once solutions are included. We describe the non-main settings further in \cref{sec:app:additional_settings}.

\subsection{Comparison of Mathematical and Textual Content in Retrieval} \label{sec:results:math_vs_words}

\begin{figure}[!hbt]
	\centering
	\includegraphics[width=\linewidth]{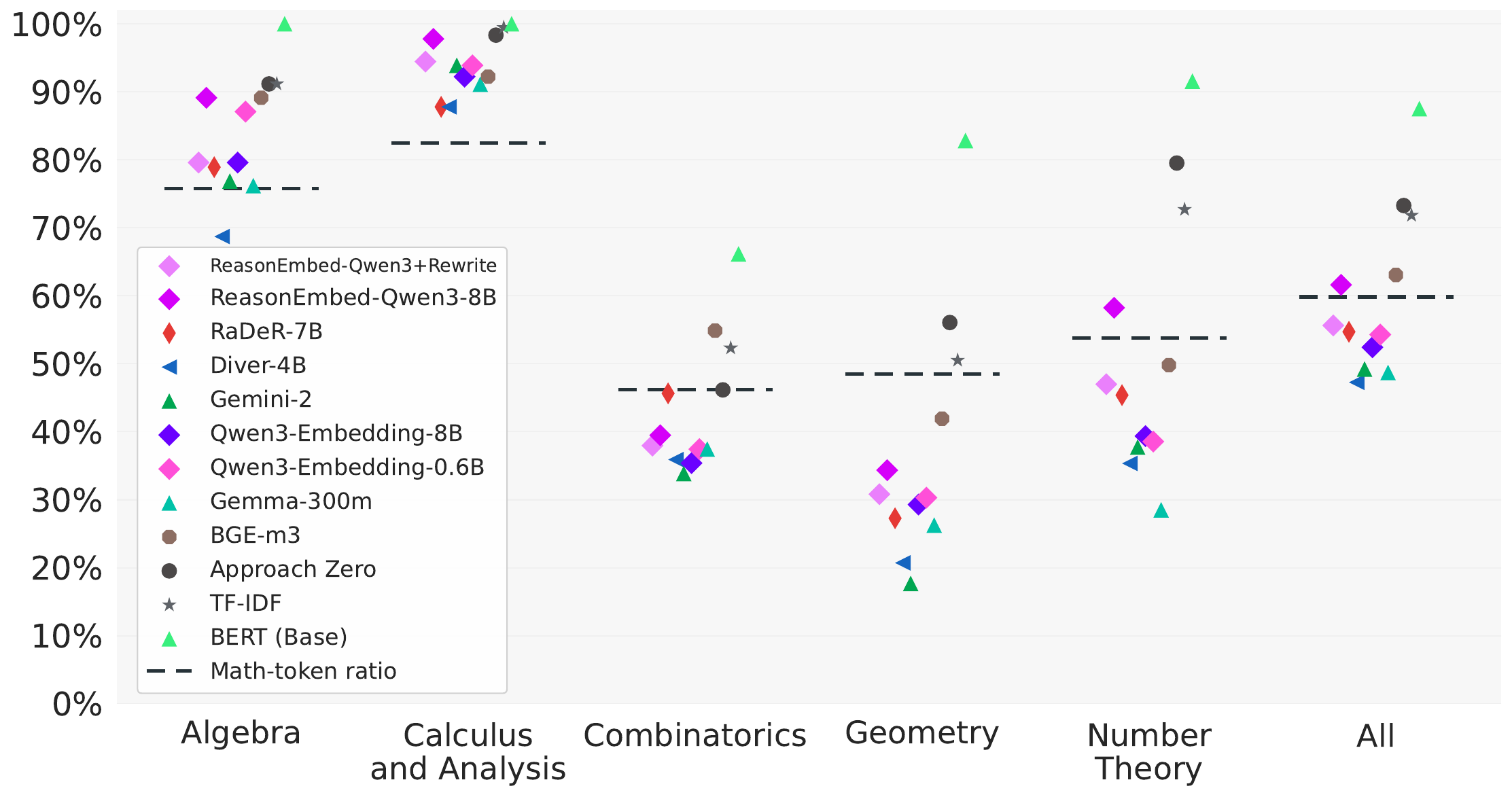}
	\vspace{-2em}
	\caption{IR model preferences for math vs. textual content, against avg. token distributions per domain.}
	\label{fig:math-vs-words-selected-models}
	\vspace{-1em}
\end{figure}

We next examine what content IR methods rely on when matching mathematics: for each target, we measure whether its math-only or word-only text better aligns with its most relevant documents. \Cref{fig:math-vs-words-selected-models} shows partial results, with full methodology in \cref{sec:app:math_vs_words} and full results in \cref{fig:math-vs-words-all-models}.

Overall, we observe that the systems' reliance on mathematical content is correlated to the math-token ratio. On top of that, we notice three trends. First, weaker IR systems tend to rely more on mathematical content across all domains, without necessarily understanding the meaning behind it.
Second, IR systems that were trained on reasoning data score highest, and rely more on mathematical content compared to generalist models, plausibly because the training data improves the representation of mathematical content. Finally, models from the same family~(\textsc{Qwen3-Embedding-8B/0.6B}) show nearly identical behaviour, implying that the preferences may be inherited from the training data. Results across all models are provided in \cref{sec:app:math_vs_words}.

\subsection{Downstream Evaluation via Deduplication} \label{sec:results:dedup}

\bench{} deliberately measures retrieval in isolation from any downstream application. This design choice, however, raises the complementary question of whether the ordering induced by \bench{} transfers to realistic tasks. To this end, we evaluate the effectiveness of each retriever on \emph{deduplication}, where IR systems are tasked with recovering a equivalent rephrasing of the query among all $71$K documents in \bench{}. In \cref{tab:dedup-all-docs}, we report the median rank assigned to the rephrasing and the top-$k$ accuracy, alongside each system's \bench{} scores for reference. As the pool is far too large to score with a cross-encoder, this experiment covers bi-encoders only. We defer the full experimental setup to \cref{sec:app:dedup_setup}.

Model ordering is largely preserved between the two tasks, with a Spearman rank correlation between \bench{} nDCG@10 and deduplication performance of $\rho=0.96$ against the (negated) median rank of the duplicate. The systems at the top of \bench{} rank at the top in this experiment, as the two \textsc{ReasonEmbed-Qwen3-8B} variants attain the best median rank of $2$, followed by the \textsc{RaDeR} and \textsc{Diver} families, with lexical baselines at the bottom of the rankings. We report the results for all systems with a more detailed analysis, as well as a reranking-scale variant of the experiment that covers all models in \cref{sec:app:dedup}.

\begin{table}[t]
	\centering
	\scriptsize
	\caption{Ability of IR systems to discover a rephrased version of the query, for a representative set of systems.}
	\vspace{-0.7em}
	\label{tab:dedup-all-docs}
	\renewcommand{\arraystretch}{1.05}
	\setlength{\tabcolsep}{2.5pt}
	\newcolumntype{R}{>{$}r<{$}}
	\resizebox{\columnwidth}{!}{%
		\begin{tabular}{@{}l R R RRRR@{}}
			\toprule
			& \multicolumn{1}{c}{\bench{}}
			& \multicolumn{1}{c}{Median}
			& \multicolumn{1}{c}{Top 1\%}
			& \multicolumn{1}{c}{Top 4\%}
			& \multicolumn{1}{c}{Top 16\%}
			& \multicolumn{1}{c}{Top 64\%} \\
			\midrule
			ReasonEmbed-Qwen3-8B-Rewrite & \mathbf{0.738} & \mathbf{2} & 45.2 & 68.0 & 84.9 & \mathbf{95.7} \\
			ReasonEmbed-Qwen3-8B & 0.710 & \mathbf{2} & \mathbf{45.4} & \mathbf{71.3} & \mathbf{86.9} & 95.1 \\
			RaDeR-14B & 0.692 & \mathbf{2} & 39.3 & 66.7 & 85.0 & 93.6 \\
			RaDeR-7B & 0.690 & 3 & 32.5 & 60.0 & 79.2 & 91.3 \\
			Diver-4B & 0.681 & 3 & 32.6 & 57.8 & 77.6 & 90.1 \\
			ReasonEmbed-Llama-3.1-8B & 0.664 & 4 & 32.6 & 52.0 & 69.6 & 84.6 \\
			Octen-8B & 0.636 & 7 & 23.5 & 43.0 & 59.8 & 77.9 \\
			Gemini-2 & 0.628 & 6 & 26.4 & 45.1 & 64.1 & 80.3 \\
			Gemini-001 & 0.616 & 3 & 34.1 & 55.8 & 75.5 & 87.7 \\
			Qwen3-Embedding-4B & 0.615 & 19 & 17.5 & 30.6 & 48.2 & 68.0 \\
			Text-Embedding-3-Large & 0.557 & 116 & 9.0 & 16.1 & 27.5 & 43.1 \\
			BM25 & 0.416 & 436 & 8.4 & 14.2 & 21.4 & 32.1 \\
			Jaccard & 0.412 & 5505 & 4.6 & 7.7 & 13.2 & 17.9 \\
			TF-IDF & 0.412 & 12800 & 2.1 & 3.9 & 5.8 & 9.9 \\
			BERT & 0.357 & 6047 & 2.9 & 4.8 & 7.1 & 10.8 \\
			RoBERTa & 0.311 & 12368 & 2.8 & 4.4 & 6.4 & 9.5 \\
			\bottomrule
		\end{tabular}%
	}
	\vspace{-2em}
\end{table}

\section{Conclusion}\label{sec:conclusion}

In this work, we introduced \bench{}, an automated benchmark for fine-grained evaluation of mathematical IR. We combine ontology-based topic matching, solution-summary similarity, and LLM pairwise judgments to construct realistic reranking tasks. Our results reveal that modern reasoning-trained embedders and rerankers outperform classical and generalist baselines, but still struggle with symbolic domains. Further, we show that general retrieval benchmarks do not reliably predict performance on mathematical retrieval tasks, but that our results transfer to realistic downstream tasks. Together, these results highlight the need for domain-specific evaluation and establish \bench{} as a valuable tool for driving progress in mathematical IR.
\newpage
\section*{Limitations and Future Work}\label{sec:app:limitations}

While \bench{} provides a scalable framework for evaluating mathematical information retrieval, several limitations remain. First, the benchmark is built primarily from high-school, olympiad, and early undergraduate-style problems, so performance may not fully transfer to research-level mathematics, formal proof libraries, textbook retrieval, or highly specialized domains. In addition, \bench{} evaluates reranking over preselected candidate sets rather than open-ended retrieval over the full corpus. This design makes it possible to assign fine-grained relevance judgments to every evaluated candidate, while complementary first-stage retrieval evaluation remains an important extension. Finally, the benchmark uses automated LLM-based processing to support construction at scale. Although this can introduce noise into individual relevance signals, our human validation indicates that the resulting judgments are reliable for comparing retrieval systems.
\section*{Acknowledgements}

This research was partially funded by the Ministry of Education and Science of Bulgaria (support for INSAIT, part of the Bulgarian National Roadmap for Research Infrastructure). This project was supported with computational resources provided by Google Cloud Platform (GCP).
\newpage
\bibliography{references}


\appendix
\section{Data Sources}\label{sec:app:data}
In this section we describe in detail what datasets we use to compile our problem corpora from \cref{sec:tech:data} and how we preprocess them.
\subsection{Gathering Data from Public Forums}\label{sec:tech:forums}
Retrieval systems often rely on large-scale curation from public web sources. To establish a broad base for our benchmark, we first collect problems and solutions from the Art of Problem Solving (AoPS) community, which hosts a vast collection of mathematical discussions. We parse nearly $500$K threads and, because forum threads frequently contain discussion fragments, incomplete arguments, or off-topic remarks, subject them to a strict filtering pipeline. We first discard every thread without any comments, as such threads cannot contain a solution. We then prompt \textsc{GPT-5-mini} to check that the opening post of the thread states a self-contained problem, discarding the thread otherwise. For the surviving threads, we use the same model to identify which comments contain substantive solutions, explicitly instructing it to reject one-line proof sketches and comments that only remark on the problem; threads left without a single feasible solution are discarded. Finally, in a separate query, we ask the model to select the most detailed of the remaining candidate solutions, while simultaneously verifying that the selected comment does address the problem stated in the thread. We deduplicate the resulting problem statements using DataTrove's MinHash deduplication pipeline~\citep{datatrove}, yielding $150$K unique entries.

\subsection{Gathering Data from High-Quality Official Sources}\label{sec:tech:officialsrc}
Although AoPS provides a large and diverse set of problems, its quality and difficulty vary substantially. Because AoPS is an open forum, extracted problems and solutions may be incomplete, noisy, or interleaved with out-of-context discussion. We therefore supplement the corpus with problems and solutions from high-quality official sources, including national and international competitions as well as preparation materials.

\begin{figure}
\centering
\begin{minipage}{\linewidth}
    \centering
    \includegraphics[width=\linewidth]{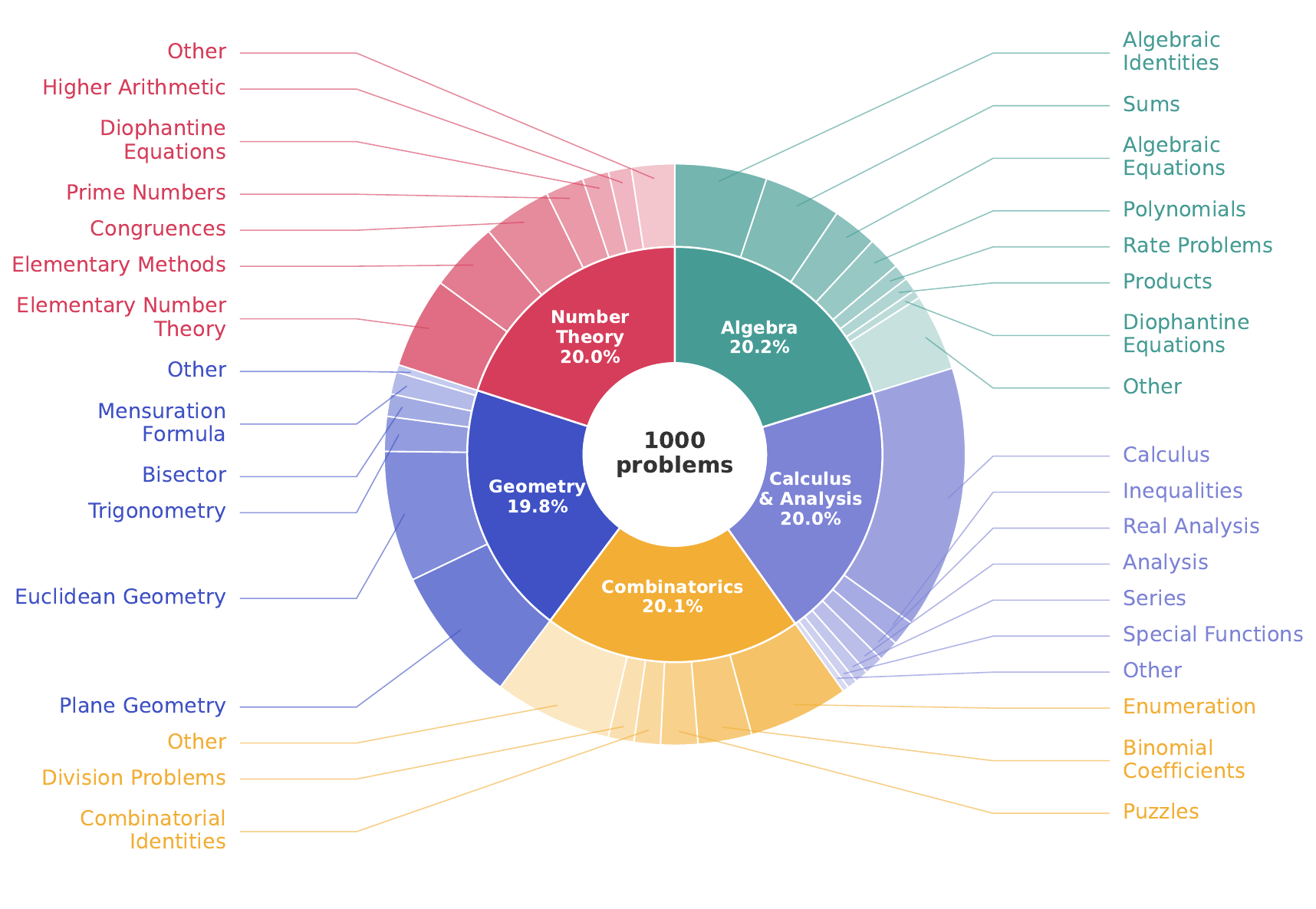}
    \caption{Domain and subdomain distribution of the query problems in \bench.}
    \label{fig:piechart-benchmark}
\end{minipage}
\hfill
\begin{minipage}{\linewidth}
    \centering
    \includegraphics[width=0.98\linewidth]{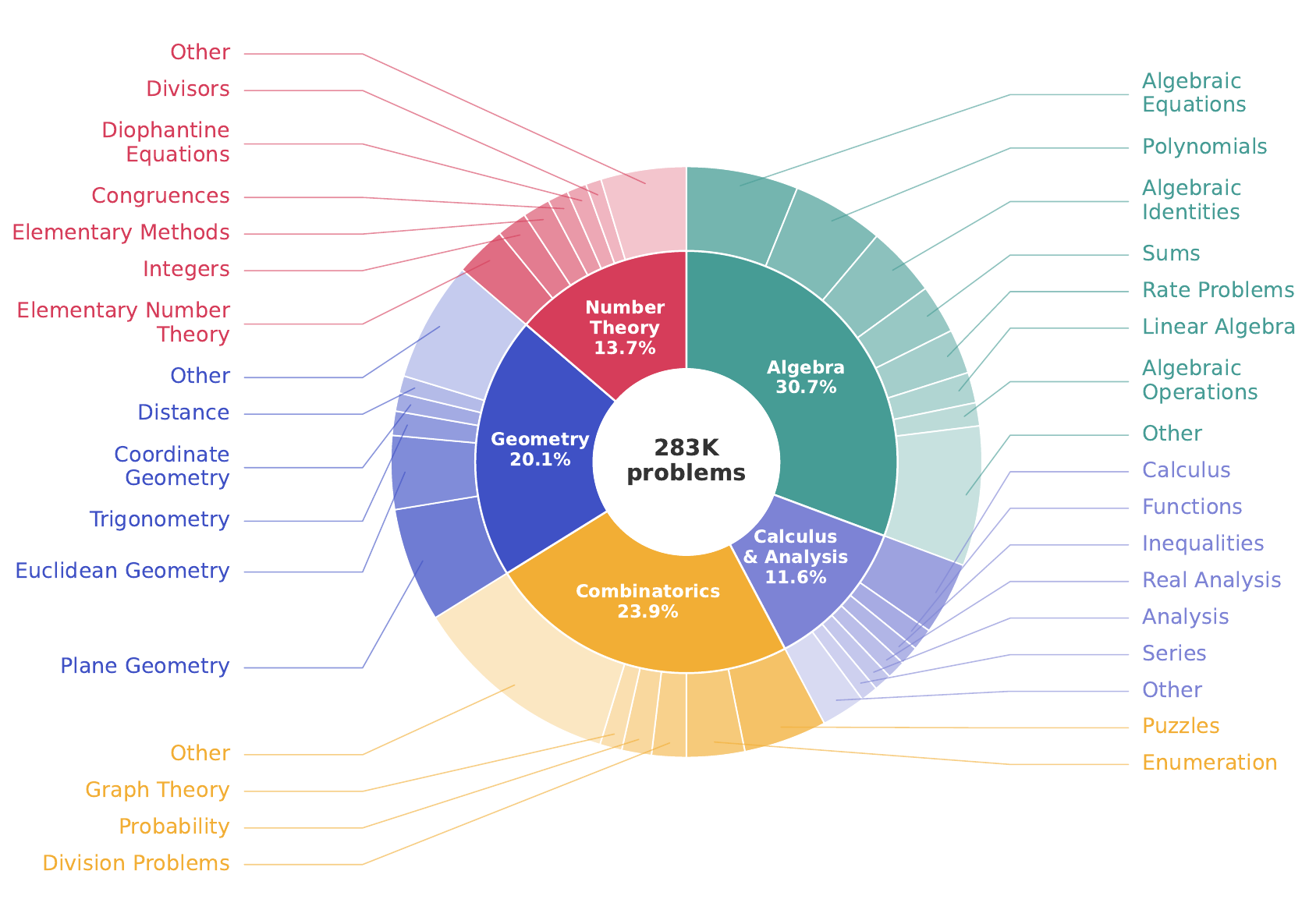}
    \caption{Domain and subdomain distribution of the sampling corpus described in \cref{sec:tech:data}.}
    \label{fig:piechart-databank}
\end{minipage}
\hfill
\begin{minipage}{\linewidth}
    \centering
    \includegraphics[width=0.97\linewidth]{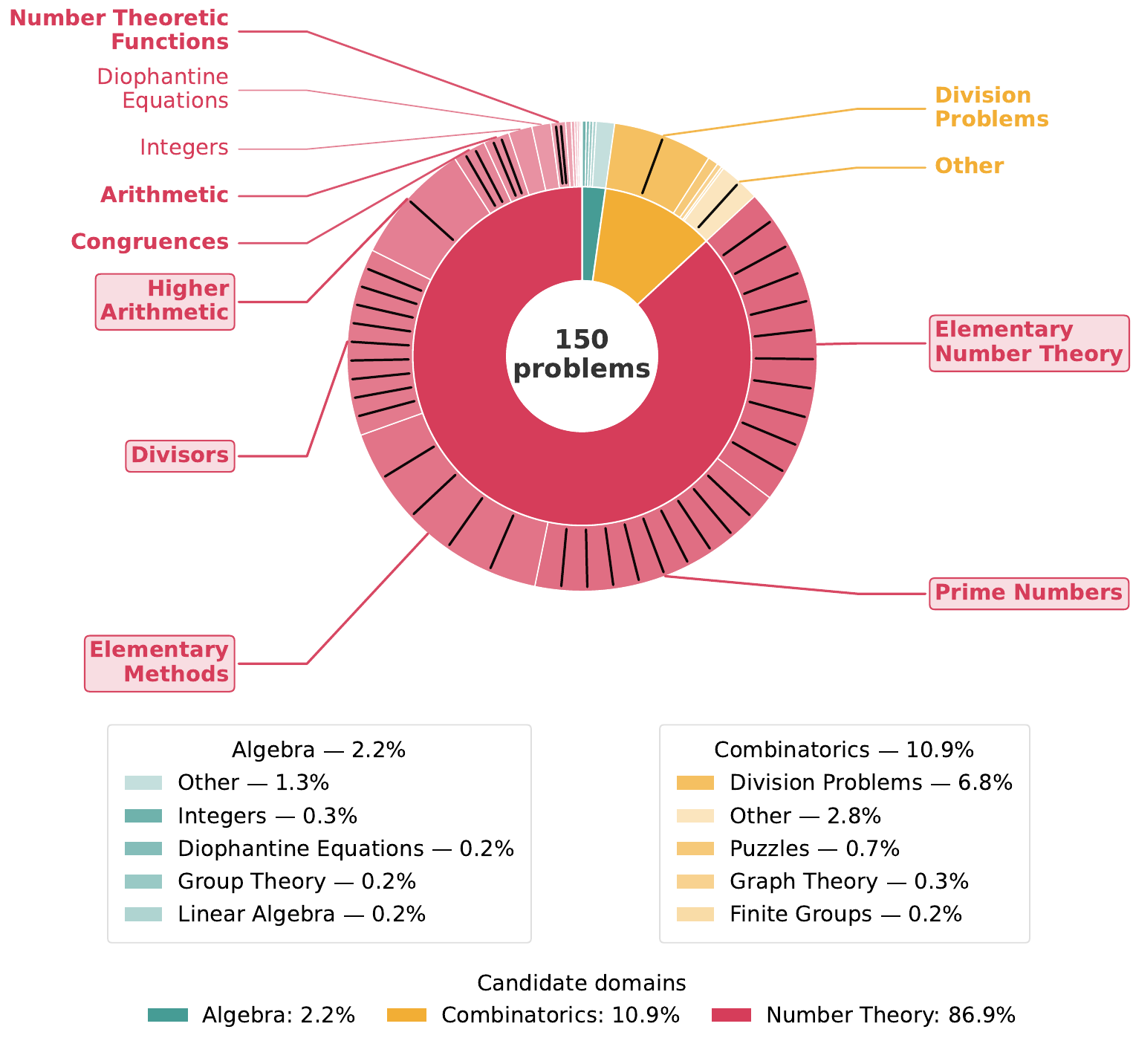}
    \caption{Domain and subdomain distribution of the documents retrieved for a representative query from \bench. Shaded subdomains indicate the subdomains associated with the query, while black lines mark documents ranked in the top 10 for the query.}
    \label{fig:piechart-candidates-NT}
\end{minipage}
\end{figure}

Starting from official PDFs manually collected from the web, we parse the documents using MathPix OCR, which is specialized for mathematical content. During the initial collection we manually inspected the resulting Markdown and \LaTeX{} against the source PDFs and found them to be generally faithful. Documents that are too long to be segmented reliably in a single pass are manually divided before further processing. We then apply a multi-step LLM-based parsing pipeline using \textsc{GPT-5-mini} as follows:

\begin{itemize}
    \item (Optional) \textbf{Translation:} For documents in languages other than English, we first translate the text into English.
    \item \textbf{Segmentation:} We segment the parsed text into individual problems and solutions, removing extraneous information such as formatting artifacts, metadata, and irrelevant text. We capture problems and solutions in XML-style tags, without modifying their content. If the number of problem and solution tags does not match, we flag the document and have a human manually review it, either correcting the segmentation or discarding the document when it cannot be reliably recovered.
    \item \textbf{Cleaning:} We apply a final cleaning step to remove any remaining formatting artifacts, problem metadata, such as the author, source, or number of the problem, and any text that is not part of the problem statement or solution.
\end{itemize}

Together with the count-matching check, these steps ensure that we retain only explicit problem--solution pairs, aligned as they appear in the source document. This process yields a high-quality set of approximately $35$K problems from challenging competitions, including the IMO, USAMO, and RMM. This subset is especially valuable for our benchmark because it contains a higher density of relevant pairs and more complex problems and solutions, making it more likely to challenge current embedding models.

\subsection{Additional Sources}\label{sec:tech:additionaldata}

Because the AoPS subset is nearly five times larger than the official-source subset, relying too heavily on AoPS could reduce the overall quality of the corpus. We therefore further supplement the corpus with problems and solutions from another high-quality, but less challenging source, \numina~\cite{numina_math_datasets}. To maintain a similar range of difficulty and mathematical domains, we sample approximately $115$K problems from the \emph{olympiads} split of \numina, discarding synthetic samples, multiple-choice problems, and entries with invalid problem statements or solutions. We restrict ourselves to this split as it is itself compiled from official mathematical sources rather than synthesized, so its reference solutions are of comparable reliability to those of our official-source subset. This completes our initial corpus of approximately $300$K mathematical problem-solution pairs.

\subsection{Cross-Dataset Deduplication}\label{sec:tech:dedup}
Although the sources are collected independently, prominent competition problems may appear in multiple datasets. To avoid including duplicate problems as both queries and documents, we perform cross-source deduplication. We detect duplicates using the same MinHash-based method described in \cref{sec:tech:forums}. When duplicates are found, we retain a single instance according to the following priority order: official sources, \numina, and then AoPS. This process removes approximately $17$K cross-source duplicates, leaving a final corpus of approximately $283$K unique problem-solution pairs.

\subsection{Data Quality Control}\label{sec:app:data_quality}
While we do not rely on a specific notion of quality, during our dataset curation we established a set of per-source measures applied throughout the pipelines described above. They filter out documents that are unusable as problem--solution pairs, while keeping the scope of the corpus broad and representative of real-world corpora. We summarize them here, grouped by the aspect of data quality we target. Faithful parsing and problem--solution alignment form the one aspect we do not revisit below, as they concern only the official-source subset, the only subset we derive from PDFs, and are addressed directly in \cref{sec:tech:officialsrc}.

\paragraph{Solution correctness} Our handling of solution reliability depends on the source. The official-source subset (\cref{sec:tech:officialsrc}) pairs every problem with the official solution published in the original competition document, and the \emph{olympiads} split of \numina (\cref{sec:tech:additionaldata}) likewise draws on official mathematical sources. We treat both as gold-standard sources with respect to correctness. The AoPS subset (\cref{sec:tech:forums}) is the only one whose solutions are community-written, and therefore the only one whose correctness we cannot inherit from an official source. This is precisely why we subject it to the stricter filtering of \cref{sec:tech:forums}, which discards threads without a well-defined problem, comments that do not constitute a substantive solution, and problems left without any feasible solution. We nonetheless retain the subset, as it contributes solutions written in a considerably more casual style than official write-ups, which is representative of the documents a mathematical retrieval system encounters in practice.

\paragraph{Duplicate leakage} We guard against duplicate leakage at two stages. We first deduplicate problem statements within each of the individual source pipelines, and then apply an additional MinHash deduplication pass across the combined corpus, as described in \cref{sec:tech:dedup}. To verify that no residual leakage survives into the final benchmark, we additionally run a targeted audit of \bench. For each of its $1000$ queries, we ask \textsc{GPT-OSS-120B}~\citep{oss} to inspect the $50$ highest-ranked documents under the final relevance ranking, i.e.\ the region in which leakage would most distort our metrics, and to flag any query--document pair stating equivalent problems. Out of the resulting $50{,}000$ audited pairs, the model flagged exactly one:

\begin{itemize}
    \item \textbf{Query:} Find the number of positive integer solutions to $a+b+c+d \leq 12$, where $a,b,c,d$ are positive integers.
    \item \textbf{Document:} How many solutions are there to $3x+y+z \leq 17$, where $x,y,z$ are nonnegative integers?
\end{itemize}

These are clearly related counting problems, but they are not duplicates, as they differ in their number of variables, coefficients, bound, and positivity constraint, and hence also in their answers. We thus find no duplicate problems in \bench.

\paragraph{Residual noise} We do not claim all $283$K problem--solution pairs in our corpus to be noise-free, as some residual noise is unavoidable at this scale. Instead, our measures aim to keep that noise within what a reasonably cleaned corpus might have. In particular, the source heterogeneity is a deliberate design choice in \bench, as realistic retrieval corpora likewise mix sources of varying quality, and drawing on three independent sources prevents the benchmark from rewarding models that latch onto the phrasing and presentation conventions of any single one.

\subsection{Mathematical Data Distribution} \label{sec:app:data_distribution}

In \cref{fig:piechart-benchmark} and \cref{fig:piechart-databank} we present visualizations of the domain and subdomain distributions of the queries in \bench{} and the sampling corpus respectively. The categories are determined by the mathematical ontology tagging described in \cref{sec:tech:ontology}. Since a single problem may belong to multiple domains or subdomains, each problem contributes a total weight of 1 distributed equally among its assigned categories.

\subsection{Example of Query-Document Topic Distribution} \label{sec:app:candidate_distribution}

To illustrate the distribution of candidate documents retrieved for a given query, \cref{fig:piechart-candidates-NT} shows the domain and subdomain distribution for a representative Number Theory query and its associated documents. The shaded subdomains correspond to those associated with the query, while the black outlines indicate the domains and subdomains of the 10 documents with the highest relevance ratings. As the figure shows, the subdomains highlighted in the query are well represented among the retrieved documents. At the same time, the pipeline also identifies relevant problems from other domains, such as Combinatorics in this example, demonstrating that mathematical relevance is not necessarily confined to the same domain or subdomain.

\subsection{Licensing}\label{sec:app:licensing}
All benchmark material that we author ourselves is released under the Creative Commons Attribution-ShareAlike 4.0 International License (CC BY-SA 4.0). This covers our mathematical ontology, the topic and solution-summary annotations, the query and document selections, the pairwise judgments, and the Bradley--Terry relevance ratings derived from them. The license allows others to share and adapt this material for any purpose, even commercially, as long as they provide appropriate credit, indicate if changes were made, and distribute their contributions under the same license.

The problem and solution text that we redistribute, in contrast, remains subject to the terms of its original source, which our release neither replaces nor relicenses. All restrictions that apply to the original sources therefore continue to apply to the corresponding portions of our compiled corpus:

\begin{itemize}
    \item The \numina-derived portion retains its Apache License 2.0 conditions, which allow for commercial use and modification, but require attribution and do not permit sublicensing.
    \item The AoPS-derived portion remains subject to the applicable Art of Problem Solving terms of service, which generally allow for non-commercial use with proper attribution, and require attribution when the material is supplied as input to an AI system. We explicitly mark this portion as unavailable for model training.
    \item The official competition problems are typically in the public domain, or released under conditions similar to those of the two sources above, which likewise forbid the use of the data for training.
\end{itemize}
\section{Additional Construction Details}\label{sec:app:technical_details}

Here, we provide additional technical details about our benchmark construction.

\subsection{Computing the Jaccard Similarity} \label{sec:app:jaccard_details}

We use a standard implementation of the Jaccard similarity. We first tokenize the input sentences by keeping only alphanumeric sequences, and convert them to lowercase. We filter out stop words using the Scikit-learn English stop word list~\citep{scikit-learn}. We additionally introduce our own list of mathematics-specific stop words, that do not significantly factor into the meaning of solution summaries. The list of these words can be seen in our supplementary materials. Finally, we compute the Jaccard similarity as the size of the intersection of the two token sets divided by the size of their union.

\subsection{Selecting the Similarity Thresholds} \label{sec:app:thresholds}

In \cref{sec:tech:seclectbench}, we described our use of threshold-based filtering to select candidate documents for our benchmark. The thresholds were determined as follows. We first computed the distribution of similarity scores across all problem pairs in our corpus per domain. Then, for each domain, we derived the minimum threshold, for which at least 150 queries had at least 50 candidate documents above the threshold for each of the three categories: topic-only, summary-only, and both. To determine a final single threshold for each signal, we took the minimal thresholds obtained across the five domains, resulting in $\tau_{topic}\approx 0.858$ and $\tau_{summ}\approx 0.211$.

\subsection{Swiss-style Tournament Details} \label{sec:app:swiss_details}

We used a Swiss-style pairing procedure to assign head-to-head comparisons among the $150$ candidate items. In the first round, documents were paired uniformly at random. After each round, each document was assigned a running score equal to the number of times the LLM judge preferred the candidate up to that point. For all subsequent rounds, pairings were constructed greedily in ascending order of these scores.

We first randomly shuffled the documents among those with equal scores, then sorted them by their current win counts. This prevents potential biases from retaining the relative document order for documents with equal scores across the entire tournament. Starting from the lowest-scoring unpaired document, we select as its opponent an unpaired document with the closest win count among those with whom it had not previously been matched. Because this constraint can make a complete pairing infeasible in later rounds, some documents may remain unpaired in a given round. The full pairing procedure is summarized in \cref{alg:swiss_pairing}.

\subsection{Deduplication Experiment Setup} \label{sec:app:dedup_setup}

Deduplication is a routine curation step for mathematical corpora, where the same competition problem is frequently restated across sources and standard lexical pipelines may only catch near-verbatim copies, without capturing nuanced rephrasings. To capture the more challenging setting, for each of the $1000$ queries in \bench{}, we prompt \textsc{GPT-OSS-120B} to rewrite the problem statement, preserving the mathematical content while changing surface form. This includes renaming variables, restating the story context, and reformulating the notation where possible. We then place this rephrasing in the retrieval pool, which also contains all $71$K documents in \bench{}, from which we remove repeated entries and the source problem itself (when it appeared as a document to another query). We report the \emph{median rank} assigned to the correct original over $1000$ queries, and \emph{top-k}, the percentage of queries for which the original is placed among the $k$ highest-scored documents, for $k \in \{1, 4, 16, 64\}$ in \cref{tab:dedup-all-docs}. The first column repeats each system's \bench{} nDCG@10 for reference.

Because the pool is far too large to score with a cross-encoder, we restrict this experiment to bi-encoders, which also reflects a key limitation of cross-encoders in downstream use, namely that they are ineligible for first-stage retrieval. We nevertheless repeat the same experiment at the reranking scale, searching only within each query's own $150$-document candidate set, which yields a less discriminative measurement, but can be applied across all models we evaluate. We report the full results of both variants in \cref{sec:app:dedup}.

\section{Additional Details on Inference}\label{sec:app:inference_details}

In this section, we provide additional details on how we ran IR systems.

\subsection{IR System Taxonomy} \label{sec:app:taxonomy}

In \cref{tab:statement-full}, the ``Type'' column labels every system with a short
code whose final letter is its matching architecture. Bi-encoders (B) represent
query and document independently and compare the two representations directly, with the most representative example being the cosine similarity between two dense vectors produced by an embedding model. Late-interaction models (L) also represent them independently, but align them part by part at scoring time. Cross-encoders (C) read query and document together and score them jointly, either one pair at a time or over a group of candidates.

The letters preceding each category act as modifiers of the base type. ``A'' marks a system reachable only through a closed API. ``S'' marks a sparse representation, in which a text is a set of weighted terms or structural fragments rather than a dense vector, and covers learned sparse encoders as well as the non-neural lexical and symbolic matchers. ``R'' marks systems that were specifically trained on high-reasoning data, such as mathematics, medicine, or other scientific branches. ``T'' marks a system that generates a reasoning trace anywhere in its pipeline. Finally, ``W'' marks a pipeline in which a dedicated model rewrites the query before it is encoded.

\begin{table*}[t]
	\centering
	\small
	\caption{Fixed input formats applied to the models that require a specific one, where \texttt{\{q\}} and \texttt{\{d\}} denote the raw query and document text.}
	\label{tab:input-envelopes}
	\begin{tabular}{@{}lll@{}}
		\toprule
		Model & Query side & Document side \\
		\midrule
		\textsc{Gemma-300m} & \texttt{task: search result | query: \{q\}} & \texttt{title: none | text: \{d\}} \\
		\textsc{Jina-v5-Nano}, \textsc{Jina-v5-Small} & \texttt{Query: \{q\}} & \texttt{Document: \{d\}} \\
		\textsc{RaDeR-3B}, \textsc{RaDeR-7B}, \textsc{RaDeR-14B} & \texttt{query: \{q\}<|im\_end|>} & \texttt{document: \{d\}<|im\_end|>} \\
		\textsc{Gemini-001} & \texttt{task\_type=RETRIEVAL\_QUERY} & \texttt{task\_type=RETRIEVAL\_DOCUMENT} \\
		\bottomrule
	\end{tabular}
\end{table*}

\subsection{Running Retrieval Systems} \label{sec:app:embedding_details}

Unless stated otherwise, every method is evaluated at its implementation's default configuration. Below we give the shared evaluation protocol, followed by every setting we deviate from or that a reimplementation would otherwise have to guess.

\paragraph{Model behaviour} For every query--document pair in \bench{}, we compute a relevance score determined by the model at hand. Embedding models rank a candidate by the cosine similarity between the query and document embeddings, while cross-encoders, lexical retrievers and generative rerankers use their own relevance score.

\paragraph{Serving and numerical precision} Gemini and OpenAI models were called through their respective embedding APIs. Local models were run with vLLM~\citep{vllm} where the architecture is supported, and SentenceTransformers~\citep{sentencetransformers} otherwise. Before adopting vLLM for a given model, we validated its embeddings against those of the model's reference implementation on a fixed sample, requiring a Spearman correlation of at least $0.999$ between the two induced candidate rankings, so that the serving backend does not affect the \bench{} results. Every model is run in the precision its checkpoint declares, defaulting to \texttt{bfloat16} where the checkpoint declares none.

\paragraph{Embedding model output dimensionality and long inputs} For embedding models, output dimensionality was never specified manually, so models that support different dimensions used the maximum output dimension. If a piece of text exceeded the model's maximum input length, we split it into non-overlapping token chunks that fit the context window. Each chunk was embedded independently, and the final embedding was computed by elementwise averaging these embeddings. Only $0.32\%$ of the documents in \bench{} exceed $2048$ tokens, so this path is exercised mainly by the models with a $512$-token context.

\paragraph{Input formats} We followed each model's documented input formatting where one exists, with one exception. In particular, we deliberately run \textsc{Multilingual-E5-Large} without the \texttt{query: } and \texttt{passage: } prefixes its model card describes as required, as on \bench{} these prefixes consistently cost more than $0.05$ nDCG@10 average score, with and without an additional instruction, on both settings whose documents contain solutions.

For Gemini-001, we specifically use the \texttt{task\_type} parameter to distinguish between the query and document. We do not use this parameter for Gemini-2, as it ignores it. The full input formats we use for all models that require one are given in \cref{tab:input-envelopes}.

\subsection{Configuration of the RaDeR Models} \label{sec:app:rader_details}

We report results for three embedding models from the RaDeR family~\citep{rader}: \textsc{RaDeR-3B}, \textsc{RaDeR-7B} and \textsc{RaDeR-14B}. As their configuration departs from their model cards in two places, we state it in full. All are served with vLLM, and use last-token pooling with the query and document formats of \cref{tab:input-envelopes}. Pooling on the last <|im\_end|> token is specifically important, as the performance of RaDeR models collapses, if the <|im\_end|> token is not included in the tokenization, or if mean pooling is applied.

\subsection{Running Classical Lexical Baselines} \label{sec:app:lexical_baselines}

We evaluate four non-neural baselines: BM25~\citep{bm25}, TF-IDF, Jaccard token overlap~\citep{jaccard}, and Approach Zero~\citep{approachzero}. All are run at their implementations' defaults, which we state explicitly:

\begin{itemize}
    \item \textbf{BM25:} the \texttt{BM25Okapi} implementation of the \texttt{rank\_bm25} library at its library defaults, $k_1 = 1.5$, $b = 0.75$, and $\epsilon = 0.25$. We rebuild the index over each query's own $150$-document candidate set, so that inverse document frequencies are computed within the pool being ranked.
    \item \textbf{TF-IDF:} Scikit-learn's \texttt{TfidfVectorizer}~\citep{scikit-learn} at its defaults, fitted on the same per-query candidate set: unigrams, raw (non-sublinear) term frequencies, smoothed inverse document frequencies, and $L_2$ normalization, scored by cosine similarity against the query vector.
    \item \textbf{Jaccard:} the size of the intersection of the query and document token sets divided by the size of their union, with no term weighting and no inverse document frequency term.
    \item \textbf{Approach Zero:} the \texttt{pya0} structure-search engine itself, unmodified and used as a retrieval system in its own right rather than as a tokenizer.
\end{itemize}

\paragraph{Tokenization} Tokenization has a strong effect on lexical retrieval, so we report BM25, TF-IDF and Jaccard under two tokenization schemes. Our main results in \cref{tab:statement-full} use a mathematics-aware tokenizer derived from Approach Zero~\cite{approachzero} for all three methods, while \cref{tab:main-results-ci} additionally reports ``-no-tok'' variants: our own implementation of standard whitespace tokenization for BM25 and Jaccard, and the default \texttt{sklearn} tokenizer for TF-IDF.

The mathematics-aware tokenizer is not uniformly beneficial. It improves BM25 in every domain, taking it from $0.398$ to $0.416$ nDCG@10 overall, and improves Jaccard in all domains but Algebra, from $0.403$ to $0.412$ overall. It nevertheless \emph{hurts} TF-IDF in every domain, lowering it from $0.430$ to $0.412$ overall. In either tokenization, however, all three baselines remain far below every modern model, so this choice affects none of our conclusions.

\begin{algorithm}[t]
\centering
\fbox{%
\begin{minipage}{\dimexpr\linewidth-2\fboxsep-2\fboxrule\relax}
\footnotesize
\caption{Swiss-style greedy pairing.}
\label{alg:swiss_pairing}
\begin{algorithmic}[1]
\Require Query $Q$, Documents $D$, Rounds $R$
\State $w_d \gets 0$ for all $d \in D$; \quad $M \gets \emptyset$
\For{$r=1,\ldots,R$}
    \If{$r=1$}
        \State $P_r \gets$ random matching of $D$
    \Else
        \State $P_r \gets \emptyset$;
        \State \quad $U \gets$ shuffle ties, then sort $D$ by $w_d$
        \While{$|U| \geq 2$}
            \State $i \gets$ lowest-scoring document in $U$
            \State $F_i \gets \{j \in U \setminus \{i\}: \{i,j\} \notin M\}$
            \If{$F_i=\emptyset$}
                \State remove $i$ from $U$
            \Else
                \State $j \gets \arg\min_{k \in F_i} |w_i-w_k|$; break ties randomly
                \State $P_r \gets P_r \cup \{\{i,j\}\}$
                \State \quad $M \gets M \cup \{\{i,j\}\}$
                \State remove $i$ and $j$ from $U$
            \EndIf
        \EndWhile
    \EndIf
    \State judge similarity of pairs in $P_r$ to $Q$ and increment each winner's score
\EndFor
\end{algorithmic}
\end{minipage}
}
\end{algorithm}

\subsection{Measuring Per-Query Latency} \label{sec:app:timing}

We report the average per-query runtime in \cref{tab:statement-full} and on the latency axis of \cref{fig:quality-latency}, using a standardized per-query task time measured under the protocol below.

\paragraph{Setup} We measure the wall-clock time from the moment a query's document set is handed to a system until a final sorted ranking is available, for every system in our results tables, on the statement--full setting, so that the reported latency describes exactly the configuration whose nDCG@10 we report, and each timed query is scored against its own full candidate list of $150$ documents. We draw a sample of $50$ queries once, with a fixed seed, and store the resulting indices, so that every system is timed on precisely the same queries.

\paragraph{Serving parity} Every system is constructed through the same code path used to produce our main results, so that timing and scoring cannot diverge in how a model is built. For each model, we ensure this is the fastest way to run inference (up to parallelization) among those that reproduce the results of its default implementation. Most neural systems are therefore served with vLLM, subject to the per-model validation described in \cref{sec:app:embedding_details}. \textsc{Jina-v5-Nano}, whose EuroBERT backbone vLLM does not support, and \textsc{INF-Retriever-v1-Pro}, whose remote-code bidirectional attention it cannot serve, run under SentenceTransformers instead. We run SPLADE and ColBERT systems on SentenceTransformers as well, with their own remote code enabled.

\paragraph{Standardized parallelism} Rather than each implementation's tuned default, we fix a budget of $16$ documents in flight for every system, and cap all CPU thread pools at $16$ to match. For vLLM-served systems, candidates are submitted in slices of $16$ per engine call. The \textsc{Rank1} rerankers, which in the main evaluation issue a single generation call spanning all candidates, are sliced identically. For \textsc{Diver-GroupRank-32B} we retain the official group size of $20$, since altering it would alter the model's scores, and slice only the generation calls. The SPLADE and ColBERT systems take the batch size at construction instead. For the closed APIs, the same budget is enforced client-side: the OpenAI embedders expose no batch endpoint and are called with $16$ concurrent requests, \textsc{Gemini-001} with one $16$-document request at a time, and \textsc{Gemini-2}, whose API accepts a single document per request, with $16$ concurrent single-document requests. The classical baselines score in slices of $16$ against an index fit on the full candidate set. Approach Zero is the sole exception, as we were unable to enforce our own parallelism in its implementation.

\paragraph{Cold cache} Embedding-based systems are timed with document and query caching disabled, which removes the amortization a bi-encoder would otherwise obtain by reusing document vectors across queries. For the composed systems that rewrite the query before encoding it, this additionally forces the rewriting step to occur inside the timed region. We note that in a practical application, a bi-encoder would be deployed with document embeddings precomputed and cached, meaning a significant speedup could be obtained over the numbers we report.

\paragraph{Hardware and timed region} Every system is timed in isolation on a single H200 GPU, with 16 CPU cores and 120GB of RAM. Model loading falls outside the timed region.

\paragraph{Interpreting the reported times} These numbers measure the reranking task as \bench{} poses it, in which every candidate is scored anew for each query. They should not be read as first-stage retrieval costs for the bi-encoders, whose document embeddings are computed once offline and reused, leaving only the query encoding and a nearest-neighbour lookup at inference time. The separation between the two frontiers in \cref{fig:quality-latency} therefore understates the difference in deployed cost between bi-encoders and rerankers for first-stage retrieval.

\section{Additional Experiments and Results} \label{sec:app:additional_experiments}

Here, we display a varying range of additional results and analyses that complement our findings from the main text.

\subsection{Relative Performance of Relevance Signals} \label{sec:app:signal_ablation}

\begin{figure}[t]
    \centering
    \includegraphics[width=\columnwidth, keepaspectratio]{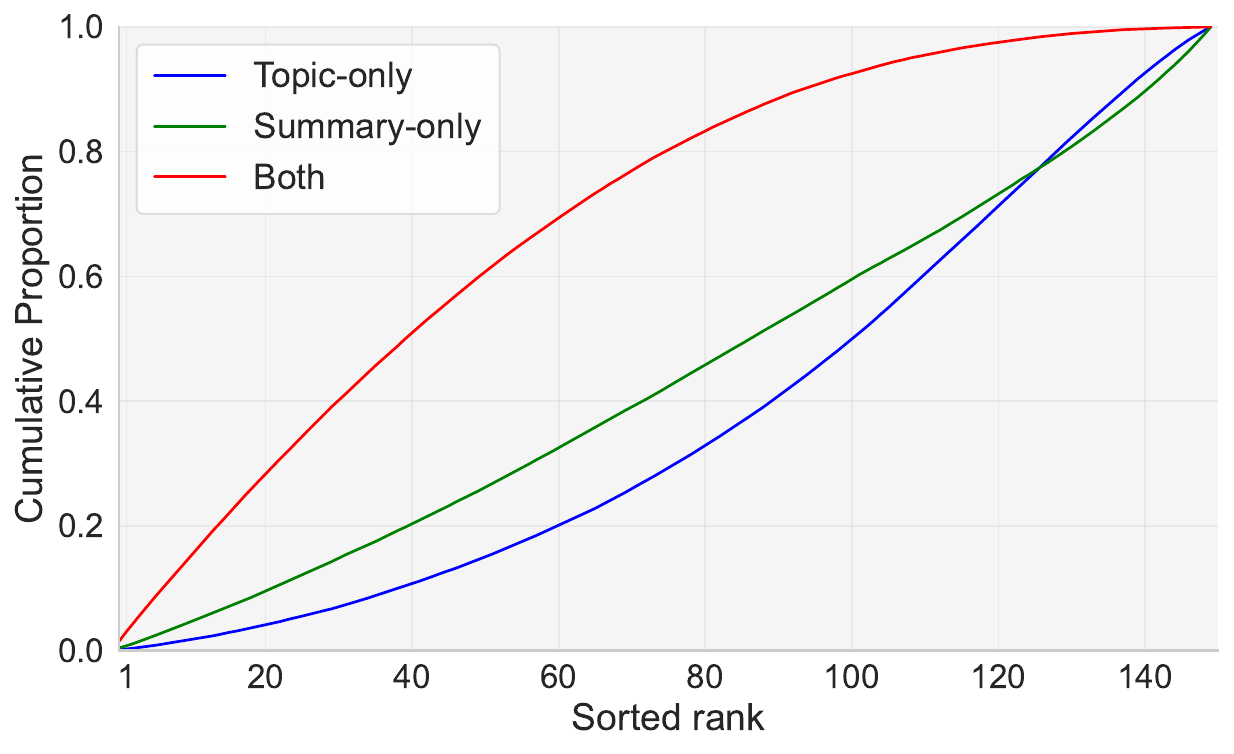}
    \vspace{-1.75em}
    \caption{Cumulative rank distribution of candidates selected by different relevance signals. Higher curves indicate that candidates from that group appear earlier in the final LLM-derived ranking.}
    \label{fig:3color-exp}
    \vspace{-1em}
\end{figure}

In \cref{sec:tech:seclectbench}, we use two proxy signals to select candidate documents: topic-based similarity and solution-summary similarity. Since each query in \bench contains candidates sampled from three groups---topic-only, summary-only, and both---we can ask how well each group fares under the final LLM-derived relevance ranking.

\cref{fig:3color-exp} plots, for each group, the cumulative fraction of candidates appearing within the top $r$ ranked positions, aggregated over all $1000$ queries. Thus, a faster rising curve corresponds to candidates that are placed higher by the final judge. Candidates selected by both signals rise fastest, showing that agreement between the topic and summary signals is a strong indicator of mathematical relevance. The single-signal groups are much closer to each other with summary-based similarity slightly ahead of topic similarity. Overall, the figure suggests that the two signals are complementary, each capturing a separate useful relevance notion.

\subsection{Selecting the Number of Rounds in the Swiss Tournament} \label{sec:app:num_rounds}

\begin{figure}[!hbt]
    \centering
    \includegraphics[width=\columnwidth, keepaspectratio]{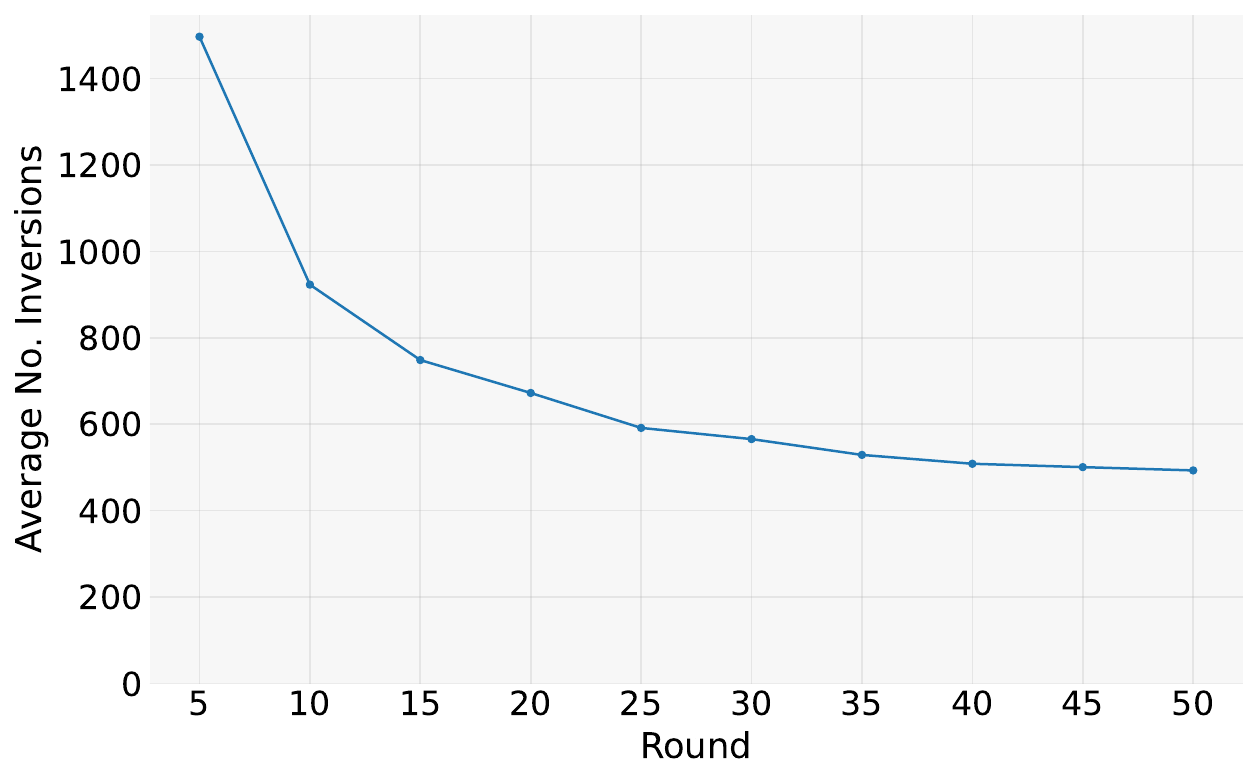}
    \vspace{-1.75em}
    \caption{Average number of inversions between orderings produced by factoring all pairs compared to a Swiss tournament.}
    \label{fig:avg-inv}
    \vspace{-1em}
\end{figure}

As described in \cref{sec:tech:rating}, we use a Swiss-style tournament to avoid exhaustive comparisons over all $\frac{n(n-1)}{2}=11175$ candidate pairs per query. To quantify the effect of the reduced comparisons, we compare the resulting Swiss-tournament rankings with rankings from exhaustive pairwise comparison. In \cref{fig:avg-inv}, we report the average number of preference inversions relative to the exhaustive ranking as a function of the number of rounds $R$.

We set $R=20$ in the final pipeline, which is near the point where additional rounds give diminishing returns. This reduces the number of comparisons from $11175$ to $1500$, a factor of $\sim 7.5$, while keeping the ranking close to the exhaustive one, with only $700$ inversions out of $11175$ possible pairs. In \cref{sec:app:human_evaluation}, we further confirm through human evaluation that this setting nearly matches the exhaustive ranking in terms of human preference.

\subsection{Final Metric Choice}\label{app:ndgccomp}
In \cref{tab:rescaling-robustness}, we show how the results of \cref{tab:statement-full} change under different versions of the nDCG metric, varying both the range of the relevance scores and the nDCG gain. The rankings are largely unaffected, with only 9 of the 861 possible model pairs reordered, and only for neighbouring models whose nDCG@10 in the original benchmark differs by less than
$0.0042$.
\begin{table}[t]
	\centering
	\footnotesize
	\renewcommand{\arraystretch}{1.05}
	\setlength{\tabcolsep}{6pt}
	\caption{Sensitivity of the \bench{} ranking to relevance score rescaling and the nDCG gain. Each variant re-scores all 42 evaluated models in \cref{tab:statement-full} and is compared against the original reported setting ([0,5] with exponential gain). Inversions count model pairs whose relative order changes, out of the $42\cdot41/2 = 861$ pairs that could. Linear gain is invariant to the rescaling, so its two rows coincide.}
	\label{tab:rescaling-robustness}
	\begin{tabular}{@{}l r r r@{}}
		\toprule
		Setting & Pearson & Spearman & Inversions \\
		\midrule
		$[0,3]$ + exponential gain & 0.9995 & 0.9995 & $3\,/\,861$ \\
		$[0,5]$ + linear gain & 0.9953 & 0.9985 & $9\,/\,861$ \\
		$[0,3]$ + linear gain & 0.9953 & 0.9985 & $9\,/\,861$ \\
		\bottomrule
	\end{tabular}
\end{table}

\subsection{Expert Validation of Benchmark and LLM-as-a-Judge} \label{sec:app:human_evaluation}

To validate our benchmark construction, we conduct a human evaluation of both the final benchmark rankings and the reliability of our LLM-as-a-judge procedure. Two authors with experience in high-school mathematical olympiads independently annotated a random sample of 200 examples. Each example consisted of a query problem and two candidate problems, together with their corresponding solutions. Annotators were asked to decide which candidate was more mathematically relevant to the query, with the option to mark a tie when the distinction was unclear or when the two candidates appeared similarly relevant. The instructions are provided in \cref{sec:app:human_evaluation_instructions}.

\paragraph{Inter-annotator agreement}

Of the 200 examples, 20 were judged by both annotators, allowing us to estimate inter-annotator agreement, where the annotators agreed on $75.0\%$ of the cases. This indicates that mathematical relevance judgments are somewhat subjective, especially after candidate prefiltering has removed clearly irrelevant documents. At the same time, the agreement rate suggests that the task contains substantial signal, as most disagreements arise from nuanced comparisons between plausibly related problems.

\paragraph{LLM-as-a-judge agreement with human annotators}

\begin{table}[t]
\centering
\footnotesize
\caption{Judge agreement rates across different document rank differences.}
\label{tab:judge-agreement-by-method-regime}

\renewcommand{\arraystretch}{1.05}
\setlength{\tabcolsep}{4pt}

\resizebox{\linewidth}{!}{%
\begin{tabular}{@{}lccc@{}}
\toprule
\multicolumn{1}{c}{Method}
& \multicolumn{1}{c}{$\Delta=0$}
& \multicolumn{1}{c}{$\Delta=25$}
& \multicolumn{1}{c}{$\Delta=50$} \\
\midrule
Inter-annotator & 75.0\% & 75.0\% & 85.7\% \\
\midrule
Pairwise Judge  & \textbf{78.0\%} & \textbf{82.8\%} & \textbf{86.3\%} \\
Majority Judge  & 77.5\% & 82.0\% & 85.6\% \\
Ordinal Ranking & 75.0\% & 78.9\% & \textbf{86.3\%} \\
\bottomrule
\end{tabular}%
}
\end{table}

We next validate our pairwise LLM judge by comparing it against two alternative judging regimes: (i) an ordinal relevance judge, which assigns each query-document pair an absolute relevance score, and (ii) a five-time majority-vote variant of the pairwise judge, where ties are assigned when the vote margin is at most one. All prompts are provided in \cref{sec:app:prompts}. To compare automatic judgments with human annotations, we assign a score of $1$ for exact agreement, $0.5$ when either judgment is a tie and the other expresses a preference, and $0$ otherwise.

\begin{table}[!htbp]
\centering
\footnotesize

\caption{Agreement between human annotators and each ranking/judging procedure, with 95\% Wilson score intervals and two-sided binomial $p$-values against chance agreement, $n=200$.}
\label{tab:human-eval-judge-result}
\renewcommand{\arraystretch}{1.05}
\setlength{\tabcolsep}{4pt}

\resizebox{\linewidth}{!}{
	\begin{tabular}{@{}lccc@{}}
	\toprule
	\multicolumn{1}{c}{Reference} & \multicolumn{1}{c}{Agreement} & \multicolumn{1}{c}{95\% Wilson CI} & \multicolumn{1}{c}{$p$ (vs.\ chance)} \\
	\midrule
	Pairwise Judge (Original)          & \textbf{0.780} & [0.718, 0.832] & $2.22\times 10^{-15}$ \\
	Full BT Winner                     & 0.790          & [0.728, 0.841] & $2.22\times 10^{-16}$ \\
	Majority Judge                     & 0.775          & [0.712, 0.827] & $7.33\times 10^{-15}$ \\
	Pairwise Judge (Confidence-based)  & 0.777          & [0.715, 0.830] & $4.22\times 10^{-15}$ \\
	DeepSeek Ranking                   & 0.775          & [0.712, 0.827] & $7.33\times 10^{-15}$ \\
	Ordinal Ranking                    & 0.750          & [0.686, 0.805] & $1.54\times 10^{-12}$ \\
	\bottomrule
	\end{tabular}
}

\end{table}

The pairwise LLM judge achieves $78.0\%$ agreement with human annotations (95\% Wilson CI $[0.718, 0.832]$), outperforming the ordinal judge, which achieves $75.0\%$ ($[0.686, 0.805]$), and slightly outperforming the majority-vote pairwise judge, which achieves $77.5\%$ ($[0.712, 0.827]$). \Cref{tab:human-eval-judge-result} additionally reports agreement for a confidence-based variant of the pairwise judge and a DeepSeek-based ranking judge, along with two-sided binomial $p$-values, which confirm that all judging regimes agree with human annotations significantly more than chance ($p < 10^{-11}$ in all cases). These results provide evidence that the pairwise LLM judge is a reliable proxy for expert relevance judgments in this setting. Moreover, the stronger performance of the pairwise judge relative to the ordinal judge supports our design choice of using comparative preferences rather than absolute relevance scores.

We further analyze judge agreement as a function of the final rank difference $\Delta_r$ between the two candidates, where $\Delta_r$ is computed from the full Bradley-Terry ranking. Larger values of $\Delta_r$ correspond to pairs whose estimated relevance differs more substantially, and should therefore be easier to judge. As shown in \cref{tab:judge-agreement-by-method-regime}, all judging regimes achieve higher agreement at larger $\Delta_r$ thresholds. In particular, the pairwise judge reaches $86.3\%$ agreement for pairs with $\Delta_r \geq 50$. This trend is consistent with the intuition that close comparisons are more subjective, while large relevance gaps are easier to identify reliably. The pairwise judge remains among the strongest methods across all $\Delta_r$ thresholds, further supporting its use in our benchmark construction.

\paragraph{Benchmark validation}

Finally, we use the same human annotations to validate our Swiss-style ranking procedure. We compare the preference implied by the final Swiss-tournament rankings against the human judgments and obtain $78.0\%$ agreement. For reference, performing full Bradley-Terry estimation over all candidate pairs yields $79.0\%$ agreement (95\% Wilson CI $[0.728, 0.841]$; \cref{tab:human-eval-judge-result}). The small gap between these two procedures suggests that the Swiss tournament recovers most of the relevant ranking information while requiring substantially fewer LLM comparisons.

We also compare against a random tournament, where each pair being compared is selected uniformly at random at each round. Using the same number of rounds as the Swiss tournament, this random scheduling baseline achieves $75.6\%$ agreement, below the Swiss-style procedure. This shows that Swiss-style scheduling is more effective than random comparison sampling at selecting informative pairwise judgments, and validates its use as a scalable approximation to exhaustive pairwise ranking.

\subsection{Confidence Intervals for \bench{} Results} \label{sec:app:confidence_intervals}

In \cref{tab:main-results-ci}, we report 95\% confidence intervals for the results presented in \cref{sec:results:main}, obtained via $1000$-fold bootstrap resampling of the queries. We observe slight inconsistencies of $\pm 0.001$ compared to the values in \cref{tab:statement-full} due to the stochastic nature of the resampling procedure.

\begin{table*}[!thbp]
	\centering
	\footnotesize
	
	\caption{Results of \bench{} with 95\% confidence intervals. All numbers are reported as nDCG@10. Additionally, we report results for the lexical benchmarks Jaccard~\citep{jaccard}, BM25~\citep{bm25} and TF-IDF where instead of using a specialized mathematical tokenizer as described in \cref{sec:app:lexical_baselines} they use standard white-space-based tokenization. These additional results are reported under the names Jaccard-no-tok, BM25-no-tok and TF-IDF-no-tok repsectively. The Type column follows the taxonomy described in \cref{sec:results:main}.}
	\label{tab:main-results-ci}
	\renewcommand{\arraystretch}{1.05}
	\setlength{\tabcolsep}{4pt}
	\newcommand{\beforerulepad}{\\[4pt]}
	\newcommand{\afterrulepad}{\rule{0pt}{\dimexpr\ht\strutbox+4pt\relax}}
	\newcommand{\tightmidrule}{\specialrule{\lightrulewidth}{0pt}{0pt}}
	\newcolumntype{R}{>{$}r<{$}}
	\resizebox{0.9\linewidth}{!}{
		\begin{tabular}{@{}l c RRRRRR@{}}
			\toprule
			& \multicolumn{1}{c}{\makecell{Type}}
			& \multicolumn{1}{c}{\makecell{Overall}}
			& \multicolumn{1}{c}{\makecell{Algebra}}
			& \multicolumn{1}{c}{\makecell{Geometry}}
			& \multicolumn{1}{c}{\makecell{Number\\Theory}}
			& \multicolumn{1}{c}{\makecell{Comb.}}
			& \multicolumn{1}{c}{\makecell[c]{Calc./\\Analysis}} \beforerulepad
			\tightmidrule
			\afterrulepad ReasonReranker-Qwen3-32B-Rewrite~\citep{reasonembed} & RTWC & \mathbf{0.741}^{+\mathbf{0.010}}_{-\mathbf{0.010}} & \mathbf{0.710}^{+\mathbf{0.020}}_{-\mathbf{0.020}} & 0.749^{+0.017}_{-0.017} & \mathbf{0.740}^{+\mathbf{0.018}}_{-\mathbf{0.018}} & \mathbf{0.783}^{+\mathbf{0.016}}_{-\mathbf{0.016}} & \mathbf{0.732}^{+\mathbf{0.017}}_{-\mathbf{0.017}} \\
			ReasonEmbed-Qwen3-8B-Rewrite~\citep{reasonembed} & RTWB & 0.738^{+0.009}_{-0.010} & \mathbf{0.710}^{+\mathbf{0.018}}_{-\mathbf{0.018}} & \mathbf{0.767}^{+\mathbf{0.016}}_{-\mathbf{0.017}} & 0.734^{+0.016}_{-0.017} & 0.781^{+0.016}_{-0.016} & 0.708^{+0.017}_{-0.018} \\
			ReasonReranker-Qwen3-32B~\citep{reasonembed} & RTC & 0.733^{+0.010}_{-0.010} & 0.699^{+0.019}_{-0.020} & 0.751^{+0.016}_{-0.015} & 0.726^{+0.018}_{-0.018} & 0.777^{+0.015}_{-0.015} & 0.712^{+0.018}_{-0.020} \\
			ReasonEmbed-Qwen3-8B~\citep{reasonembed} & RB & 0.710^{+0.010}_{-0.010} & 0.689^{+0.018}_{-0.017} & 0.744^{+0.018}_{-0.017} & 0.697^{+0.018}_{-0.018} & 0.738^{+0.016}_{-0.017} & 0.682^{+0.017}_{-0.017} \\
			Diver-GroupRank-32B~\citep{grouprank} & RTC & 0.693^{+0.010}_{-0.009} & 0.645^{+0.017}_{-0.018} & 0.716^{+0.015}_{-0.015} & 0.674^{+0.018}_{-0.017} & 0.729^{+0.014}_{-0.015} & 0.696^{+0.017}_{-0.018} \\
			RaDeR-14B~\citep{rader} & RB & 0.692^{+0.009}_{-0.010} & 0.669^{+0.018}_{-0.018} & 0.727^{+0.017}_{-0.016} & 0.682^{+0.017}_{-0.018} & 0.714^{+0.018}_{-0.017} & 0.665^{+0.018}_{-0.019} \\
			RaDeR-7B~\citep{rader} & RB & 0.690^{+0.009}_{-0.010} & 0.666^{+0.018}_{-0.019} & 0.725^{+0.016}_{-0.017} & 0.675^{+0.018}_{-0.018} & 0.715^{+0.017}_{-0.017} & 0.666^{+0.018}_{-0.018} \\
			Qwen3-Reranker-4B~\citep{qwen3embedding} & C & 0.683^{+0.010}_{-0.009} & 0.649^{+0.018}_{-0.019} & 0.709^{+0.016}_{-0.015} & 0.672^{+0.017}_{-0.017} & 0.712^{+0.016}_{-0.016} & 0.671^{+0.017}_{-0.018} \\
			Diver-4B~\citep{diver} & RB & 0.681^{+0.009}_{-0.010} & 0.665^{+0.017}_{-0.017} & 0.711^{+0.016}_{-0.018} & 0.672^{+0.018}_{-0.018} & 0.708^{+0.017}_{-0.018} & 0.645^{+0.017}_{-0.018} \\
			Qwen3-Reranker-8B~\citep{qwen3embedding} & C & 0.666^{+0.010}_{-0.010} & 0.632^{+0.018}_{-0.018} & 0.701^{+0.016}_{-0.016} & 0.658^{+0.019}_{-0.019} & 0.697^{+0.017}_{-0.017} & 0.639^{+0.017}_{-0.018} \\
			ReasonEmbed-Llama-3.1-8B~\citep{reasonembed} & RB & 0.664^{+0.009}_{-0.010} & 0.641^{+0.017}_{-0.017} & 0.695^{+0.016}_{-0.016} & 0.652^{+0.018}_{-0.018} & 0.678^{+0.016}_{-0.017} & 0.650^{+0.017}_{-0.016} \\
			RaDeR-3B~\citep{rader} & RB & 0.654^{+0.010}_{-0.010} & 0.629^{+0.019}_{-0.019} & 0.694^{+0.016}_{-0.016} & 0.648^{+0.017}_{-0.018} & 0.680^{+0.018}_{-0.018} & 0.624^{+0.018}_{-0.018} \\
			Qwen3-Reranker-0.6B~\citep{qwen3embedding} & C & 0.652^{+0.009}_{-0.009} & 0.629^{+0.018}_{-0.019} & 0.690^{+0.016}_{-0.016} & 0.638^{+0.018}_{-0.017} & 0.667^{+0.016}_{-0.016} & 0.634^{+0.017}_{-0.017} \\
			SPLADE-Code-8B~\citep{splade} & SB & 0.650^{+0.010}_{-0.009} & 0.610^{+0.016}_{-0.017} & 0.689^{+0.016}_{-0.016} & 0.644^{+0.017}_{-0.017} & 0.694^{+0.016}_{-0.016} & 0.618^{+0.016}_{-0.016} \\
			Octen-8B~\citep{octen2025rteb} & B & 0.636^{+0.010}_{-0.009} & 0.594^{+0.017}_{-0.017} & 0.664^{+0.016}_{-0.017} & 0.629^{+0.017}_{-0.018} & 0.665^{+0.017}_{-0.017} & 0.630^{+0.017}_{-0.018} \\
			Diver-0.6B~\citep{diver} & RB & 0.632^{+0.010}_{-0.009} & 0.619^{+0.018}_{-0.018} & 0.668^{+0.017}_{-0.016} & 0.617^{+0.018}_{-0.017} & 0.655^{+0.017}_{-0.018} & 0.609^{+0.016}_{-0.017} \\
			Octen-4B~\citep{octen2025rteb} & B & 0.632^{+0.009}_{-0.010} & 0.586^{+0.017}_{-0.017} & 0.673^{+0.016}_{-0.016} & 0.627^{+0.019}_{-0.017} & 0.667^{+0.016}_{-0.017} & 0.609^{+0.016}_{-0.017} \\
			Gemini-2~\citep{geminiembedding} & AB & 0.628^{+0.009}_{-0.009} & 0.599^{+0.016}_{-0.016} & 0.647^{+0.016}_{-0.017} & 0.622^{+0.017}_{-0.016} & 0.656^{+0.016}_{-0.017} & 0.614^{+0.015}_{-0.017} \\
			Gemini-001~\citep{geminiembedding} & AB & 0.616^{+0.010}_{-0.009} & 0.585^{+0.018}_{-0.017} & 0.651^{+0.015}_{-0.015} & 0.617^{+0.018}_{-0.017} & 0.635^{+0.016}_{-0.017} & 0.596^{+0.018}_{-0.017} \\
			Qwen3-Embedding-4B~\citep{qwen3embedding} & B & 0.615^{+0.009}_{-0.009} & 0.576^{+0.017}_{-0.018} & 0.652^{+0.016}_{-0.017} & 0.610^{+0.018}_{-0.017} & 0.642^{+0.016}_{-0.017} & 0.597^{+0.017}_{-0.017} \\
			Qwen3-Embedding-8B~\citep{qwen3embedding} & B & 0.611^{+0.009}_{-0.010} & 0.569^{+0.017}_{-0.017} & 0.647^{+0.016}_{-0.017} & 0.606^{+0.017}_{-0.018} & 0.633^{+0.017}_{-0.016} & 0.598^{+0.017}_{-0.017} \\
			Rank1-32B~\citep{rank1} & RTC & 0.610^{+0.010}_{-0.010} & 0.583^{+0.019}_{-0.019} & 0.617^{+0.018}_{-0.018} & 0.613^{+0.018}_{-0.018} & 0.673^{+0.017}_{-0.018} & 0.574^{+0.018}_{-0.018} \\
			Harrier-27b~\citep{harrier} & B & 0.608^{+0.008}_{-0.010} & 0.569^{+0.016}_{-0.016} & 0.649^{+0.016}_{-0.016} & 0.601^{+0.016}_{-0.016} & 0.620^{+0.018}_{-0.016} & 0.596^{+0.016}_{-0.016} \\
			SPLADE-Code-0.6B~\citep{splade} & SB & 0.595^{+0.009}_{-0.010} & 0.565^{+0.017}_{-0.017} & 0.646^{+0.016}_{-0.017} & 0.591^{+0.017}_{-0.018} & 0.625^{+0.016}_{-0.017} & 0.552^{+0.016}_{-0.016} \\
			INF-Retriever-v1-Pro~\citep{infretriever} & RB & 0.594^{+0.009}_{-0.009} & 0.573^{+0.017}_{-0.017} & 0.618^{+0.016}_{-0.017} & 0.585^{+0.017}_{-0.017} & 0.607^{+0.017}_{-0.016} & 0.592^{+0.016}_{-0.016} \\
			KaLM-12B-2511~\citep{kalm} & B & 0.584^{+0.009}_{-0.009} & 0.548^{+0.015}_{-0.017} & 0.617^{+0.016}_{-0.015} & 0.583^{+0.015}_{-0.018} & 0.606^{+0.017}_{-0.016} & 0.569^{+0.016}_{-0.016} \\
			LLaMa-Nemotron-8B~\citep{nemotron} & B & 0.579^{+0.009}_{-0.009} & 0.542^{+0.015}_{-0.016} & 0.610^{+0.015}_{-0.016} & 0.580^{+0.016}_{-0.016} & 0.600^{+0.017}_{-0.016} & 0.562^{+0.016}_{-0.018} \\
			Qwen3-Embedding-0.6B~\citep{qwen3embedding} & B & 0.575^{+0.009}_{-0.009} & 0.545^{+0.017}_{-0.016} & 0.629^{+0.017}_{-0.016} & 0.564^{+0.017}_{-0.017} & 0.589^{+0.017}_{-0.017} & 0.546^{+0.016}_{-0.016} \\
			Harrier-0.6b~\citep{harrier} & B & 0.572^{+0.009}_{-0.009} & 0.538^{+0.016}_{-0.016} & 0.613^{+0.016}_{-0.016} & 0.566^{+0.017}_{-0.017} & 0.581^{+0.017}_{-0.017} & 0.557^{+0.016}_{-0.017} \\
			Jina-v5-Small~\citep{jina} & B & 0.565^{+0.009}_{-0.010} & 0.521^{+0.016}_{-0.016} & 0.611^{+0.016}_{-0.017} & 0.557^{+0.017}_{-0.017} & 0.589^{+0.018}_{-0.018} & 0.545^{+0.016}_{-0.015} \\
			Reason-ColBERT~\citep{reasonir} & RL & 0.558^{+0.009}_{-0.009} & 0.533^{+0.018}_{-0.017} & 0.594^{+0.016}_{-0.016} & 0.555^{+0.018}_{-0.016} & 0.567^{+0.017}_{-0.016} & 0.537^{+0.015}_{-0.015} \\
			Text-Embedding-3-Large~\citep{openaiembedding} & AB & 0.557^{+0.009}_{-0.009} & 0.535^{+0.015}_{-0.017} & 0.571^{+0.015}_{-0.016} & 0.560^{+0.016}_{-0.017} & 0.574^{+0.016}_{-0.017} & 0.552^{+0.016}_{-0.015} \\
			Rank1-7B~\citep{rank1} & RTC & 0.551^{+0.010}_{-0.010} & 0.517^{+0.019}_{-0.019} & 0.575^{+0.016}_{-0.017} & 0.548^{+0.019}_{-0.019} & 0.615^{+0.018}_{-0.018} & 0.512^{+0.019}_{-0.018} \\
			Jina-v5-Nano~\citep{jina} & B & 0.530^{+0.009}_{-0.009} & 0.488^{+0.015}_{-0.016} & 0.568^{+0.017}_{-0.017} & 0.532^{+0.017}_{-0.017} & 0.551^{+0.017}_{-0.017} & 0.510^{+0.016}_{-0.016} \\
			GTE-ColBERT~\citep{gte} & L & 0.527^{+0.009}_{-0.009} & 0.506^{+0.018}_{-0.016} & 0.559^{+0.016}_{-0.016} & 0.519^{+0.016}_{-0.016} & 0.538^{+0.017}_{-0.018} & 0.519^{+0.015}_{-0.015} \\
			Gemma-300m~\citep{embeddinggemma} & B & 0.527^{+0.009}_{-0.009} & 0.505^{+0.017}_{-0.016} & 0.559^{+0.016}_{-0.015} & 0.522^{+0.017}_{-0.016} & 0.543^{+0.018}_{-0.017} & 0.504^{+0.015}_{-0.016} \\
			Text-Embedding-3-Small~\citep{openaiembedding} & AB & 0.512^{+0.010}_{-0.009} & 0.487^{+0.016}_{-0.015} & 0.557^{+0.016}_{-0.016} & 0.504^{+0.017}_{-0.016} & 0.526^{+0.017}_{-0.017} & 0.491^{+0.016}_{-0.016} \\
			BGE-m3~\citep{bge-m3} & B & 0.511^{+0.010}_{-0.008} & 0.484^{+0.017}_{-0.016} & 0.549^{+0.016}_{-0.017} & 0.502^{+0.019}_{-0.017} & 0.518^{+0.018}_{-0.017} & 0.500^{+0.014}_{-0.014} \\
			ReasonIR-8B~\citep{reasonir} & RB & 0.507^{+0.009}_{-0.009} & 0.495^{+0.015}_{-0.016} & 0.511^{+0.016}_{-0.016} & 0.500^{+0.017}_{-0.016} & 0.534^{+0.016}_{-0.017} & 0.489^{+0.014}_{-0.014} \\
			Harrier-270m~\citep{harrier} & B & 0.498^{+0.009}_{-0.009} & 0.470^{+0.016}_{-0.016} & 0.545^{+0.016}_{-0.017} & 0.491^{+0.016}_{-0.017} & 0.512^{+0.016}_{-0.016} & 0.469^{+0.014}_{-0.015} \\
			INF-X-Retriever~\citep{infretriever} & RWB & 0.494^{+0.010}_{-0.011} & 0.493^{+0.016}_{-0.017} & 0.468^{+0.022}_{-0.023} & 0.450^{+0.018}_{-0.017} & 0.535^{+0.019}_{-0.019} & 0.522^{+0.017}_{-0.017} \\ 
			Multilingual-E5-Large~\citep{e5} & B & 0.488^{+0.008}_{-0.009} & 0.455^{+0.015}_{-0.014} & 0.529^{+0.016}_{-0.018} & 0.471^{+0.017}_{-0.016} & 0.500^{+0.016}_{-0.016} & 0.477^{+0.014}_{-0.014} \\
			Approach Zero~\citep{approachzero} & SL & 0.468^{+0.009}_{-0.010} & 0.485^{+0.017}_{-0.016} & 0.443^{+0.016}_{-0.016} & 0.454^{+0.017}_{-0.017} & 0.480^{+0.018}_{-0.017} & 0.490^{+0.016}_{-0.017} \\
			TF-IDF-no-tok & SB & 0.430^{+0.008}_{-0.008} & 0.416^{+0.015}_{-0.014} & 0.430^{+0.015}_{-0.015} & 0.399^{+0.014}_{-0.014} & 0.459^{+0.016}_{-0.016} & 0.430^{+0.014}_{-0.015} \\
			BM25~\citep{bm25} & SB & 0.416^{+0.009}_{-0.008} & 0.405^{+0.015}_{-0.015} & 0.448^{+0.016}_{-0.016} & 0.392^{+0.015}_{-0.014} & 0.429^{+0.016}_{-0.015} & 0.393^{+0.014}_{-0.013} \\
			Jaccard~\citep{jaccard} & SB & 0.412^{+0.008}_{-0.007} & 0.381^{+0.013}_{-0.013} & 0.448^{+0.015}_{-0.014} & 0.397^{+0.015}_{-0.015} & 0.442^{+0.015}_{-0.014} & 0.383^{+0.012}_{-0.012} \\
			TF-IDF & SB & 0.412^{+0.009}_{-0.009} & 0.414^{+0.016}_{-0.015} & 0.402^{+0.018}_{-0.018} & 0.383^{+0.016}_{-0.016} & 0.424^{+0.016}_{-0.016} & 0.414^{+0.014}_{-0.014} \\
			Jaccard-no-tok~\citep{jaccard} & SB & 0.403^{+0.009}_{-0.008} & 0.387^{+0.015}_{-0.014} & 0.426^{+0.017}_{-0.017} & 0.394^{+0.015}_{-0.015} & 0.443^{+0.016}_{-0.016} & 0.358^{+0.012}_{-0.013} \\
			BM25-no-tok~\citep{bm25} & SB & 0.398^{+0.008}_{-0.008} & 0.394^{+0.015}_{-0.015} & 0.420^{+0.017}_{-0.016} & 0.381^{+0.015}_{-0.014} & 0.422^{+0.016}_{-0.017} & 0.368^{+0.013}_{-0.013} \\
			Rank1-0.5B~\citep{rank1} & RTC & 0.365^{+0.007}_{-0.007} & 0.376^{+0.012}_{-0.012} & 0.390^{+0.013}_{-0.012} & 0.340^{+0.012}_{-0.011} & 0.353^{+0.014}_{-0.013} & 0.374^{+0.012}_{-0.012} \\
			BERT~\citep{bert} & B & 0.357^{+0.008}_{-0.007} & 0.369^{+0.014}_{-0.013} & 0.342^{+0.015}_{-0.014} & 0.344^{+0.013}_{-0.012} & 0.389^{+0.017}_{-0.014} & 0.335^{+0.012}_{-0.012} \\
			RoBERTa~\citep{roberta} & B & 0.311^{+0.006}_{-0.007} & 0.306^{+0.011}_{-0.010} & 0.293^{+0.013}_{-0.013} & 0.314^{+0.013}_{-0.012} & 0.342^{+0.013}_{-0.014} & 0.287^{+0.011}_{-0.011} \\
			\bottomrule
		\end{tabular}
	}
	
\end{table*}

\subsection{Results on \bench{} on Additional Tasks} \label{sec:app:additional_settings}

We include in \cref{tab:additional_settings} the results for the supplementary settings of \bench{}, including statement-statement and full-full retrieval. For the most part, the same trends observed in the main setting hold across these additional settings, and their respective domains. In general, rerankers seem to make better use of the additional context, as they on average fall behind in the statement-statement setting, but perform slightly better on the full-full one. However, the overall ranking of models remains relatively consistent across settings, with the strongest models in the main setting generally remaining strong in the additional settings.

\begin{table*}[!htbp]
	\centering
	\scriptsize
	\caption{Results across the non-main settings of \bench{}. All numbers are reported as nDCG@10. The Type column follows the taxonomy described in \cref{sec:results:main}.}
	\label{tab:additional_settings}
	\renewcommand{\arraystretch}{1.05}
	\setlength{\tabcolsep}{2.5pt}
	\newcolumntype{R}{>{$}r<{$}}
	\resizebox{\textwidth}{!}{%
		\begin{tabular}{@{}l c RRRRRR RRRRRR@{}}
			\toprule
			&
			& \multicolumn{6}{c}{\textbf{Statement--Statement}}
			& \multicolumn{6}{c}{\textbf{Full--Full}} \\
			\cmidrule(lr){3-8}
			\cmidrule(lr){9-14}
			& \multicolumn{1}{c}{\makecell{Type}}
			& \multicolumn{1}{c}{\makecell{Overall}}
			& \multicolumn{1}{c}{\makecell{Algebra}}
			& \multicolumn{1}{c}{\makecell{Geometry}}
			& \multicolumn{1}{c}{\makecell{Number\\Theory}}
			& \multicolumn{1}{c}{\makecell{Comb.}}
			& \multicolumn{1}{c}{\makecell[c]{Calc./\\Analysis}}
			& \multicolumn{1}{c}{\makecell{Overall}}
			& \multicolumn{1}{c}{\makecell{Algebra}}
			& \multicolumn{1}{c}{\makecell{Geometry}}
			& \multicolumn{1}{c}{\makecell{Number\\Theory}}
			& \multicolumn{1}{c}{\makecell{Comb.}}
			& \multicolumn{1}{c}{\makecell[c]{Calc./\\Analysis}} \\
			\midrule
			ReasonEmbed-Qwen3-8B & RB
			& \mathbf{0.685} & \mathbf{0.668} & 0.706 & \mathbf{0.675} & \mathbf{0.724} & \mathbf{0.655}
			& 0.795 & 0.768 & \mathbf{0.819} & 0.793 & 0.823 & 0.775 \\
			RaDeR-14B & RB
			& 0.677 & 0.654 & \mathbf{0.708} & 0.666 & 0.709 & 0.645
			& 0.741 & 0.715 & 0.769 & 0.736 & 0.767 & 0.723 \\
			RaDeR-7B & RB
			& 0.671 & 0.644 & 0.705 & 0.657 & 0.704 & 0.642
			& 0.738 & 0.716 & 0.766 & 0.739 & 0.762 & 0.712 \\
			Diver-4B & RB
			& 0.659 & 0.641 & 0.696 & 0.647 & 0.691 & 0.619
			& 0.737 & 0.713 & 0.771 & 0.737 & 0.759 & 0.711 \\
			ReasonEmbed-Qwen3-8B-Rewrite & RTWB
			& 0.659 & 0.646 & 0.691 & 0.637 & 0.690 & 0.628
			& 0.769 & 0.736 & 0.796 & 0.764 & 0.804 & 0.749 \\
			Qwen3-Reranker-4B & C
			& 0.653 & 0.631 & 0.670 & 0.640 & 0.691 & 0.633
			& 0.730 & 0.708 & 0.745 & 0.727 & 0.745 & 0.722 \\
			Diver-GroupRank-32B & RTC
			& 0.651 & 0.609 & 0.685 & 0.632 & 0.686 & 0.639
			& 0.729 & 0.688 & 0.740 & 0.721 & 0.748 & 0.745 \\
			ReasonReranker-Qwen3-32B & RTC
			& 0.647 & 0.610 & 0.674 & 0.624 & 0.700 & 0.622
			& \mathbf{0.800} & \mathbf{0.782} & 0.798 & \mathbf{0.803} & \mathbf{0.823} & \mathbf{0.799} \\
			RaDeR-3B & RB
			& 0.644 & 0.622 & 0.677 & 0.628 & 0.677 & 0.611
			& 0.705 & 0.675 & 0.740 & 0.707 & 0.726 & 0.684 \\
			ReasonEmbed-Llama-3.1-8B & RB
			& 0.643 & 0.624 & 0.667 & 0.632 & 0.671 & 0.619
			& 0.751 & 0.718 & 0.782 & 0.749 & 0.771 & 0.735 \\
			ReasonReranker-Qwen3-32B-Rewrite & RTWC
			& 0.638 & 0.609 & 0.664 & 0.614 & 0.688 & 0.616
			& 0.796 & 0.776 & 0.802 & 0.797 & 0.816 & 0.797 \\
			SPLADE-Code-8B & SB
			& 0.634 & 0.592 & 0.665 & 0.633 & 0.675 & 0.600
			& 0.693 & 0.649 & 0.725 & 0.691 & 0.729 & 0.673 \\
			Octen-8B & B
			& 0.623 & 0.586 & 0.648 & 0.624 & 0.654 & 0.606
			& 0.672 & 0.631 & 0.685 & 0.670 & 0.700 & 0.673 \\
			Qwen3-Reranker-0.6B & C
			& 0.621 & 0.601 & 0.661 & 0.609 & 0.638 & 0.598
			& 0.717 & 0.695 & 0.744 & 0.723 & 0.730 & 0.699 \\
			Octen-4B & B
			& 0.619 & 0.581 & 0.659 & 0.615 & 0.651 & 0.591
			& 0.663 & 0.616 & 0.695 & 0.658 & 0.695 & 0.653 \\
			Diver-0.6B & RB
			& 0.611 & 0.589 & 0.639 & 0.601 & 0.640 & 0.586
			& 0.695 & 0.660 & 0.727 & 0.701 & 0.718 & 0.673 \\
			Qwen3-Reranker-8B & C
			& 0.609 & 0.578 & 0.623 & 0.607 & 0.644 & 0.587
			& 0.714 & 0.684 & 0.722 & 0.722 & 0.742 & 0.698 \\
			Gemini-2 & AB
			& 0.603 & 0.569 & 0.623 & 0.602 & 0.630 & 0.587
			& 0.658 & 0.630 & 0.665 & 0.659 & 0.675 & 0.652 \\
			Qwen3-Embedding-8B & B
			& 0.600 & 0.562 & 0.643 & 0.600 & 0.622 & 0.569
			& 0.662 & 0.621 & 0.693 & 0.660 & 0.679 & 0.657 \\
			Gemini-001 & AB
			& 0.599 & 0.558 & 0.636 & 0.605 & 0.622 & 0.573
			& 0.667 & 0.622 & 0.690 & 0.677 & 0.685 & 0.661 \\
			Qwen3-Embedding-4B & B
			& 0.597 & 0.554 & 0.638 & 0.591 & 0.626 & 0.573
			& 0.667 & 0.624 & 0.693 & 0.670 & 0.693 & 0.656 \\
			Harrier-27b & B
			& 0.585 & 0.554 & 0.612 & 0.586 & 0.609 & 0.562
			& 0.659 & 0.615 & 0.700 & 0.664 & 0.678 & 0.640 \\
			KaLM-12B-2511 & B
			& 0.579 & 0.546 & 0.605 & 0.578 & 0.597 & 0.565
			& 0.599 & 0.557 & 0.628 & 0.607 & 0.618 & 0.585 \\
			SPLADE-Code-0.6B & SB
			& 0.578 & 0.550 & 0.618 & 0.583 & 0.613 & 0.531
			& 0.648 & 0.609 & 0.692 & 0.652 & 0.677 & 0.611 \\
			Qwen3-Embedding-0.6B & B
			& 0.566 & 0.529 & 0.611 & 0.564 & 0.586 & 0.538
			& 0.625 & 0.586 & 0.676 & 0.624 & 0.634 & 0.603 \\
			LLaMa-Nemotron-8B & B
			& 0.565 & 0.535 & 0.584 & 0.578 & 0.588 & 0.540
			& 0.616 & 0.569 & 0.641 & 0.629 & 0.638 & 0.602 \\
			INF-Retriever-v1-Pro & RB
			& 0.564 & 0.540 & 0.593 & 0.553 & 0.591 & 0.546
			& 0.660 & 0.617 & 0.686 & 0.665 & 0.675 & 0.657 \\
			Rank1-32B & RTC
			& 0.556 & 0.539 & 0.541 & 0.549 & 0.619 & 0.532
			& 0.629 & 0.594 & 0.645 & 0.624 & 0.682 & 0.605 \\
			Jina-v5-Small & B
			& 0.549 & 0.508 & 0.593 & 0.544 & 0.571 & 0.528
			& 0.611 & 0.558 & 0.662 & 0.611 & 0.631 & 0.597 \\
			Harrier-0.6b & B
			& 0.549 & 0.521 & 0.589 & 0.538 & 0.568 & 0.523
			& 0.632 & 0.588 & 0.676 & 0.637 & 0.638 & 0.613 \\
			Text-Embedding-3-Large & AB
			& 0.540 & 0.522 & 0.554 & 0.541 & 0.560 & 0.526
			& 0.594 & 0.564 & 0.609 & 0.598 & 0.611 & 0.588 \\
			Reason-ColBERT & RL
			& 0.539 & 0.519 & 0.568 & 0.536 & 0.554 & 0.516
			& 0.611 & 0.574 & 0.648 & 0.613 & 0.618 & 0.602 \\
			Jina-v5-Nano & B
			& 0.519 & 0.481 & 0.561 & 0.522 & 0.545 & 0.488
			& 0.572 & 0.526 & 0.607 & 0.578 & 0.585 & 0.560 \\
			ReasonIR-8B & RB
			& 0.513 & 0.504 & 0.529 & 0.503 & 0.539 & 0.482
			& 0.617 & 0.584 & 0.635 & 0.618 & 0.629 & 0.613 \\
			Gemma-300m & B
			& 0.511 & 0.491 & 0.540 & 0.500 & 0.536 & 0.487
			& 0.585 & 0.548 & 0.631 & 0.594 & 0.598 & 0.553 \\
			Multilingual-E5-Large & B
			& 0.508 & 0.485 & 0.549 & 0.501 & 0.523 & 0.475
			& 0.545 & 0.505 & 0.580 & 0.549 & 0.544 & 0.536 \\
			Rank1-7B & RTC
			& 0.506 & 0.483 & 0.507 & 0.491 & 0.569 & 0.483
			& 0.467 & 0.444 & 0.499 & 0.456 & 0.531 & 0.410 \\
			BGE-m3 & B
			& 0.505 & 0.483 & 0.539 & 0.498 & 0.518 & 0.485
			& 0.563 & 0.513 & 0.616 & 0.575 & 0.567 & 0.543 \\
			Harrier-270m & B
			& 0.497 & 0.466 & 0.537 & 0.489 & 0.517 & 0.473
			& 0.557 & 0.525 & 0.590 & 0.565 & 0.565 & 0.539 \\
			Text-Embedding-3-Small & AB
			& 0.497 & 0.470 & 0.540 & 0.483 & 0.519 & 0.473
			& 0.554 & 0.520 & 0.597 & 0.552 & 0.565 & 0.536 \\
			GTE-ColBERT & L
			& 0.494 & 0.477 & 0.507 & 0.488 & 0.510 & 0.490
			& 0.577 & 0.544 & 0.605 & 0.577 & 0.577 & 0.583 \\
			Approach Zero & SL
			& 0.469 & 0.489 & 0.446 & 0.456 & 0.483 & 0.484
			& 0.481 & 0.493 & 0.471 & 0.461 & 0.505 & 0.487 \\
			Jaccard & SB
			& 0.448 & 0.425 & 0.476 & 0.445 & 0.477 & 0.406
			& 0.469 & 0.438 & 0.500 & 0.472 & 0.496 & 0.419 \\
			INF-X-Retriever & RWB
			& 0.437 & 0.446 & 0.409 & 0.393 & 0.470 & 0.461
			& 0.508 & 0.516 & 0.479 & 0.461 & 0.527 & 0.553 \\
			TF-IDF & SB
			& 0.434 & 0.445 & 0.435 & 0.427 & 0.453 & 0.410
			& 0.447 & 0.447 & 0.457 & 0.442 & 0.447 & 0.433 \\
			BM25 & SB
			& 0.426 & 0.421 & 0.450 & 0.409 & 0.446 & 0.397
			& 0.437 & 0.420 & 0.454 & 0.425 & 0.455 & 0.417 \\
			BERT & B
			& 0.416 & 0.432 & 0.399 & 0.394 & 0.440 & 0.416
			& 0.429 & 0.435 & 0.428 & 0.425 & 0.453 & 0.406 \\
			RoBERTa & B
			& 0.406 & 0.415 & 0.394 & 0.399 & 0.423 & 0.392
			& 0.397 & 0.409 & 0.390 & 0.382 & 0.416 & 0.381 \\
			Rank1-0.5B & RTC
			& 0.371 & 0.377 & 0.384 & 0.342 & 0.375 & 0.376
			& 0.363 & 0.355 & 0.402 & 0.336 & 0.353 & 0.371 \\
			\bottomrule
		\end{tabular}%
	}
\end{table*}

\subsection{The Quality--Latency Frontier} \label{sec:app:latency_frontier}

\begin{figure}[!tbp]
	\centering
	\includegraphics[width=\linewidth]{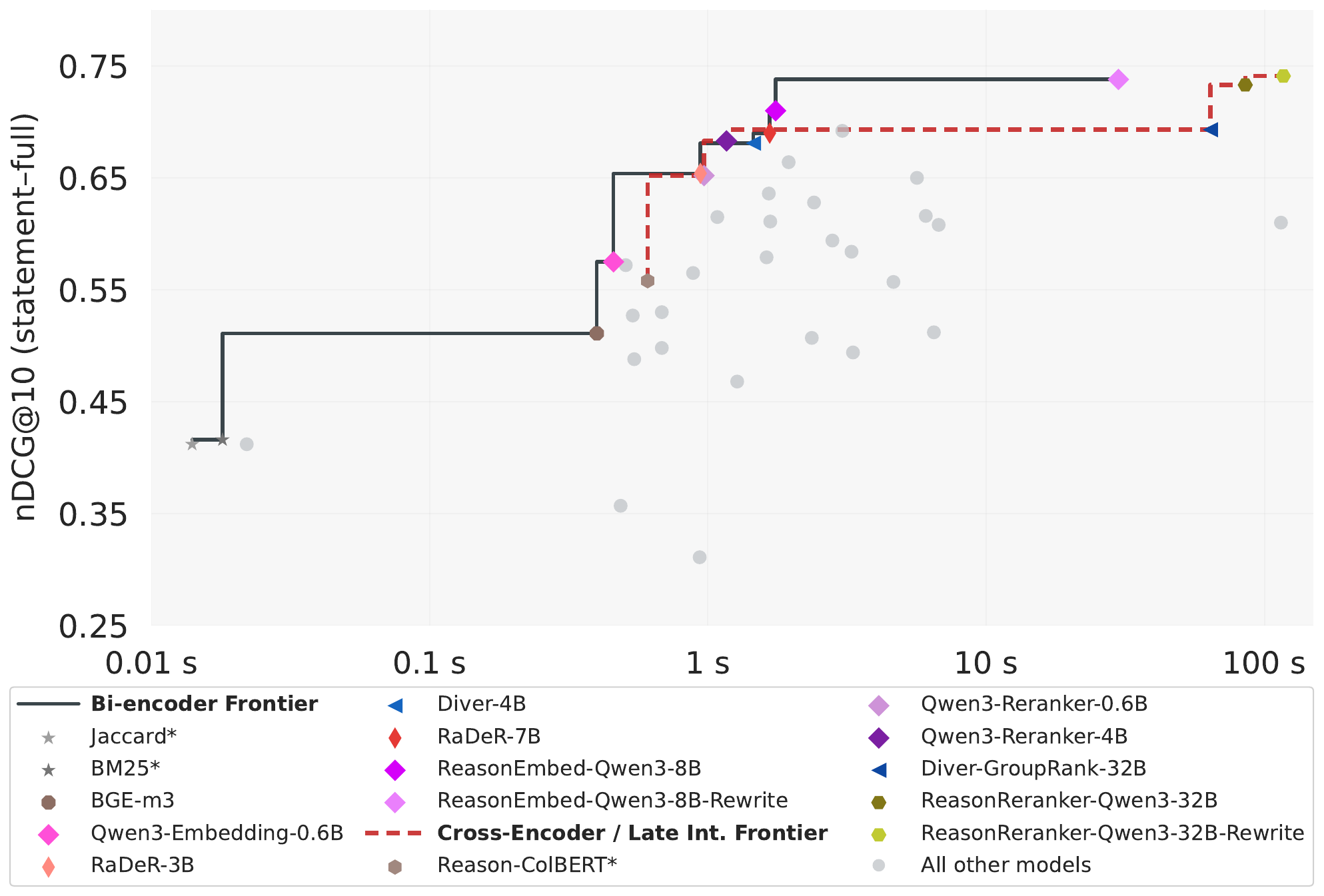}
	\caption{nDCG@10 in the
			statement--full setting against median wall-clock time per query.
			Systems marked \textbf{*} are incompatible with vLLM and were run in their
			own frameworks instead. We trace the frontiers for both cross-encoders and late-interaction models.}
	\label{fig:quality-latency}
\end{figure}

\cref{fig:quality-latency} plots each system's \bench{} score in the statement--full setting against its median wall-clock time per query, both measured under the protocol of \cref{sec:app:timing}. We say a system lies on the \emph{quality--latency Pareto frontier} if no other system is simultaneously at least as accurate and at least as fast. We trace the bi-encoders and cross-encoders frontiers separately, since the two architecture classes solve the reranking task with fundamentally different amounts of computation. The corresponding names are marked in \cref{tab:statement-full}, \bifront{underlined} for the bi-encoder frontier and \crossfront{italicized} for the cross-encoder and late-interaction one.

\paragraph{The bi-encoder frontier is narrow}

The bi-encoder frontier spans more than three orders of magnitude in latency, from \textsc{Jaccard} ($0.014$s, $0.412$) to \textsc{ReasonEmbed-Qwen3-8B-Rewrite} ($29.93$s, $0.738$), but is almost empty in between. The only systems faster than $0.03$s are the lexical baselines, and no system at all occupies the range between $0.022$s and $0.399$s. The interesting region is therefore a narrow band: \textsc{BGE-m3} ($0.399$s) reaches $0.511$ and \textsc{Qwen3-Embedding-0.6B} ($0.472$s) reaches $0.575$, after which the reasoning-trained \textsc{RaDeR-3B}, \textsc{Diver-4B}, \textsc{RaDeR-7B} and \textsc{ReasonEmbed-Qwen3-8B} climb from $0.654$ to $0.710$ between $0.951$s and $1.751$s. The frontier thus gains $0.056$ nDCG@10 within less than a factor of two in latency, consistent with \cref{sec:results:main}.

\paragraph{Joint scoring is cheap at small scale and very expensive at large scale}

On the other frontier, \textsc{Reason-ModernColBERT} anchors the cheap end ($0.607$s, $0.558$), followed by \textsc{Qwen3-Reranker-0.6B} ($0.957$s, $0.652$) and \textsc{Qwen3-Reranker-4B} ($1.167$s, $0.683$). \textsc{Qwen3-Reranker-4B} is both faster than \textsc{ReasonEmbed-Qwen3-8B} and only $0.027$ behind it, so reading query and document together is not intrinsically costly at moderate scale. Above $0.683$, however, the frontier is held exclusively by $32$B systems two orders of magnitude slower: \textsc{Diver-GroupRank-32B} adds $0.010$ over \textsc{Qwen3-Reranker-4B} for $55\times$ its latency, and \textsc{ReasonReranker-Qwen3-32B} and its rewriting variant reach $0.733$ and $0.741$ at $85.44$s and $117.06$s per query.

\paragraph{The strongest system is not the practical one}

Taken together, the two frontiers show that the top of \cref{tab:statement-full} is bought at a steep price. \textsc{ReasonReranker-Qwen3-32B-Rewrite} scores $0.741$ at $117.06$s per query, while \textsc{ReasonEmbed-Qwen3-8B} comes within $0.031$ of it at $1.751$s, a $67\times$ difference in cost. As noted in \cref{sec:app:timing}, this understates the gap in deployed cost, since a bi-encoder's document embeddings are computed once offline and reused, whereas a reranker rescores every candidate for every query.

\subsection{\bench{} Score over Time}

\begin{figure}[!tbp]
	\centering
	\includegraphics[width=\linewidth]{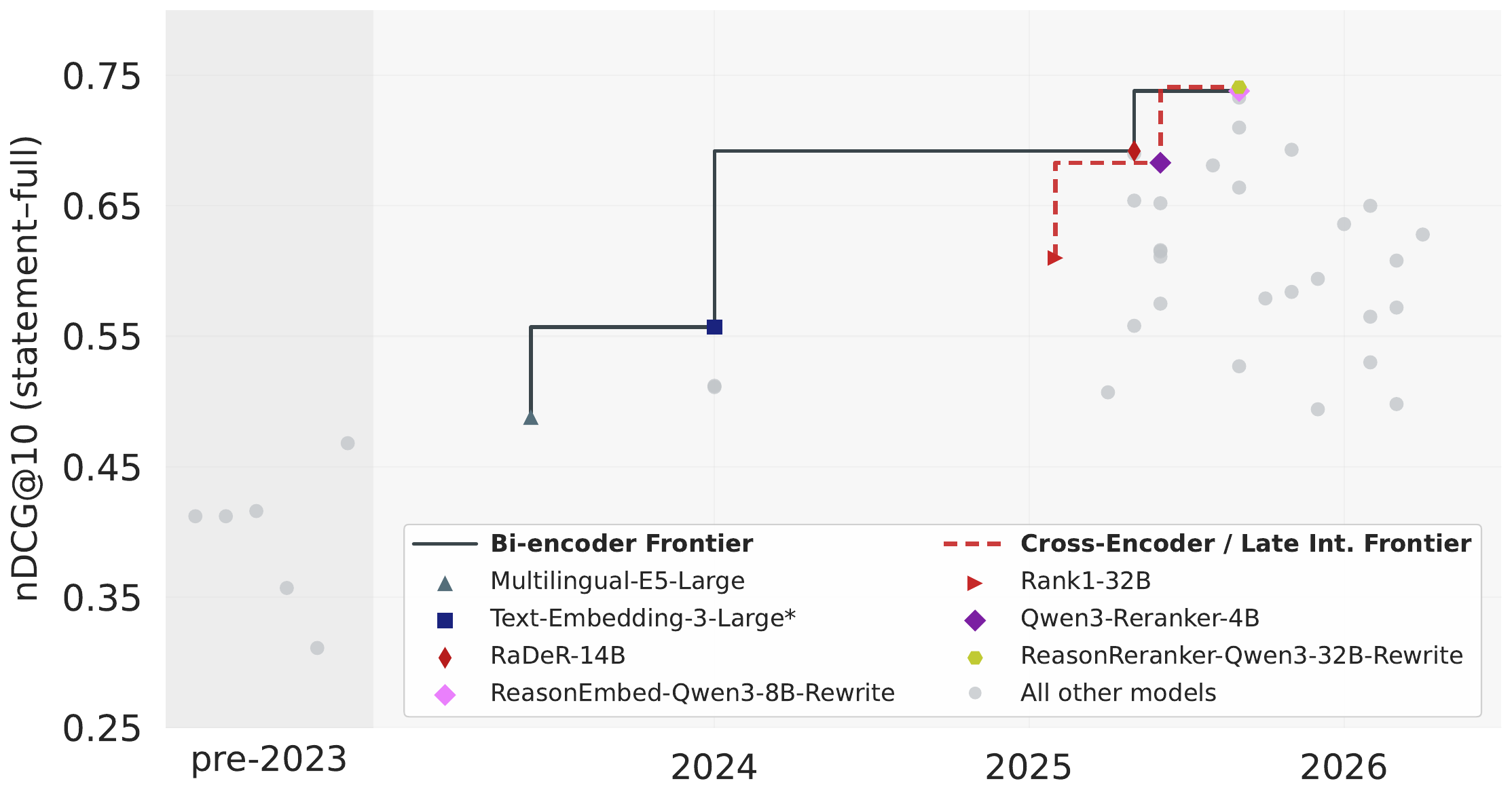}
	\caption{\textbf{\bench{} score against release date.} nDCG@10 in the
	statement--full setting against the month a model was released publicly;
	models predating 2023 share the shaded band on the left. Two
	state-of-the-art traces are shown: solid black for bi-encoders, dashed
	red for cross-encoders and late-interaction models. }
	\label{fig:quality-release}
\end{figure}

In \cref{fig:quality-release} we aim to show how \bench{} scores change over time. To do this, for each model, we plot its score in the statement--full setting against its public release date.

We track two types of models separately: on one hand, retrieval (bi-encoder) models and on the other, cross-encoder and late-interaction models. Most models do not sit on the frontier line -- \cref{fig:quality-release} shows these as grey circles. We observe that many recent models score lower than considerably older ones, so a recent release date alone does not imply strong \bench{} performance.

\subsection{How instructions influence \bench{}'s scores} \label{sec:app:instruction_effect}

By default, we evaluate every model on \bench{} with a default task instruction (when available), inside its documented input format, as described in \cref{sec:app:embedding_details}, so as to capture each system's native representation of mathematical content rather than its instruction-following ability. To measure the effect of instructions, we further re-evaluate every model that supports them -- embedding models and rerankers alike -- under the three instructions of \cref{sec:app:instruction_prompts}. We use the model's dedicated instruction field, if one exists, and we prepend the instruction to the query otherwise. In both cases the instruction is applied to the query only, and documents are encoded without one. We exclude \textsc{Approach Zero}, \textsc{BM25}, \textsc{Jaccard}, and \textsc{TF-IDF}, which have no instruction mechanism and for which lexical matching of the instruction text could only add noise.

We observe that the best prompt varies across models, and for most models the three prompts sit within a few points of nDCG@10 of the no-instruction baseline. The clearest patterns are at the extremes, as models predating instruction-tuned retrieval are hurt by instructions -- \textsc{BERT}, \textsc{RoBERTa}, \textsc{Multilingual-E5-Large} and both \textsc{Text-Embedding-3} models score best without one. On the other hand, generalist instruction-tuned models tend to gain modestly from being told the task. Reasoning-specialised systems benefit from instructions less often, as they are already trained for reasoning tasks.

\begin{table}[t]
	\centering
	\footnotesize
	\caption{Effect of task instructions on \bench{} in the statement--full setting. All numbers are reported as nDCG@10 over the 1000 queries. ``No Instr.'' is the setting reported in \cref{tab:statement-full}. I1--I3 are the instruction prompts of \cref{sec:app:instruction_prompts}}
	\label{tab:instruction-arms}
	\renewcommand{\arraystretch}{1.05}
	\setlength{\tabcolsep}{4pt}
	\newcommand{\beforerulepad}{\\[4pt]}
	\newcommand{\afterrulepad}{\rule{0pt}{\dimexpr\ht\strutbox+4pt\relax}}
	\newcommand{\tightmidrule}{\specialrule{\lightrulewidth}{0pt}{0pt}}
	\newcolumntype{R}{>{$}r<{$}}
	\resizebox{\linewidth}{!}{
		\begin{tabular}{@{}l c RRRR@{}}
			\toprule
			& \multicolumn{1}{c}{\makecell{Type}}
			& \multicolumn{1}{c}{\makecell{No Instr.}}
			& \multicolumn{1}{c}{\makecell{I1}}
			& \multicolumn{1}{c}{\makecell{I2}}
			& \multicolumn{1}{c}{\makecell{I3}} \beforerulepad
			\tightmidrule
			\afterrulepad ReasonReranker-Qwen3-32B-Rewrite & RTWC & 0.741 & 0.747 & \mathbf{0.748} & 0.743 \\
			ReasonEmbed-Qwen3-8B-Rewrite & RTWB & \mathbf{0.738} & 0.733 & 0.729 & 0.735 \\
			ReasonReranker-Qwen3-32B & RTC & 0.733 & 0.750 & \mathbf{0.756} & 0.732 \\
			ReasonEmbed-Qwen3-8B & RB & 0.710 & 0.726 & 0.725 & \mathbf{0.727} \\
			Diver-GroupRank-32B & RTC & 0.693 & 0.677 & 0.675 & \mathbf{0.694} \\
			RaDeR-14B & RB & \mathbf{0.692} & 0.674 & 0.666 & 0.687 \\
			RaDeR-7B & RB & \mathbf{0.690} & 0.683 & 0.669 & 0.678 \\
			Qwen3-Reranker-4B & C & 0.683 & 0.705 & \mathbf{0.706} & 0.686 \\
			Diver-4B & RB & \mathbf{0.681} & 0.675 & 0.674 & 0.678 \\
			Qwen3-Reranker-8B & C & 0.666 & \mathbf{0.702} & 0.699 & 0.675 \\
			ReasonEmbed-Llama-3.1-8B & RB & 0.664 & \mathbf{0.674} & 0.669 & 0.674 \\
			RaDeR-3B & RB & \mathbf{0.654} & 0.643 & 0.631 & 0.645 \\
			Qwen3-Reranker-0.6B & C & \mathbf{0.652} & 0.648 & 0.648 & 0.647 \\
			SPLADE-Code-8B & SB & \mathbf{0.650} & 0.643 & 0.639 & 0.635 \\
			Octen-8B & B & 0.636 & \mathbf{0.648} & 0.644 & 0.644 \\
			Diver-0.6B & RB & \mathbf{0.633} & 0.628 & 0.632 & 0.632 \\
			Octen-4B & B & 0.632 & \mathbf{0.644} & 0.639 & 0.636 \\
			Gemini-2 & AB & 0.628 & 0.636 & \mathbf{0.637} & 0.636 \\
			Gemini-001 & AB & \mathbf{0.616} & 0.604 & 0.602 & 0.603 \\
			Qwen3-Embedding-4B & B & 0.615 & \mathbf{0.633} & 0.629 & 0.621 \\
			Qwen3-Embedding-8B & B & 0.611 & \mathbf{0.628} & 0.627 & 0.616 \\
			Rank1-32B & RTC & 0.610 & \mathbf{0.685} & 0.674 & 0.613 \\
			Harrier-27b & B & 0.608 & \mathbf{0.634} & 0.625 & 0.614 \\
			SPLADE-Code-0.6B & SB & \mathbf{0.595} & 0.577 & 0.575 & 0.587 \\
			INF-Retriever-v1-Pro & RB & 0.594 & \mathbf{0.617} & 0.609 & 0.610 \\
			KaLM-12B-2511 & B & 0.584 & \mathbf{0.594} & 0.586 & 0.593 \\
			LLaMa-Nemotron-8B & B & 0.579 & 0.586 & \mathbf{0.600} & 0.577 \\
			Qwen3-Embedding-0.6B & B & 0.575 & 0.579 & \mathbf{0.581} & 0.579 \\
			Harrier-0.6b & B & \mathbf{0.572} & 0.571 & 0.571 & 0.565 \\
			Jina-v5-Small & B & 0.565 & \mathbf{0.566} & 0.564 & 0.563 \\
			Reason-ColBERT & RL & \mathbf{0.558} & 0.545 & 0.546 & 0.553 \\
			Text-Embedding-3-Large & AB & \mathbf{0.557} & 0.553 & 0.539 & 0.545 \\
			Rank1-7B & RTC & 0.551 & 0.585 & \mathbf{0.596} & 0.553 \\
			Jina-v5-Nano & B & \mathbf{0.530} & 0.523 & 0.522 & 0.526 \\
			GTE-ColBERT & L & 0.527 & 0.524 & 0.524 & \mathbf{0.530} \\
			Gemma-300m & B & \mathbf{0.527} & 0.523 & 0.517 & 0.525 \\
			Text-Embedding-3-Small & AB & \mathbf{0.512} & 0.492 & 0.478 & 0.496 \\
			BGE-m3 & B & \mathbf{0.511} & 0.498 & 0.492 & 0.505 \\
			ReasonIR-8B & RB & 0.507 & \mathbf{0.568} & 0.562 & 0.567 \\
			Harrier-270m & B & 0.498 & 0.493 & \mathbf{0.499} & 0.495 \\
			INF-X-Retriever & RWB & \mathbf{0.494} & 0.454 & 0.456 & 0.472 \\
			Multilingual-E5-Large & B & \mathbf{0.488} & 0.412 & 0.473 & 0.485 \\
			Rank1-0.5B & RTC & \mathbf{0.365} & 0.353 & 0.344 & 0.318 \\
			BERT & B & \mathbf{0.357} & 0.330 & 0.336 & 0.352 \\
			RoBERTa & B & \mathbf{0.311} & 0.290 & 0.290 & 0.300 \\
			\bottomrule
		\end{tabular}
	}
	
\end{table}

\subsection{Comparison of Mathematical vs. Textual Content in Retrieval} \label{sec:app:math_vs_words}

\begin{table}[!tp]
	\centering
	\footnotesize

	\caption{Percentage of targets where the equation-only representation is more similar to the relevant candidates than the word-only representation (M$>$W, all domains), per instruction prompt. I1, I2, and I3 correspond to the three respective instruction prompts described in \cref{sec:app:instruction_prompts}.}
	\label{tab:math-vs-words-instruct}
	\renewcommand{\arraystretch}{1.05}
	\setlength{\tabcolsep}{4pt}

	\resizebox{\linewidth}{!}{
		\begin{tabular}{@{}lcccc@{}}
			\toprule
			Model & No Instr. & I1 & I2 & I3 \\
			\midrule
			BERT & 87.5 & 92.6 & 93.8 & 89.5 \\
			Multilingual-E5-Large & 70.9 & 61.8 & 63.9 & 60.2 \\
			Reason-ColBERT & 70.7 & 64.8 & 63.9 & 68.4 \\
			GTE-ColBERT & 69.7 & 69.6 & 69.6 & 69.6 \\
			ReasonIR-8B & 66.5 & 66.9 & 66.3 & 66.2 \\
			RoBERTa & 66.3 & 86.6 & 86.3 & 81.4 \\
			BGE-m3 & 63.1 & 58.7 & 59.6 & 56.6 \\
			Diver-0.6B & 62.0 & 58.9 & 61.9 & 60.7 \\
			SPLADE-Code-0.6B & 62.0 & 51.2 & 51.0 & 49.1 \\
			LLaMa-Nemotron-8B & 61.8 & 60.2 & 63.7 & 57.8 \\
			SPLADE-Code-8B & 61.8 & 59.2 & 56.7 & 54.3 \\
			ReasonEmbed-Qwen3-8B & 61.6 & 55.9 & 57.3 & 58.3 \\
			ReasonEmbed-Llama-3.1-8B & 60.8 & 57.8 & 57.7 & 59.3 \\
			Text-Embedding-3-Large & 60.2 & 58.4 & 56.9 & 56.3 \\
			Text-Embedding-3-Small & 60.1 & 50.6 & 49.7 & 51.5 \\
			INF-Retriever-v1-Pro & 58.5 & 61.2 & 59.6 & 56.9 \\
			Harrier-0.6b & 58.0 & 57.3 & 57.5 & 57.8 \\
			KaLM-12B-2511 & 57.8 & 52.4 & 53.7 & 56.3 \\
			Qwen3-Embedding-4B & 56.9 & 56.8 & 52.0 & 52.5 \\
			Harrier-270m & 56.9 & 54.2 & 50.6 & 52.9 \\
			Jina-v5-Nano & 56.3 & 53.4 & 52.4 & 54.9 \\
			Jina-v5-Small & 56.0 & 55.3 & 54.7 & 57.4 \\
			ReasonEmbed-Qwen3-8B-Rewrite & 55.6 & 56.2 & 55.8 & 54.8 \\
			Octen-8B & 54.9 & 54.9 & 53.7 & 54.5 \\
			Harrier-27b & 54.7 & 56.9 & 59.2 & 55.6 \\
			RaDeR-7B & 54.7 & 49.5 & 52.4 & 53.9 \\
			RaDeR-14B & 54.6 & 55.1 & 55.2 & 53.1 \\
			RaDeR-3B & 54.6 & 43.4 & 44.7 & 48.6 \\
			Qwen3-Embedding-0.6B & 54.3 & 51.2 & 53.4 & 50.8 \\
			Qwen3-Embedding-8B & 52.4 & 55.8 & 56.1 & 57.5 \\
			Octen-4B & 51.8 & 53.1 & 51.2 & 52.2 \\
			Gemini-2 & 49.2 & 60.8 & 61.4 & 56.7 \\
			Gemma-300m & 48.7 & 47.6 & 42.4 & 47.5 \\
			Qwen3-Reranker-0.6B & 48.6 & 47.6 & 47.4 & 49.3 \\
			Qwen3-Reranker-4B & 48.2 & 49.9 & 42.4 & 45.1 \\
			Diver-4B & 47.3 & 51.0 & 53.7 & 50.6 \\
			INF-X-Retriever & 45.6 & 51.5 & 52.8 & 49.0 \\
			Rank1-0.5B & 45.1 & 26.3 & 28.2 & 41.2 \\
			Gemini-001 & 40.4 & 52.7 & 53.1 & 50.3 \\
			Qwen3-Reranker-8B & 39.5 & 48.8 & 39.3 & 42.1 \\
			ReasonReranker-Qwen3-32B-Rewrite & 38.9 & 37.5 & 39.5 & 41.3 \\
			Diver-GroupRank-32B & 36.4 & 37.7 & 42.6 & 32.4 \\
			RaDeR-Reranker-7B & 27.1 & 27.7 & 33.1 & 29.4 \\
			ReasonReranker-Qwen3-32B & 20.6 & 22.0 & 23.9 & 21.5 \\
			Rank1-7B & 20.5 & 34.6 & 36.1 & 22.8 \\
			Rank1-32B & 16.2 & 31.6 & 33.5 & 17.6 \\
			\bottomrule
		\end{tabular}%
	}
\end{table}

We now provide complementary details and results for the experiment comparing the importance of mathematical versus textual content in the retrieval task, as described in \cref{sec:results:math_vs_words}.

\paragraph{Methodology and motivation}

To investigate whether mathematical notation or surrounding textual content contributes more to retrieving relevant documents, we conduct the following experiment using our benchmark.

For each query, we extract its mathematical expressions and its remaining textual content from the problem statement alone first using \textsc{GPT-OSS-120B} to ensure every mathematical expression is properly delimited, then applying regex-based segmentation. We discard queries for which either component is empty, leaving 969 of the original 1000 query problems. For each retrieval method, including both embedding-based and classical approaches, we then compute two relevance estimates for every remaining query.

First, we compare the query's mathematical content with the problem-and-solution text of each of its five highest-rated relevant candidate problems according to the benchmark. We compute the relevance score assigned by the retrieval method for each of these five candidates and take their average. We repeat the same procedure using only the query's textual content. Thus, each query receives two average relevance scores: one based on its mathematical content and one based on its textual content.

\begin{figure*}
    \centering
    \includegraphics[width=\linewidth]{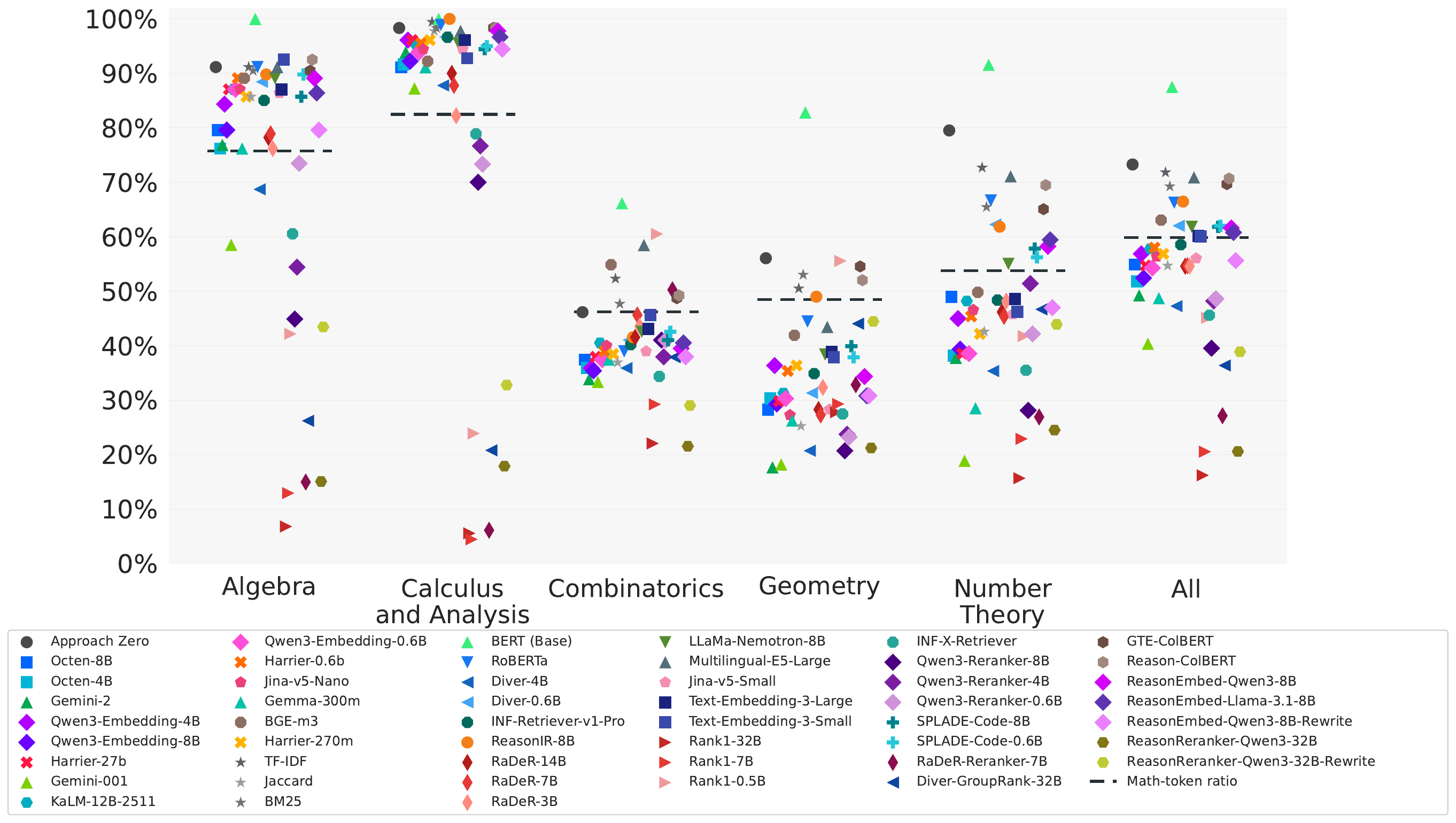}
    \caption{Percentage of query problems for which the mathematical component receives a higher average relevance score than the textual component, across retrieval methods.}
    \label{fig:math-vs-words-all-models}
\end{figure*}

Finally, for each query and retrieval method, we record whether the mathematical or textual component receives the higher average relevance score. Since these scores are averaged real-valued quantities, ties are extremely rare and do not affect the analysis. For each retrieval method, we report the percentage of queries for which the mathematical component is judged more relevant than the textual component. The results for all evaluated embedding models and classical retrieval methods are shown in \cref{fig:math-vs-words-all-models}.

We tokenize mathematical expressions using the Approach Zero mathematics tokenizer, since it captures the deeper mathematical meaning of an expression rather than treating it as plain, whitespace-delimited text. This choice also benefits token-based methods such as Jaccard similarity and TF-IDF, for which appropriate tokenization is essential.

Finally, in \cref{tab:math-vs-words-instruct} we present the results for the same experiment when the models (where applicable) are given the specific instructions as described in \cref{sec:app:instruction_prompts}.

\subsection{Deduplication Results} \label{sec:app:dedup}

\begin{table}[!tp]
	\centering
	\scriptsize
	\caption{Ability of IR systems to discover a rephrased, equivalent version of the query, over all evaluated bi-encoders. \cref{tab:dedup-all-docs} shows a representative subset of these rows.}
	\label{tab:dedup-all-docs-full}
	\renewcommand{\arraystretch}{1.05}
	\setlength{\tabcolsep}{2.5pt}
	\newcolumntype{R}{>{$}r<{$}}
	\resizebox{\columnwidth}{!}{%
		\begin{tabular}{@{}l R R RRRR@{}}
			\toprule
			& \multicolumn{1}{c}{\bench{}}
			& \multicolumn{1}{c}{Median}
			& \multicolumn{1}{c}{Top 1\%}
			& \multicolumn{1}{c}{Top 4\%}
			& \multicolumn{1}{c}{Top 16\%}
			& \multicolumn{1}{c}{Top 64\%} \\
			\midrule
			ReasonEmbed-Qwen3-8B-Rewrite & \mathbf{0.738} & \mathbf{2} & 45.2 & 68.0 & 84.9 & \mathbf{95.7} \\
			ReasonEmbed-Qwen3-8B & 0.710 & \mathbf{2} & \mathbf{45.4} & \mathbf{71.3} & \mathbf{86.9} & 95.1 \\
			RaDeR-14B & 0.692 & \mathbf{2} & 39.3 & 66.7 & 85.0 & 93.6 \\
			RaDeR-7B & 0.690 & 3 & 32.5 & 60.0 & 79.2 & 91.3 \\
			Diver-4B & 0.681 & 3 & 32.6 & 57.8 & 77.6 & 90.1 \\
			ReasonEmbed-Llama-3.1-8B & 0.664 & 4 & 32.6 & 52.0 & 69.6 & 84.6 \\
			RaDeR-3B & 0.654 & 6 & 24.3 & 44.7 & 67.1 & 83.8 \\
			Octen-8B & 0.636 & 7 & 23.5 & 43.0 & 59.8 & 77.9 \\
			Diver-0.6B & 0.632 & 5 & 28.3 & 49.1 & 68.9 & 83.5 \\
			Octen-4B & 0.632 & 12 & 21.4 & 38.5 & 53.9 & 72.1 \\
			Gemini-2 & 0.628 & 6 & 26.4 & 45.1 & 64.1 & 80.3 \\
			Gemini-001 & 0.616 & 3 & 34.1 & 55.8 & 75.5 & 87.7 \\
			Qwen3-Embedding-4B & 0.615 & 19 & 17.5 & 30.6 & 48.2 & 68.0 \\
			Qwen3-Embedding-8B & 0.611 & 48 & 12.2 & 23.3 & 37.4 & 54.6 \\
			Harrier-27b & 0.608 & 182 & 7.2 & 14.2 & 23.5 & 37.2 \\
			INF-Retriever-v1-Pro & 0.594 & 143 & 9.1 & 18.1 & 27.9 & 41.1 \\
			KaLM-12B-2511 & 0.584 & 145 & 6.9 & 14.0 & 23.8 & 39.4 \\
			LLaMa-Nemotron-8B & 0.579 & 175 & 8.8 & 16.3 & 27.4 & 38.6 \\
			Qwen3-Embedding-0.6B & 0.575 & 70 & 11.1 & 21.8 & 33.9 & 48.9 \\
			Harrier-0.6b & 0.572 & 119 & 9.1 & 17.2 & 28.6 & 43.4 \\
			Jina-v5-Small & 0.565 & 235 & 6.1 & 12.2 & 22.4 & 36.1 \\
			Text-Embedding-3-Large & 0.557 & 116 & 9.0 & 16.1 & 27.5 & 43.1 \\
			Jina-v5-Nano & 0.530 & 619 & 4.8 & 9.3 & 17.1 & 26.7 \\
			Gemma-300m & 0.527 & 242 & 7.3 & 13.8 & 23.0 & 35.6 \\
			Text-Embedding-3-Small & 0.512 & 2729 & 1.9 & 4.2 & 7.8 & 14.1 \\
			BGE-m3 & 0.511 & 814 & 3.5 & 7.7 & 13.4 & 22.2 \\
			ReasonIR-8B & 0.507 & 1598 & 3.7 & 7.6 & 13.1 & 19.5 \\
			Harrier-270m & 0.498 & 765 & 4.9 & 9.3 & 16.4 & 25.4 \\
			INF-X-Retriever & 0.494 & 739 & 2.6 & 5.7 & 10.1 & 19.5 \\
			Multilingual-E5-Large & 0.488 & 913 & 4.4 & 8.8 & 15.3 & 25.4 \\
			BM25 & 0.416 & 436 & 8.4 & 14.2 & 21.4 & 32.1 \\
			Jaccard & 0.412 & 5505 & 4.6 & 7.7 & 13.2 & 17.9 \\
			TF-IDF & 0.412 & 12800 & 2.1 & 3.9 & 5.8 & 9.9 \\
			BERT & 0.357 & 6047 & 2.9 & 4.8 & 7.1 & 10.8 \\
			RoBERTa & 0.311 & 12368 & 2.8 & 4.4 & 6.4 & 9.5 \\
			\bottomrule
		\end{tabular}%
	}
\end{table}

We give the setup of this experiment in \cref{sec:app:dedup_setup}, and report here its full results, in both the retrieval and the reranking variant.

\paragraph{Retrieval over the full corpus}

\cref{tab:dedup-all-docs-full} reports the results for all $35$ evaluated bi-encoders, searching over the entire $71$K-document pool, of which \cref{tab:dedup-all-docs} shows a representative subset. The agreement with \bench{} holds across every metric we report. The Spearman rank correlation between \bench{} nDCG@10 and deduplication performance is $\rho = 0.96$, when measured against the models' negated median rank and $\rho = 0.92$ against top-$1$ accuracy. It further rises to $\rho = 0.96$ at top-$64$.

We further note that deduplication separates systems far more sharply than \bench{} itself. Over the same set of models, \bench{} nDCG@10 spans a factor of $2.4$, from $0.311$ to $0.738$, whereas the median rank of the duplicate spans more than three orders of magnitude, from $2$ for the strongest systems to $12800$ for TF-IDF. This is expected, as recovering a single rewritten problem out of $71$K candidates is a far more demanding query than ranking a $150$-document pool. It does mean, however, that modest differences on \bench{} can correspond to substantial differences in downstream utility.

Lexical matching behaves inconsistently on this task, as \textsc{TF-IDF} and \textsc{Jaccard} sit at the very bottom by median rank, at $12800$ and $5505$ respectively, which is expected given that our rewrites deliberately alter surface form. \textsc{BM25}, however, attains a median rank of $436$, ahead of dense retrievers such as \textsc{Multilingual-E5-Large} ($913$), \textsc{ReasonIR-8B} ($1598$) and \textsc{Text-Embedding-3-Small} ($2729$), all of which clearly outrank it on \bench{}. This suggests that our rewrites preserve enough rare mathematical notation for inverse-document-frequency-weighted term matching to remain partially effective, while the weaker dense models fail to represent that content at all.

Finally, there are some minor exceptions to the overall agreement between the two tasks. In particular, \textsc{Harrier-27b} and \textsc{Qwen3-Embedding-8B} score higher on \bench{} than \textsc{Harrier-0.6b} and \textsc{Qwen3-Embedding-4B} respectively, yet have a worse median duplicate rank, while \textsc{Gemini-001} is $12$th on \bench{} but tied for $4$th by median rank. Such disagreements are nonetheless mild, as they mostly reorder models whose \bench{} confidence intervals already overlap (\cref{sec:app:confidence_intervals}), and reflect the tasks' inherent differences.

\paragraph{Reranking within the candidate set}

Restricting the search to each query's own $150$ candidate documents makes the task tractable for cross-encoders and generative rerankers as well, at the cost of a much coarser measurement. \cref{tab:dedup-per-query} reports the median rank assigned to the rephrase in this setting, together with the percentage of queries for which it falls within the top-$\{1,4,16,64\}$ candidates, over all $49$ systems we evaluate.

\begin{table}[!tp]
	\centering
	\scriptsize
	\caption{Ability of IR systems to discover a rephrased, equivalent version of the query, when the search is restricted to each query's own $150$-document candidate set, over all evaluated systems. \cref{tab:dedup-all-docs-full} reports the same experiment over the full $71$K-document pool.}
	\label{tab:dedup-per-query}
	\renewcommand{\arraystretch}{1.05}
	\setlength{\tabcolsep}{2.5pt}
	\newcolumntype{R}{>{$}r<{$}}
	\resizebox{\columnwidth}{!}{%
		\begin{tabular}{@{}l R R RRRR@{}}
			\toprule
			& \multicolumn{1}{c}{\bench{}}
			& \multicolumn{1}{c}{Median}
			& \multicolumn{1}{c}{Top 1\%}
			& \multicolumn{1}{c}{Top 4\%}
			& \multicolumn{1}{c}{Top 16\%}
			& \multicolumn{1}{c}{Top 64\%} \\
			\midrule
			ReasonReranker-Qwen3-32B-Rewrite & \mathbf{0.741} & \mathbf{1} & 80.3 & 96.1 & \mathbf{99.9} & \mathbf{100.0} \\
			ReasonEmbed-Qwen3-8B-Rewrite & 0.738 & \mathbf{1} & 66.6 & 88.6 & 97.8 & 99.8 \\
			ReasonReranker-Qwen3-32B & 0.733 & \mathbf{1} & \mathbf{91.3} & \mathbf{98.5} & \mathbf{99.9} & \mathbf{100.0} \\
			ReasonEmbed-Qwen3-8B & 0.710 & \mathbf{1} & 67.5 & 89.5 & 96.8 & 99.5 \\
			Diver-GroupRank-32B & 0.693 & \mathbf{1} & 90.4 & 95.2 & 98.9 & \mathbf{100.0} \\
			RaDeR-14B & 0.692 & \mathbf{1} & 63.2 & 86.7 & 96.4 & 99.5 \\
			RaDeR-7B & 0.690 & \mathbf{1} & 56.9 & 83.1 & 95.6 & 99.3 \\
			Qwen3-Reranker-4B & 0.683 & \mathbf{1} & 68.3 & 91.0 & 98.1 & \mathbf{100.0} \\
			Diver-4B & 0.681 & \mathbf{1} & 56.3 & 80.8 & 94.5 & 99.5 \\
			Qwen3-Reranker-8B & 0.666 & \mathbf{1} & 84.6 & 96.5 & 99.3 & \mathbf{100.0} \\
			ReasonEmbed-Llama-3.1-8B & 0.664 & \mathbf{1} & 52.5 & 75.4 & 92.0 & 99.0 \\
			RaDeR-3B & 0.654 & 2 & 45.2 & 73.5 & 91.7 & 99.4 \\
			Qwen3-Reranker-0.6B & 0.652 & \mathbf{1} & 73.4 & 91.0 & 98.0 & 99.7 \\
			SPLADE-Code-8B & 0.650 & 2 & 45.8 & 69.5 & 89.7 & 99.0 \\
			Octen-8B & 0.636 & 2 & 44.1 & 68.7 & 88.4 & 99.0 \\
			Diver-0.6B & 0.632 & \mathbf{1} & 50.7 & 76.3 & 92.7 & 99.4 \\
			Octen-4B & 0.632 & 2 & 39.8 & 63.6 & 86.0 & 99.0 \\
			Gemini-2 & 0.628 & 2 & 46.4 & 73.7 & 91.9 & 99.7 \\
			Gemini-001 & 0.616 & \mathbf{1} & 58.1 & 83.2 & 95.5 & 99.8 \\
			Qwen3-Embedding-4B & 0.615 & 3 & 32.6 & 57.6 & 83.3 & 98.4 \\
			Qwen3-Embedding-8B & 0.611 & 6 & 26.4 & 45.4 & 70.5 & 94.1 \\
			Rank1-32B & 0.610 & \mathbf{1} & 86.3 & 97.0 & 99.3 & \mathbf{100.0} \\
			Harrier-27b & 0.608 & 15 & 15.9 & 30.3 & 52.6 & 86.4 \\
			SPLADE-Code-0.6B & 0.595 & 5 & 27.6 & 49.1 & 73.2 & 96.2 \\
			INF-Retriever-v1-Pro & 0.594 & 11 & 19.5 & 34.7 & 58.8 & 87.4 \\
			KaLM-12B-2511 & 0.584 & 11 & 15.7 & 31.5 & 57.1 & 90.7 \\
			LLaMa-Nemotron-8B & 0.579 & 12 & 17.8 & 33.8 & 56.2 & 84.5 \\
			Qwen3-Embedding-0.6B & 0.575 & 7 & 23.5 & 42.4 & 65.6 & 91.0 \\
			Harrier-0.6b & 0.572 & 8 & 22.1 & 37.7 & 61.7 & 91.7 \\
			Jina-v5-Small & 0.565 & 15 & 15.5 & 29.4 & 52.6 & 82.9 \\
			Reason-ModernColBERT & 0.558 & 11 & 19.9 & 36.2 & 56.9 & 84.7 \\
			Text-Embedding-3-Large & 0.557 & 8 & 19.9 & 39.3 & 65.0 & 91.8 \\
			Rank1-7B & 0.552 & \mathbf{1} & 81.0 & 94.4 & 99.2 & \mathbf{100.0} \\
			Jina-v5-Nano & 0.530 & 26 & 12.4 & 24.7 & 42.4 & 70.9 \\
			GTE-ModernColBERT & 0.527 & 7 & 26.6 & 44.1 & 63.4 & 90.9 \\
			Gemma-300m & 0.527 & 13 & 17.9 & 33.4 & 55.7 & 83.8 \\
			Text-Embedding-3-Small & 0.512 & 56 & 6.1 & 13.3 & 25.2 & 54.2 \\
			BGE-m3 & 0.511 & 24 & 11.2 & 21.9 & 41.7 & 76.7 \\
			ReasonIR-8B & 0.507 & 44 & 9.7 & 17.0 & 32.2 & 61.5 \\
			Harrier-270m & 0.498 & 25 & 12.9 & 24.3 & 41.8 & 73.8 \\
			INF-X-Retriever & 0.494 & 25 & 8.6 & 20.8 & 43.0 & 79.9 \\
			Multilingual-E5-Large & 0.488 & 26 & 11.6 & 23.0 & 41.4 & 71.3 \\
			Approach Zero & 0.468 & \mathbf{1} & 61.8 & 77.6 & 86.8 & 95.6 \\
			BM25 & 0.416 & 11 & 21.6 & 38.6 & 57.5 & 85.6 \\
			Jaccard & 0.412 & 46 & 11.5 & 19.8 & 33.7 & 57.4 \\
			TF-IDF & 0.412 & 68 & 6.1 & 12.4 & 23.4 & 48.7 \\
			Rank1-0.5B & 0.364 & 26 & 8.8 & 18.2 & 36.9 & 80.0 \\
			BERT & 0.357 & 44 & 8.2 & 15.2 & 26.9 & 60.5 \\
			RoBERTa & 0.311 & 62 & 7.6 & 13.1 & 24.4 & 51.3 \\
			\bottomrule
		\end{tabular}%
	}
\end{table}

The broad agreement with \bench{} survives, but the measurement is much less discriminative. The Spearman rank correlation between \bench{} nDCG@10 and the negated median rank drops to $\rho = 0.83$, from $\rho = 0.96$ over the full pool. The main reason is saturation, as $17$ of the $49$ systems tie at a median rank of $1$, $21$ place the rephrase in the top-$64$ for at least $99\%$ of queries, and the median rank spans only $1$ to $68$, against more than three orders of magnitude over the $71$K-document pool. With only $150$ candidates and a target that is a paraphrase of the query itself, the task is simply too easy to separate the stronger half of the field, which is why we treat the full-corpus variant as the primary measurement.

The one systematic effect the restricted setting does expose is that cross-encoders are unusually well suited to this task, which makes sense as this is suited to the reranking scale. The eight highest top-$1$ accuracies all belong to cross-encoders or generative rerankers, led by \textsc{ReasonReranker-Qwen3-32B} at $91.3\%$ and \textsc{Diver-GroupRank-32B} at $90.4\%$, ahead of the best bi-encoder, \textsc{ReasonEmbed-Qwen3-8B}, at $67.5\%$. The effect is strong enough to override \bench{} ordering: \textsc{Rank1-32B} reaches $86.3\%$ top-$1$ and \textsc{Rank1-7B} $81.0\%$ despite \bench{} scores of only $0.610$ and $0.552$, both far above bi-encoders that outscore them on \bench{} by a wide margin. Reading query and document together lets a reranker check whether the two describe the same problem, rather than having to place a paraphrase near its source in a fixed embedding geometry. This advantage does not extend to the smallest reranker, as \textsc{Rank1-0.5B} attains a median rank of $26$ and $8.8\%$ top-$1$, in line with its \bench{} score.

Two further outliers are worth noting. \textsc{Approach Zero} attains a median rank of $1$ and $61.8\%$ top-$1$ while placing $43$rd of $49$ on \bench{}, by far the largest disagreement in either variant, which we attribute to its structural formula matching: our rewrites alter prose and variable names but leave much of the expression structure intact, and identifying a near-identical formula is precisely what the system is built for. Conversely, query rewriting, which helps on \bench{} itself, hurts here, as the rewriting variants of \textsc{ReasonReranker-Qwen3-32B} and \textsc{ReasonEmbed-Qwen3-8B} lose $11.0$ and $0.9$ points of top-$1$ accuracy relative to their base versions. This is expected, since rewriting replaces the query with a generated restatement and thereby discards the surface cues that make its duplicate identifiable.

\section{Compute Resources}\label{sec:app:compute}
This section details the computational and financial resources required to generate and evaluate our benchmark.

\subsection{Data Collation and Processing}

For the AoPS and NuminaMath datasets, the data was for the most part, readily available in a structured format, requiring only minor cleaning and deduplication. The official sources, however, required a more intensive processing pipeline, costing around \$300 in compute for OpenAI Batch API usage over the course of a total of 72 hours.

\subsection{Benchmark Generation}

After deriving the large corpus of problems as described in \cref{sec:tech:data}, the first stage of benchmark generation consists of extracting topic-based and solution-summary-based signals for each problem in the corpus. Topic extraction for all $\sim$283K problems took approximately two days using 8 H200 GPUs. Extracting summaries for all solutions in the corpus took approximately 12 hours using 4 H200 GPUs. Computing the BMA and Jaccard scores was comparatively efficient, requiring 1 hour on 25 CPU cores. Storing the resulting signals for all problem pairs in the corpus required around 300GB of disk space, where we used a more efficient byte-level storage format.

After filtering for relevant problems, the next stage uses a Swiss-style tournament together with the Bradley--Terry algorithm. Pairwise judging for the 1000 filtered queries took approximately 160 hours on 4 H200 GPU nodes.

\subsection{Applying the Benchmark}

Computing the benchmark results in \cref{tab:statement-full,tab:additional_settings} took approximately 200 hours using 4 H200 GPUs for the open-weight embedding models. For the OpenAI and Gemini models, our usage totalled approximately \$100 in API costs.


\section{Instructions and Prompts} \label{sec:app:prompts}
In this section, we make available the prompts used to generate our benchmark.

\subsection{Ontology Topic Assignment Prompt}
\begin{prompt}{Ontology topic assignment}
[SYSTEM]

You are an Expert Mathematical Annotator. You will be provided with a mathematical text(problem, solution, theorem, lesson, etc.) and a list of possible tags. You must assign relevance score for each of the tags with respect to the provided text.

# INPUT
You will be provided with:
1. A mathematical text
2. A comma-separated list of tags

# TASK
For each of the provided tags assign a real number from the interval [0.0, 1.0]. The most relevant tag to the text must have a relevance score of 1.0 and the most irrelevant must have a score of 0.0. Align the remaining tags with respect to those.

# OUTPUT SCHEMA
Constrain your output to a pure JSON with no explanations, markdown or comments. Return only pure JSON. Follow the schema - each tag name is a key in the returned JSON object with a value the relevenace score:
```json
{
	"<tagName>": <float | tag relevance score 0.0-1.0 to the provided mathematical text>,
	...
}
[USER]

# Mathematical Text:
{problem_and_solution}

With respect to the above text, assign relevance scores to the following tags and return the results in JSON:

{list_of_comma_separaed_tags}
\end{prompt}

\subsection{Extracting Short Solution Summaries Prompt}

\begin{prompt}{Extracting short solution summaries}
# INSTRUCTION
You are an expert mathematical annotator tasked with identifying the *core idea*-the central mathematical insight-from a math problem.

# GOALS
1. Identify what makes the solution work conceptually, not how to carry it out. Capture the untrial step or idea that is the greatest hint for the solution.
2. Never include any multi-step reasoning, equations, or numeric computations. Don't include any annotations that are in the solution but not in the original problem statement.
3. Never try to solve the problem on your own, and don't include your reasoning or thoughts.
4. Output a single valid JSON object matching the schema below.
5. Structure the ideas imperatively so they look like you are giving a hint to someone.
6. If the problem seems too easy or straightforward, or you can't identify a core idea, store its value as 'null' and set the 'noCoreIdea' to 'true'.

# SCHEMA
```json
{
    "noCoreIdea": <true|false>,
    "coreIdea": "<string - one short sentence (up to 30 words) naming the main insight to the problem>",
    "supportingIdeas": ["<strings - 0-3 short technique phrases>"],
    "keywords": ["<strings - 1-2 word phrases summarizing the ideas, theorems, etc. in the solution>"],
    "confidence": <0.0-1.0>
}

Here are the problem statement and solution:

Statement: {problem}

Solution: {solution}
\end{prompt}

\subsection{Pairwise Relevance Judging Prompt}
\begin{prompt}{Pairwise relevance judging}
You are an expert mathematician and analytical-reasoning engine. Your job is to decide which of two sample problems (Sample 1 or Sample 2) is *conceptually closer* to a target problem, judged by the *mathematical ideas and solution techniques actually used* (not surface wording).

  You will receive 3 complete items, each with BOTH a problem statement and its correct solution:

  [Target]
  - Problem: {target_problem}
  - Correct Solution: {target_solution}

  [Sample 1]
  - Problem: {sample1_problem}
  - Correct Solution: {sample1_solution}

  [Sample 2]
  - Problem: {sample2_problem}
  - Correct Solution: {sample2_solution}

  Your output must follow the steps and formatting rules below exactly.

  --------
  STEP 1 - Technique Extraction (do this for Target, Sample 1, Sample 2)
  For each item, read the Problem + Correct Solution together, then produce a structured list containing:

  A. Core topic tags (2-6 tags)
    Examples: "modular arithmetic", "similar triangles", "graph invariants", "generating functions", "optimization via calculus", "AM-GM", "inclusion-exclusion".

  B. Key tools / theorems / lemmas explicitly used
    Name the theorem/identity/strategy when possible (e.g., "Pigeonhole Principle", "Cauchy-Schwarz", "Euclidean algorithm", "double counting", "invariant/monovariant", "casework", "recurrence + induction").

  C. Primary solution moves (3-8 bullets)
    Describe the actual moves that drive the solution, e.g.:
    - "rewrite expression into telescoping sum"
    - "introduce substitution of variable to linearize"
    - "construct auxiliary line and apply angle chase"
    - "count objects in two ways; equate counts"
    - "prove monotonicity; then bound endpoints"

  D. Structural signature (1-3 sentences)
    Summarize the solution's backbone: what is introduced, what is transformed, and what final step seals the result.

  Label these sections clearly as:
  - Target: Technique Profile
  - Sample 1: Technique Profile
  - Sample 2: Technique Profile

  --------
  STEP 2 - Comparative Analysis
  Compare each sample to the target using *deep* similarity signals:

  1) Technique overlap:
    - Are the same kinds of tools central (not merely mentioned)?
    - Are the same transformations used (e.g., factoring vs. symmetry/invariants vs. counting)?

  2) Problem structure alignment:
    - Are the same intermediate objects introduced (auxiliary point, substitution, generating function, invariant quantity, etc.)?
    - Do both require the same "shape" of argument (e.g., construction + chase, or setup of recurrence + induction, or extremal argument + contradiction)? This criterion should be weighted lower than the technique overlap but can be a tie-breaker when techniques are similar.

  3) Difficulty is NOT the criterion:
    - Prefer shared method and structure over "hard/easy".

  4) Penalize superficial similarity:
    - Do NOT reward matching variable names, story context, or domain language if the underlying method differs. Further, algebraic computations do not imply similarity if the core method is different, or if the algebra is just a technical detail rather than the main driver.

  Produce two subsections:
  - Sample 1 vs Target -- Similarities / Differences / Overall Match Score (qualitative)
  - Sample 2 vs Target -- Similarities / Differences / Overall Match Score (qualitative)

  --------
  STEP 3 - Decision + Tie-break rules
  Choose the sample that is fundamentally closer in methods and structure.

  If it's close, use these tie-breakers in order:
  1) The sample whose *main* technique is the same as the target's main technique.
  2) The sample whose *sequence of moves* (setup -> transformation -> key lemma -> finish) matches more closely.
  3) The sample that relies on the same representation (e.g., algebraic manipulation vs. geometric configuration vs. combinatorial counting vs. graph reasoning).

  --------
  OUTPUT FORMAT (must follow)
  - You may include your extracted profiles and comparisons above.
  - Output your final decision on the last line, formatted as either:

  $\\boxed{{1}}$ -> corresponding to the first sample being more similar to the target,
  or
  $\\boxed{{2}}$ -> corresponding to the second sample being more similar to the target.
\end{prompt}

\subsection{Ordinal Judge Prompt}

\begin{prompt}{Ordinal ranking prompt for additional ordering experiments \cref{sec:app:human_evaluation}}

You are a helpful mathematical expert and technique-similarity evaluator. You will be given two mathematical items: a Target problem with its correct solution, and a Sample problem with its correct solution.

Your task is to determine how similar the two problems are based on the mathematical techniques, ideas, and solution structure actually used in their solutions --- not based on superficial wording, notation, topic labels, or story context.

You will receive:

[Target]
Problem: {target_problem}
Correct Solution: {target_solution}

[Sample]
Problem: {sample_problem}
Correct Solution: {sample_solution}

Follow the instructions below exactly.

STEP 1 - Technique Extraction
For each item, read the Problem and Correct Solution together and identify all techniques genuinely used in the solution.

For both Target and Sample, produce:

A. Core topic tags (2-8 tags)
Examples: "modular arithmetic", "similar triangles", "double counting", "generating functions", "optimization", "AM-GM", "inclusion-exclusion", "induction", "graph invariants".

B. Key tools / theorems / identities / lemmas used
Name explicit or implicit tools when possible, such as:
- Pigeonhole Principle
- Cauchy-Schwarz
- Euclidean algorithm
- Vieta's formulas
- telescoping
- contradiction
- casework
- recurrence + induction
- symmetry
- invariant / monovariant
- angle chase
- substitution
- extremal argument
- bounding

C. Primary solution moves (3-10 bullets)
Describe the actual moves that drive the solution, for example:
- rewrite the expression into a factorized form
- introduce a substitution to simplify the equation
- construct an auxiliary object
- split into cases
- count the same set in two ways
- derive a recurrence and solve it
- apply a standard inequality after normalization
- use parity or modular reduction
- reduce to a previously established identity

D. Structural signature (1-3 sentences)
Summarize the backbone of the argument: how the solution starts, what key transformation or idea unlocks it, and what final step completes the proof.

Label these sections clearly as:

Target: Technique Profile
Sample: Technique Profile

STEP 2 - Comparative Analysis
Compare the Sample to the Target using the extracted technique profiles.

Your comparison must include:

1. Technique overlap
- Which techniques are shared?
- Which shared techniques are central rather than incidental?
- Are the same theorems, transformations, or proof strategies doing the main work?

2. Technique differences
- Which important techniques appear in one but not the other?
- Do the problems rely on different core ideas even if they are in the same broad subject?

3. Structural alignment
- Do the two solutions have a similar shape, such as:
  - setup -> substitution -> simplification -> conclusion
  - construction -> theorem application -> chase -> finish
  - recurrence -> induction
  - extremal argument -> contradiction
  - counting representation -> double count -> algebraic cleanup
- This matters, but it should be weighted less than overlap in core techniques.

Important rules:
- Focus on the methods actually used in the solutions.
- Do NOT reward superficial similarity such as matching notation, similar phrasing, same answer form, or broad subject area alone.
- Do NOT treat routine algebraic manipulation as strong similarity unless it is a central technique in both.
- Weight main techniques much more heavily than minor technical steps.
- If one problem uses a tool only incidentally but the other depends on it centrally, count that as weak overlap.

STEP 3 - Similarity Score
Assign a single integer score from 0 to 100 that reflects how similar the Sample is to the Target in terms of mathematical method and solution structure.

Use this scale as guidance:

0-10: essentially no meaningful similarity in method
11-25: very weak similarity; mostly superficial or broad-topic overlap
26-40: limited similarity; some secondary techniques overlap but main approach differs
41-55: moderate similarity; at least one important method overlaps
56-70: fairly strong similarity; core ideas overlap substantially, with notable differences
71-85: strong similarity; main methods and argument structure are closely aligned
86-95: very strong similarity; nearly the same technique and solution pattern
96-100: essentially the same core method, with only minor presentation differences

Give a brief justification for the score in 2-5 sentences.

OUTPUT FORMAT
Use exactly this structure:

Target: Technique Profile
- Core topic tags: ...
- Key tools / theorems / identities / lemmas: ...
- Primary solution moves:
  - ...
  - ...
  - ...
- Structural signature: ...

Sample: Technique Profile
- Core topic tags: ...
- Key tools / theorems / identities / lemmas: ...
- Primary solution moves:
  - ...
  - ...
  - ...
- Structural signature: ...

Comparison
- Technique overlap: ...
- Technique differences: ...
- Structural alignment: ...
- Score justification: ...

Final Score:
\boxed{{N}}

where N is a single integer from 0 to 100.
\end{prompt}

\subsection{Mathematical Equation Formatting Prompt}

\begin{prompt}{Formatting mathematical equations into LaTeX}
Role: You are a LaTeX Formatting Expert specializing in mathematical notation standardization.

Task: You will be provided with mathematical text (problems or solutions). Your task is to standardize the LaTeX formatting and enclose the entire final result within a \boxed{{...}} command.

Strict Constraints:
1. DELIMITER CONVERSION: Replace all instances of \( ... \) and \\( ... \\) with standard dollar signs.
   - Use $...$ for inline text.
   - Use $$...$$ for display equations.
2. UNIVERSAL MATH TAGGING: Apply math mode ($...$) to every single mathematical element without exception.
3. CONTENT INTEGRITY: Do not solve the problem or edit the prose.
4. FINAL WRAPPING: The entire output must be contained within \boxed{{ <your_formatted_text_here> }}.
5. NO VERBOSITY: Provide ONLY the \boxed{{...}} block.

Example Transformation:
Input: If the radius r is 5, find the area. Use \( \pi \).
Output: \boxed{{If the radius $r$ is $5$, find the area. Use $\pi$.}}
\end{prompt}

\subsection{Adding Specific Task Instruction Prompts} \label{sec:app:instruction_prompts}

Here, we describe the three instruction prompts used in our experiments in \cref{sec:app:instruction_effect}. The first instruction targets the notion of similarity we study in this work:

\begin{prompt}{Instruction 1: mathematical ideas and solution techniques}
You are an expert mathematical retrieval engine. Your job is to retrieve sample problems that are *conceptually close* to a target problem, judged by the *mathematical ideas and solution techniques actually used* (not surface wording).
{query}
\end{prompt}

The second instruction instead instructs the model explicitly against the common failure modes we observe in \cref{sec:experimental}:

\begin{prompt}{Instruction 2: invariance to surface form}
Given a mathematical query, retrieve documents that share identical problem-solving structure and logical steps, even if they use entirely different variables, symbols, or natural language phrasing.
{query}
\end{prompt}

The third instruction is a generic retrieval instruction, so as to isolate the effect of general guidance.

\begin{prompt}{Instruction 3: generic retrieval control}
Find the most relevant document based on the query.
{query}
\end{prompt}

\subsection{Human Annotator Instructions} \label{sec:app:human_evaluation_instructions}

\begin{prompt}{Human Annotator Instructions}
Goal: Compare Candidate 1 and Candidate 2 against the Target problem.

Candidate 1: Choose this if Candidate 1 is more similar to the Target.

Candidate 2: Choose this if Candidate 2 is more similar to the Target.

Tie: Choose this if both candidates are equally good / bad / equally similar.

Save: Records your judgment. Completed tasks turn green.

Remove judgment: Deletes your previous answer after confirmation. This removes the task from the list of "Finished Tasks".
\end{prompt}

\section{Use of LLMs in This Work}

In this work, we have not used LLMs for research design, ideation, or decision-making. We have used LLMs to write code for assisting with plot creation and formatting, but all code was reviewed and edited by the authors. We have used LLMs to assist with polishing the writing, but all content was written by the authors and reviewed for accuracy. We have not relied on LLMs for generating any part of the bibliography, and all references were added manually by the authors.

\end{document}